\documentclass{lmcs}
\usepackage{hyperref}
\usepackage{enumerate}
\usepackage{enumitem}
\usepackage{amssymb,amsmath,amsthm} %must be before unicode-math
\usepackage[mathletters]{ucs}
\usepackage[utf8x]{inputenc}
\usepackage{bussproofs}
\usepackage{graphicx}
\usepackage{stmaryrd}
\usepackage[autosize]{dot2texi}
\usepackage{multicol}
\usepackage{tikz}
\usetikzlibrary{shapes,arrows,tikzmark,decorations.pathreplacing,calc}
\usepackage{adjustbox}

\usepackage{mathtools}
\mathtoolsset{centercolon}

\newcommand{\type}[3][]{\ensuremath{#1 \vdash #2 : #3}}
\newcommand{\subtype}[4]{\ensuremath{#1 \vdash #2 \in #3 \subset #4}}
\newcommand{\subt}[2]{\ensuremath{#1\subset#2}}
\newcommand{\lsubt}[3]{\ensuremath{#1 \in #2 \subset #3}}
\newcommand{\nsubt}[3]{\ensuremath{#1 \vdash #2 \subset #3}}
\newcommand{\case}[2]{\ensuremath{[#1 \mid #2]}}
\newcommand{\casearray}[2]{\ensuremath{\left[#1\,\middle| #2\!\!\right]}}
\newcommand{\record}[1]{\{#1\}}
\newcommand{\bnfeq}{\mathrel{::=}\;}
\newcommand{\bnfor}{\mathrel{\vert}}
\newcommand{\sem}[1]{\ensuremath{\llbracket #1 \rrbracket}}

\newcommand{\pure}{\sem{\Lambda}}
\newcommand{\formsem}{\sem{\mathcal{F}}}
\newcommand{\ord}{\mathcal{O}}
\newcommand{\ordsem}{\sem{\ord}}

\newcommand{\calN}{\ensuremath{\mathcal{N}}}
\newcommand{\calW}{\ensuremath{\mathcal{W}}}
\newcommand{\calWZ}{\ensuremath{\mathcal{W}_0}}
\newcommand{\calWZS}{\ensuremath{\overline{\mathcal{W}_0}}}
\newcommand{\calF}{\ensuremath{\mathcal{F}}}

\newcommand{\calM}{\ensuremath{\mathcal{M}}}
\newcommand{\N}{\ensuremath{\mathbb{N}}}
\newcommand{\calNZ}{\ensuremath{\mathcal{N}_{0}}}
\newcommand{\calNZS}{\ensuremath{\overline{\mathcal{N}}_{0}}}
\newcommand{\calNS}{\ensuremath{\overline{\mathcal{N}}}}
\newcommand{\of}{\hbox{ of }}
\newcommand{\parterm}{\Lambda^{\!*}}
\newcommand{\parform}{\mathcal{F}^{*}}
\newcommand{\parord}{\mathcal{O}^{*}}
\newcommand{\sysname}{SubML}
\newcommand{\st}{\mid}
\newcommand{\LET}{\mathrm{let}\;}
\newcommand{\ST}{\;\mathrm{such\;that}\;}
\newcommand{\IN}{\;\mathrm{in}\;}
\renewcommand{\epsilon}{\varepsilon}
\newcommand{\eps}{\varepsilon}
\newcommand{\I}[1]{\text{I}_{#1}}
\newcommand{\IP}[1]{\text{I}^+_{#1}}
\newcommand{\G}{\text{G}}
\newcommand{\GP}{\text{G}^+}
\renewcommand{\H}[2]{[#1]_{#2}}
\renewcommand{\vec}[1]{\overline{#1}}
\newcommand{\OMaxi}{\ensuremath{\mathcal{O}}}
\DeclareUnicodeCharacter{8339}{_x}
\DeclareUnicodeCharacter{8345}{_n}
\DeclareUnicodeCharacter{7522}{_i}
\DeclareUnicodeCharacter{11388}{_j}
\DeclareUnicodeCharacter{7523}{_r}
\allowdisplaybreaks

\theoremstyle{definition}
\newtheorem{nota}[thm]{Convention}

\begin{document}

\title[Practical Subtyping for System F]{Practical Subtyping for System F\\
       with Sized (Co-)Induction}

\author[R. Lepigre and C. Raffalli]{Rodolphe Lepigre and Christophe Raffalli}
\address{LAMA, UMR 5127 CNRS - Université Savoie Mont Blanc}
\email{\{rodolphe.lepigre$\,\mid\,$christophe.raffalli\}@univ-smb.fr}

%\author[R. Lepigre and C. Raffalli]{Rodolphe Lepigre}
%\address{LAMA, UMR 5127 CNRS - Université Savoie Mont Blanc}
%\email{rodolphe.lepigre@univ-smb.fr}
%
%\author[]{Christophe Raffalli}
%\address{LAMA, UMR 5127 CNRS - Université Savoie Mont Blanc}
%\email{christophe.raffalli@univ-smb.fr}


\begin{abstract}
We present a rich type system with subtyping for an extension of
System F. Our type constructors include sum and product types,
universal and existential quantifiers, inductive and coinductive
types. The latter two may carry annotations allowing the encoding of
size invariants that are used to ensure the termination of recursive
programs.  For example, the termination of quicksort can be derived by
showing that partitioning a list does not increase its size.
The system deals with complex programs involving mixed induction and
coinduction, or even mixed polymorphism and (co-)induction (as for
Scott-encoded data types). One of the key ideas is to completely separate the
notion of size from recursion. We do not check the termination of programs
directly, but rather show that their (circular) typing proofs are
well-founded. We then obtain termination using a standard semantic proof of
normalisation.
To demonstrate the practicality of our system, we provide an implementation
which accepts all the examples discussed in the paper.

\end{abstract}

\maketitle

\section{Introduction}
Polymorphism and subtyping allow for a more generic programming style. They
lead to programs that are shorter, easier to understand and hence more
reliable. Although polymorphism is widespread among programming languages,
only limited forms of subtyping are used in practice. They usually focus on
product types like records or modules \cite{modulesub}, or on sum types like
polymorphic variants \cite{polyvar}. The main reason why subtyping failed to
be fully integrated in practical languages like Haskell or OCaml is that it
does not mix well with their complex type systems. Moreover, they were not
conceived with the aim of supporting a general form of subtyping.

In this paper, we propose a new framework for the construction of type systems
with subtyping. Our goal being the design of a practical programming language,
we consider a very expressive calculus based on System F. It provides records,
polymorphic variants, existential types, inductive types and coinductive
types. The latter two carry ordinal numbers which can be used to encode size
invariants into the type system \cite{hugues}. For example we can show that
the usual \emph{map} function on lists is size-preserving.
The system can be implemented using standard unification techniques thanks to
the syntax-directed nature of its typing and subtyping rules (Figures
\ref{sntypingrules} and \ref{subtypingrules}). In particular, only one typing
rule applies for each term constructor, and at most one subtyping rule applies
for every two type constructors.
As a consequence, all of the difficulties are focused in the handling of
unification variables and in the construction of circular proofs (see Sections
\ref{sct} and \ref{algorithm}).

\subsection*{Local subtyping and choice operators for terms.} %%%%%%%%%%%%%%%%

To obtain syntax-directed rules, several technical innovations are required.
Most notably, a finer notion of subtyping has to be considered: we generalise
the usual relation $\subt{A}{B}$ using a new \emph{local subtyping} relation
$\lsubt{t}{A}{B}$. It is interpreted as ``if $t$ has type $A$ then it also has
type $B$''.  Usual subtyping is then recovered using choice operators inspired
from Hilbert's Epsilon and Tau functions\nocite{hilbert}.
In our system, the choice operator $ε_{x∈A}(t∉B)$ denotes a term of type $A$
such that $t[x := ε_{x∈A}(t∉B)]$ does not have type $B$. If no such term
exists, then an arbitrary term of type $A$ can be chosen.\footnote{Our model
being based on reducibility candidates \cite{girard, pat}, the interpretation
of a type is never empty.}
The usual subtyping relation $\subt{A}{B}$ can then be defined as
$\lsubt{ε_{x∈A}(x∉B)}{A}{B}$. Indeed, $ε_{x∈A}(x∉B)$ denotes a counterexample
to $\subt{A}{B}$, if it exists. Therefore, if we can derive $\subt{A}{B}$ then
such a counterexample cannot exist, which exactly means that $A$ is a subtype
of $B$ in the usual sense.

More generally, choice operators can be used to replace free variables, thus
suppressing the need for typing contexts.\footnote{We will still use a form of
context to store ordinals assumed to be nonzero (see Section \ref{language}).}
Intuitively, the term $ε_{x∈A}(t∉B)$ denotes a counterexample to the fact that
$λx.t$ has type $A → B$, if it exists. We can thus use this choice operator to
build the following unusual typing rule for $λ$-abstractions.
\begin{prooftree}
\AxiomC{$\type{t[x := ε_{x∈A}(t∉B)]}{B}$}
\UnaryInfC{$\type{λx.t}{A → B}$}
\end{prooftree}
\smallskip
It can be read as a proof by contradiction as its premise is only valid when
there is no term $u$ of type $A$ such that $t[x := u]$ does not have type $B$.
Note that this exactly corresponds to the usual realisability interpretation
of the arrow type.
Thanks to this new typing rule, terms remain closed throughout typing
derivations. In particular, the choice operator $ε_{x∈A}(t∉B)$ binds the
variable $x$ in the term $t$. As a consequence, the axiom rule is replaced
by the following typing rule for choice operators.
\begin{prooftree}
\AxiomC{}
\UnaryInfC{$\type{ε_{x∈A}(t∉B)}{A}$}
\end{prooftree}
\smallskip
The other typing rules, including the rule for application given below, are
not affected by the introduction of choice operators and they remain usual.
\begin{prooftree}
\AxiomC{$\type{t}{A → B}$}
\AxiomC{$\type{u}{A}$}
\BinaryInfC{$\type{t u}{B}$}
\end{prooftree}
Note however that the typing rules of the system (Figure \ref{sntypingrules})
are presented in a slightly more general way. In particular, most of them
include a local subtyping judgment.

\subsection*{Choice operators for types.} %%%%%%%%%%%%%%%%%%%%%%%%%%%%%%%%%%%%

Thanks to local subtyping, the typing rules of the system can be formulated
in such a way that connectives without algorithmic contents are only handled
in local subtyping judgments (see Figure \ref{sntypingrules}).
To manage quantifiers, we introduce two new type constructors $ε_X(t∈A)$ and
$ε_X(t∉A)$ corresponding to choice operators satisfying the denoted
properties. For example, $ε_X(t∉B)$ is interpreted as a type such that $t$
does not have type $B[X := ε_X(t∉B)]$. Intuitively, $ε_X(t∉B)$ is a
counterexample to the fact that $t$ has type $∀ X.B$. Thus, to show that $t$
has type $∀ X.B$, it will be enough to show that it has type
$B[X := ε_X(t∉B)]$.
As a consequence, the usual introduction rule for the universal quantifier
is subsumed by the following local subtyping rule.
\begin{prooftree}
\AxiomC{$\subtype{}{t}{A}{B[X:=ε_X(t∉B)]}$}
\UnaryInfC{$\subtype{}{t}{A}{∀ X.B}$}
\end{prooftree}
\smallskip
Note that this rule does not carry a (usually required) freshness constraint,
as there are no free variable thanks to the use of choice operators.

In conjunction with local subtyping, our choice operators for types allow the
derivation of valid permutations of quantifiers and connectors. For instance,
Mitchell's containment axiom \cite{mitchell} can be easily derived in the
system.
$$∀ X.(A → B) ⊂ (∀ X.A) → (∀ X.B)$$
Another important consequence of these innovations is that our system does not
rely on a transitivity rule for local subtyping. In practice, type annotations
like $((t : A) : B) : C$ can be used to force the decomposition of a proof of
$t : C$ into proofs of $t : A$, $t : A ⊂ B$ and $t : B ⊂ C$, which may help
the system to find the right instantiation for unification variables.
As such annotations are seldom required, we conjecture that a transitivity rule for
local subtyping is admissible in the system.

\subsection*{Implicit covariance condition for (co-)inductive types.} %%%%%%%%

Inductive and coinductive types are generally handled using types $μX. F(X)$
and $ν X. F(X)$ denoting the least and greatest fixpoint of a covariant
parametric type $F$. In our system, the subtyping rules are so fine-grained
that no syntactic covariance condition is required on such types. In fact,
the covariance condition is obtained automatically when traversing the types.
For instance, if $F$ is not covariant then it will not be possible to
derive $μX.F(X) ⊂ νX.F(X)$ or $μX.F(X) ⊂ F(μX.F(X))$. As far as the authors
know, this is the first work in which covariance is not explicitly required
for inductive and coinductive types.

\subsection*{Well-founded ordinal induction and size change principle.} %%%%%%

In this paper, inductive and coinductive types carry an ordinal number $κ$
to form sized types $μ_{κ} X. F(X)$ and $ν_{κ} X. F(X)$
\cite{abel06,hugues,sacchini}.
Intuitively, they correspond to $κ$ iterations of $F$ on the types $⊥$
and $⊤$ respectively. In particular, if $t$ has type $μ_{κ} X. F(X)$ then
there must be $τ < κ$ such that $t$ has type $F(μ_{τ} X. F(X))$. Dually, if
$t$ has type $ν_{κ} X. F(X)$ then $t$ has type $F(ν_{τ} X. F(X))$ for all
${τ < κ}$. More precisely, $μ_{κ} X. F(X)$ is interpreted as the union of all
the $F(μ_{τ} X. F(X))$ for $τ < κ$, and $ν_{κ} X. F(X)$ is interpreted as
the intersection of all the $F(ν_{τ} X. F(X))$ for $τ < κ$.
These definitions are monotonous in $κ$, even if $F$ is not
covariant. This implies that there exists an ordinal $∞$ from which
the constructions are stationary. As a consequence, we have
$F(μ_{∞} X. F(X)) ⊂ μ_{∞} X. F(X)$ and
$ν_{∞} X. F(X) ⊂ F(ν_{∞} X. F(X))$, which are sufficient for the
correctness of our subtyping rules. In particular, $μ_{∞} X. F(X)$ and
$ν_{∞} X. F(X)$ only correspond to the least and greatest fixpoint of
$F$ when it is covariant. If $F$ is not covariant, then these
stationary points are not fixpoints.

In this paper, we introduce a uniform induction rule for local subtyping. It
is able to deal with many inductive and coinductive types at once, but accepts
proofs that are not well-founded. To solve this problem, we rely on the size
change principle \cite{scp}, which allows us to check for well-foundedness a
posteriori. Our system is able to deal with subtyping relations between mixed
inductive and coinductive. For example, it is able to derive subtyping
relations like $μ X. ν Y. F(X,Y) ⊂ ν Y. μ X. F(X,Y)$ for a given covariant
type $F$ with two parameters.
When we restrict ourselves to types without universal and existential
quantifiers, our experiments tend to indicate that our system is in some sense
complete. However, we failed to prove completeness in the presence of function
types, the main problem being the mere definition of completeness in this
setting.

\subsection*{Totality of recursive functions.} %%%%%%%%%%%%%%%%%%%%%%%%%%%%%%%

As for local subtyping judgments, it is possible to use circular proofs
for typing recursive programs. General recursion is enabled by extending
the language with a fixpoint combinator $Yx.t$, reduced using the rule
$Yx.t ≻ t[x := Yx.t]$. It is handled using the following, very simple
typing rule.
\smallskip
\begin{prooftree}
\AxiomC{$\type{t [x := Yx.t]}{A}$}
\UnaryInfC{$\type{Y x.t}{A}$}
\end{prooftree}
\medskip
It is clear that it induces circularity as a proof of $⊢ Y x. t : A$
will require a proof of $⊢ Y x. t : A$. As there is no guarantee that such
circular proofs are well-founded, we need to rely on the size change principle
again. Given its simplicity, our system is surprisingly powerful. In
particular, a fixpoint may be unfolded several times to obtain a well-founded
circular proof (see Section \ref{applications}).

One of the major advantages of our presentation is that it allows for a good
integration of the termination check to the type system, both in the theory
and in the implementation. Indeed, we do not prove the termination of a
program directly, but rather show that its circular typing proof is
well-founded. Normalisation is then established indirectly, using a
standard semantic proof based on a well-founded induction on the typing
derivation.
To show that a circular typing proof is well-founded we rely on the size
change principle \cite{scp}. It is run on size informations that are
extracted from the circular structure of our proofs in a precisely defined
way (see Section \ref{sct}).

\subsection*{Quantification over ordinals.} %%%%%%%%%%%%%%%%%%%%%%%%%%%%%%%%%%

As types can carry ordinal sizes, it is natural to allow quantification over
the ordinals themselves. We can thus use the following type for the usual \emph{map}
function, where
$\mathrm{List}(A,α)$ denotes the type of lists of size $α$ with
elements of type $A$ (it is defined as
$μ_{α} L.[\mathrm{Nil} \st \mathrm{Cons} \of A×L]$).
$$ ∀A.∀B.∀α.(A → B) → \mathrm{List}(A,α) → \mathrm{List}(B,α) $$
Thanks to the quantification on the ordinal $α$, which links the size of the
input list to the size of the output list, we can express the fact that the
output is not greater than the input. This means that the system will allow
us to make recursive calls through the \emph{map} function, without loosing
size information (and thus termination information). This technique also
applies to other relevant functions such as insertion sort.

Using size preserving functions and ordinal quantification is important for
showing the termination of more complex algorithms. For instance, proving the
termination of \emph{quicksort} requires showing that partitioning a list of
size $α$ produces two lists of size at most $α$. To do so, the partitioning
function must be defined with the following type.
$$ ∀A.∀α.\mathrm{List}(A,α) → \mathrm{List}(A,α) × \mathrm{List}(A,α) $$
It is then possible to define \emph{quicksort} in the usual way, without any
other modification. Note that the termination of simple functions is derived
automatically by the implementation (i.e., without specific size annotations).

In this paper, the language of the ordinals that can be represented in the
syntax is very limited. As in \cite{sacchini13}, it only contains a constant
$∞$, a successor symbol and variables for quantification. Working with such a
small language allows us to keep things simple while still allowing the
encoding of many size invariants. Nonetheless, it is clear that the system
could be improved by extending the language of ordinals with function symbols
such as, for example, maximum or addition.

\subsection*{Properties of the system.} %%%%%%%%%%%%%%%%%%%%%%%%%%%%%%%%%%%%%%

A first version of the language without general recursion (i.e., without the
fixpoint combinator) is defined in Section~\ref{language}. It has three main
properties: strong normalisation, type safety and logical consistency
(Theorems \ref{th:strongnorm}, \ref{th:safety} and \ref{th:consistency}).
These results follow from the construction of a realisability model presented
in Section~\ref{semantics}. They are consequences of the adequacy lemma
(Theorem~\ref{th:snadequacy}), which establishes the compatibility of the
model with the language and type system.

After the introduction of the fixpoint combinator in Section \ref{fixpoint},
the properties of the system are mostly preserved (Theorems
\ref{th:swnadequacy} and \ref{th:finalprops}).
However, the definition of the model needs to be changed
slightly as strong normalisation (in the usual sense) is compromised by the
fixpoint combinator. Indeed, the reduction rule $Y x.t ≻ t[x := Y x.t]$ is
obviously non-terminating. Nonetheless, we can still prove normalisation for
all the weak reduction strategies (i.e., those that do not reduce under
$λ$-abstractions).

\subsection*{Implementation.} %%%%%%%%%%%%%%%%%%%%%%%%%%%%%%%%%%%%%%%%%%%%%%%%

Typing and subtyping are likely to be undecidable in our system. Indeed, it
contains Mitchell's variant of System F \cite{mitchell}, for which both typing
and subtyping are undecidable \cite{urzyczyn,wells94,wells99}. Moreover, we
believe that there are no practical, complete semi-algorithms for extensions
of System F like ours.\footnote{It is an open problem whether
  every normalising extensions of system F is undecidable.}
Instead, we propose an incomplete semi-algorithm that may fail or even
diverge on a typable program. In practice we almost never meet non
termination, but even in such an eventuality, the user can interrupt
the program to obtain a relevant error message. Indeed, type-checking
can only diverge when checking a local subtyping judgment. In this
case, a reasonable error message can be built using the last applied
typing rule.

As a proof of concept, we implemented a toy programming language based on our
system. It is called \sysname{} and is available online \cite{proto}. Aside
from a few subtleties described in Section \ref{algorithm}, the implementation
is straightforward and remains very close to the typing rules of Figure
\ref{typingrules}\footnote{The rules of Figure \ref{sntypingrules} need to be
modified slightly to handle fixpoints.} and to the subtyping rules of Figures
\ref{subtypingrules} and \ref{subtypingrules2}.
Although the system has a great expressive power, its simplicity allows for a
very concise implementation. The main functions (type-checking and subtyping)
require less than 600 lines of OCaml code. The current implementation,
including parsing, evaluation and \LaTeX{} pretty printing contains less than
6500 lines of code.

We conjecture that our implementation is complete (i.e., it may succeed on all
typable programs), provided that enough type annotations are given. On
practical instances, the required amount of annotations seems to be
reasonably small (see Section \ref{applications}).
Overall, the system provides a similar user experience to statically typed
functional languages like OCaml or Haskell. In fact, such languages also
require type annotations for advanced features like polymorphic recursion.

\sysname{} provides literate programming features inspired by the PhoX
language \cite{phox}. They can notably be used to generate \LaTeX{} documents.
In particular, the examples presented in Sections~\ref{language},
\ref{scottrec} and \ref{applications} (including proof trees) have been
generated using \sysname{}, and are therefore machine checked.
Many other program examples (more than 4000 lines of code) are
provided with the implementation of \sysname{}. They can be used to check that
the system is indeed usable in practice.
\sysname{} can either be installed from its source code or tried online at
\url{https://lama.univ-smb.fr/subml}.\footnote{The online version is compiled
to Javascript using js\_of\_ocaml (\url{https://ocsigen.org/js_of_ocaml/}).}

\subsection*{Applications.} %%%%%%%%%%%%%%%%%%%%%%%%%%%%%%%%%%%%%%%%%%%%%%%%%%

In addition to classical examples, our system allows for applications that we
find very interesting (see Sections \ref{scottrec} and \ref{applications}).
As a first example, we can program with the Church encoding of algebraic data
types. Although this has little practical interest (if any), it requires the
full power of System F and is a good test suite for polymorphism.
As Church encoding is known for having a bad time complexity, Dana Scott
proposed a better alternative using a combination of polymorphism and
inductive types \cite{abadi}. For instance, the type of natural numbers can be
defined as follows.
$$\mathbb{N}_S = μ X. ∀ Y. ((X → Y) → Y → Y)$$
Unlike Church numerals, Scott numerals admit a constant time predecessor
function with the expected type $\mathbb{N}_S → \mathbb{N}_S$.

In standard systems, recursion on inductive data types requires
specific typing rules for recursors, like in Gödel's System T. In contrast,
our system is able to type a recursor encoded as a $λ$-term, without having
to extend the language. This recursor was shown to the second author by Michel
Parigot \cite{parigotX}. We then adapted it to other algebraic data types,
showing that Scott encoding can be used to program in a strongly normalisable
system with the expected asymptotic complexity.

We also discovered a surprising $λ$-calculus coiterator for streams encoded
as follows, using an existentially quantified type $S$ as an internal state.
$$\mathrm{Stream}(A) =  ν X. ∃ S. S × (S → A × X)$$
An element of type $S$ must be provided to progress in the computation of
the stream.
Note that here, the product type does not have to be encoded using
polymorphism as for Church or Scott encoded data types. As a consequence, the
above definition of streams may have a practical interest.

\subsection*{Curry style and type annotations.} %%%%%%%%%%%%%%%%%%%%%%%%%%%%%%

For our incomplete type checking algorithm to be usable in practice, the user
has to guide the system using type annotations. However, the language is Curry
style, which means that polymorphic types are interpreted as intersections
(and existential types as unions) in the semantics. As a consequence, the
terms do not include type abstractions and type applications as in Church
style, where polymorphic types are interpreted as functions (and existential
types as pairs). This means that it is not possible to introduce a name for
a type variable in a term, which is necessary for annotating subterms of
polymorphic functions with their types.

As our system relies on choice operators for types, it never manipulates type
variables. However, we found a way to name choice operators corresponding to
local types using a pattern matching syntax. It can be used to extract the
definition of choice operators from types and make it available to the user
for giving type annotations. As an example, we can fully annotate the
polymorphic identity function as follows.
$$
  \mathrm{Id} : ∀ X.X → X
  = {λ x. \hbox{ let } X \hbox{ such that } x:X \hbox{ in } (x:X)}
$$
Note that such annotations are not part of the theoretical type system. They
are only provided in the implementation to allow the user to guide the system
toward guessing the correct instantiation of unification variables.

Another interesting application of choice operators for types is the dot
notation for existential types, which allows the encoding of a module system
based on records. As an example, we can encode a signature for isomorphisms
with the following type.
$$ \mathrm{Iso} = ∃ T.∃ U. \{ \mathrm{f} : T → U; \mathrm{g} : U → T\} $$
Given a term $h$ of type $\mathrm{Iso}$, we can then define the following
syntactic sugars to access the abstract types corresponding to $T$ and $U$.
\begin{align*}
  h.T &= ε_T(h ∈ ∃ U. \{ \mathrm{f} :  T → U; \mathrm{g} : U → T  \}) \\
  h.U &= ε_U(h ∈ \{ \mathrm{f} :  h.T → U; \mathrm{g} : U → h.T  \})
\end{align*}
The first choice operator denotes a type $T$ such that $h$ has type
$∃ U. \{ \mathrm{f} : T → U; \mathrm{g} : U → T \}$. As our system never
infers polymorphic or existential types, we can rely on the name that was
chosen by the user for the bound variable.
This new approach to abstract types seems simpler than previous work like
\cite{courant}.

\subsection*{Related work.} %%%%%%%%%%%%%%%%%%%%%%%%%%%%%%%%%%%%%%%%%%%%%%%%%%

The language presented in this paper is an extension of John Mitchell's System
F$_{η}$ \cite{mitchell}, which itself extends Jean-Yves Girard and John
Reynolds's System F \cite{girard, reynolds} with subtyping. Unlike previous
work \cite{amadio,pierce}, our system supports mixed induction and
coinduction with polymorphic and existential types. In particular, we improve
on an unpublished work of the second author \cite{raffalli}. Our type system
also strongly relates to sized types \cite{hugues} as our inductive and
coinductive types carry ordinal numbers. Such a technique is widespread for
handling induction \cite{blanqui06,blanqui09,gregoiresacchini10,blanquiRiba06,
sacchini} and even coinduction \cite{abel06,abel13,sacchini13}, in settings
where termination is required.

The most important difference difference between this paper and previous work
precisely lies in the handling of inductive and coinductive types. In all the
systems that the authors are aware of, inductive and coinductive types are
strongly linked to recursion, and thus to termination. In particular, they
rely on specific rules for checking size relations between ordinal parameters
when using recursion. In this paper, inductive and coinductive types are
handled in a way that is completely orthogonal to recursion. Ordinal sizes are
only manipulated in the subtyping rules related to inductive and coinductive
types, while recursion is handled separately using a simple typing rule for
the fixpoint combinator. This leads to a system that has a rather simple
presentation compared to previous work, even if it relies on unusual concepts
such as choice operators and circular proofs.

We believe that our system is simpler than previous work for two main reasons.
First, the complete distinction between (general) recursion and inductive and
coinductive types allows for simpler, more natural typing and subtyping rules.
In particular, we do not need to rely on syntactic conditions such as the
semi-continuity used by Andreas Abel \cite{abel06}, or even the standard
covariance condition. Instead, we consider a
formalism of potentially not well-founded circular proofs. We then make a
singular use of the size change principle of Lee, Jones and Ben-Amram
\cite{scp}, which is usually used to prove the termination of programs.
For example, it is used in this way in the work of Pierre Hyvernat
\cite{pierre} and Andreas Abel \cite{foetus}, but also in the implementations
of Agda \cite{agda} and PML \cite{pml}. Here however,
the size change principle is not used to prove the termination of programs
directly, but to show that typing proofs are well-founded. Termination is
then obtained using a semantic proof by well-founded induction on the
structure of our typing and subtyping derivations.

To our knowledge, techniques from the termination checking community have
never been used to check the correctness of circular proofs before. The
literature on circular proofs in general seems to be limited to the work of
Luigi Santocanale \cite{luigi4,luigi3}, where circular proofs are related to
parity games \cite{parity} and given a category-theoretic semantics. However,
the considered language is based on the modal $μ$-calculus \cite{arnold}. Its
expressiveness is thus limited and it does not include subtyping.

Subtyping has been extensively studied in the context of ML-like languages,
starting with the work of Roberto Amadio and Luca Cardelli \cite{amadio}.
Recent work includes the MLsub system \cite{mlsub}, which extends unification
to handle subtyping constraints. Unlike our system, it relies on a flow
analysis between the input and output types, borrowed from the work of
François Pottier \cite{pottier}. However, we are not aware of any work on
subtyping that leads to a system as expressive as ours for a Curry-style
extension of System F. In particular, no other system seems to be able to
handle the permutation of quantifiers with other connectives as well as
mixed inductive and coinductive types (see Sections~\ref{language} and
\ref{scottrec}).

From a more practical perspective, we chose to trade the decidability of
type-checking for simplicity. Indeed, we chose not to look for (and prove)
a decidability result, unlike most work on programming languages. We are
happy to work with a semi-algorithm as our experiments showed that
this is perfectly acceptable in practice. In particular, the user experience
is not different from working with meta-variables or implicit arguments as
in Coq or Agda \cite{coq,agda}. Nevertheless, this is not completely
satisfactory, and we would like to prove that our semi-algorithm is complete
for the quantifier-free fragment of our calculus. We believe that we could
achieve completeness using a specific algorithm to solve size constraints.
Such an algorithm has already been used by Frédéric Blanqui for a language
with only a successor symbol \cite{blanqui17}. This could hopefully be
adapted to our setting.

%%% Local Variables:
%%% ispell-personal-dictionary: "~/.hunspell-en"
%%% ispell-local-dictionary: "british"
%%% End:

\section{Syntactic ordinals}\label{ordinals}
In this section, we introduce a syntax for representing ordinals. It will be
used to equip the types of our language with a notion of size, as is usually
done for sized types \cite{hugues}. Here, ordinals will also be used to show
that infinite typing derivations are well-founded.
\begin{nota}
  We will use the vector notation $\vec{e}$ for a tuple $(e₁,\dots,e_n)$
  which length will be denoted $|\vec{e}| = n$. The concatenation of two
  vectors $\vec{x}$ and $\vec{y}$ will be denoted $\vec{x}.\vec{y}$.
  Note that there will sometimes be implicit constraints on the length of
  vectors (e.g., when working with substitutions such as
  $E[\vec{x} := \vec{e}]$).
\end{nota}
\begin{defi}\label{def:synt_ord}
  Let $\mathcal{P} = \{P,Q,\dots\}$ be a set of predicate symbols (of mixed
  arities) ranging over ordinals. The sets of \emph{syntactic ordinals}
  $\ord$ is defined by the first category of the following BNF grammar
  using a set of ordinal variables $\mathcal{V}_\ord = \{α,β,\dots\}$.
  \begin{align*}
    κ, τ, υ \bnfeq & α
            \bnfor ∞
            \bnfor τ+1
            \bnfor {ε_{\vec{α} < \vec{w}} P(\vec{α}.\vec{κ})}_i \\
    w \bnfeq & κ \bnfor \OMaxi
  \end{align*}
  In syntactic ordinals of the form
  ${ε_{\vec{α} < \vec{w}} P(\vec{α}.\vec{κ})}_i$, the variables of
  $\vec{α} = (α₁,\dots,α_n)$ are bound in $P(\vec{α}.\vec{κ})$ but not in
  $\vec{w}$. Moreover, we enforce $1 ≤ i ≤ |\vec{α}| = |\vec{w}|$ and
  $|P| = |\vec{α}| + |\vec{κ}|$, where $|P|$ denotes the arity of the
  predicate $P$. Note that $\mathcal{O}$ may itself appear in the syntax
  as an upper bound for ordinal variables.
\end{defi}

Syntactic ordinals are built using the constant $∞$, a successor symbol
and \emph{ordinal choice operators} (or \emph{ordinal witnesses}) of the form
${ε_{\vec{α} < \vec{w}} P(\vec{α}.\vec{κ})}_i$. Intuitively, the vector
$\vec{τ}$ defined as ${τ_i = {ε_{\vec{α} < \vec{w}} P(\vec{α}.\vec{κ})}_i}$
denotes syntactic ordinals that are point-wise smaller than $\vec{w}$, and
such that ``$P(\vec{\tau}.\vec{κ})$ is true'' (this will be made formal in
Definition \ref{def:ordsem}).
In the upper bound $\vec{w}$, one can use the notation $α<\OMaxi$ in the case
where there is no constraint on the variable $α$. In other words, $\OMaxi$
denotes an ordinal that is bigger than all the syntactic ordinals, and as a
consequence it is not a syntactic ordinal itself.

In the semantics, the symbol $∞$ will be interpreted using the ordinal
$2^{2^ω}$, where $ω$ denotes the cardinal of the natural numbers. This
ordinal will be large enough to ensure the convergence of all the fixpoints
corresponding to inductive and coinductive types. However, $2^{2^ω}$
cannot be the biggest ordinal of our semantics since larger ones may be
represented in the syntax using the successor symbol.\footnote{We will
in fact never use $∞+1$ (or other successors of $∞$) in practice.}
\begin{defi}
  We denote $\ordsem$ the ordinal $2^{2^ω}+ω$, which is also the set of all
  the ordinals of our semantics. Note that it can be thought of as the
  interpretation of $\OMaxi$.
\end{defi}

We will now extend the syntax of syntactic ordinals with (actual) ordinals,
thus embedding the elements of the semantics into the syntax. This common
technique will allow us to substitute variables using ordinals directly,
without having to rely on a semantical map for interpreting variables. This
will allow us to only manipulate closed (parametric) syntactic ordinals.
\begin{defi}\label{def:par_synt_ord}
  The set of \emph{parametric syntactic ordinals} $\ord^*$ is obtained
  by extending the language of syntactic ordinals with (actual) ordinals
  $o ∈ \ordsem$.
  \begin{align*}
    κ, τ, υ \bnfeq & α
            \bnfor ∞
            \bnfor τ+1
            \bnfor o
            \bnfor {ε_{\vec{α} < \vec{w}} P(\vec{α}.\vec{κ})}_i \\
    w \bnfeq & κ \bnfor \OMaxi
  \end{align*}
  We will denote $κ[α:=o]$ the syntactic ordinal $κ$ in which the free
  occurrences of the variable $α$ have been replaced by the ordinal
  $o ∈ \ordsem$. We will also use the notation $κ[\vec{α} := \vec{o}]$
  for multiple simultaneous substitution of ordinal variables.
\end{defi}

\begin{nota}
  We will use the notation $\vec{ε}_{\vec{α}<\vec{w}}P(\vec{α}.\vec{κ})$ for
  the vector $({ε_{\vec{α} < \vec{w}} P(\vec{α}.\vec{κ})}_i)_{1≤i≤|α|}$.
  When $|\vec{α}| = |\vec{κ}| = 1$, we will write $ε_{α < w}P(α.\vec{κ})$ for
  both $\vec{ε}_{\vec{α}<\vec{w}}P(\vec{α}.\vec{κ})$ and
  ${ε_{\vec{α} < \vec{w}} P(\vec{α}.\vec{κ})}_1$.
\end{nota}

We will now give the semantical interpretation of the closed parametric
syntactic ordinals, using (actual) ordinals of $\ordsem$. As syntactic
ordinals contain predicate symbols, they will need to be interpreted as well.
\begin{defi}\label{def:ordsem}
  To interpret predicate symbols, we require
  an interpretation function (or valuation) $\sem{{-}}$ such that for
  all $P ∈ \mathcal{P}$ we have $\sem{P} ∈ \ordsem^{|P|} → \{0,1\}$.
  The semantics of closed (vectors of) parametric syntactic ordinals is defined
  inductively as follows.
  $$
    \sem{∞} = 2^{2^{ω}}
    \hspace{2em}
    \sem{\OMaxi} = 2^{2^{ω}} + ω
    \hspace{2em}
    \sem{κ+1} = \sem{κ} + 1
    \hspace{2em}
    \sem{o} = o
    \hspace{2em}
    \sem{\vec{κ}} = (\sem{κ₁},\dots,\sem{κ_n})
  $$
  $$
    \sem{\vec{ε}_{\vec{α} < \vec{w}} P(\vec{α}.\vec{κ})} =
      \left\{
        \begin{array}{l}
           \vec{o} ∈ \ordsem^{|\vec{w}|}
             \text{ with } \vec{o} < \sem{\vec{w}} \text{ and }
                \sem{P}(\vec{o}.\sem{\vec{κ}}) = 1 \text{ if it exists}\\
           \vec{0} \text{ otherwise}
        \end{array}
      \right.
  $$
  Here, $\vec{o} < \sem{\vec{w}}$ denotes point-wise ordering on vectors of
  ordinals, and $\vec{0}$ denotes a vector of $0$ ordinals. Note that there
  may be several possible choices for $\vec{o}$ in the case of an ordinal
  witness. We will thus consider different \emph{models}, for which the
  choice of $\vec{o}$ will be made differently. If $\calM$ is such a model,
  we will denote $\sem{κ}^\calM$ the induced interpretation.
\end{defi}
\begin{nota}
  We will most of the time omit to mention the model $\calM$. In this case,
  we will assume that it is fixed, but arbitrary.
\end{nota}
\begin{lem}\label{lem:changemodel}
  Let $\calM$ be a model and $\vec{ε}_{\vec{α} < \vec{w}} P(\vec{α}.\vec{κ})$
  be a vector of ordinal choice operators of size $n$. If $\vec{o}∈\ordsem^n$
  is a vector of ordinals such that $\vec{o} < \sem{\vec{w}}^\calM$ and
  $\sem{P}(\vec{o}.\sem{\vec{κ}}^\calM) = 1$, then there is a model $\calM'$
  such that $\sem{\vec{w}}^{\calM'} = \sem{\vec{w}}^\calM$,
  $\sem{\vec{κ}}^{\calM'} = \sem{\vec{κ}}^\calM$ and
  $\sem{\vec{ε}_{\vec{α} < \vec{w}} P(\vec{α}.\vec{κ})}^{\calM'} = \vec{o}$.
\end{lem}
\begin{proof}
  We define the height $h(τ)$ of a syntactic ordinal $τ$ as follows.
  $$
    h(∞) = 0
    \hspace{1cm}
    h(\OMaxi) = 0
    \hspace{1cm}
    h(o) = 0
    \hspace{1cm}
    h(κ+1) = 1 + h(k)
  $$
  $$
    h(κ₁,\dots,κ_n) = \max(h(κ₁),\dots,h(κ_n))
    \hspace{1cm}
    h({ε_{\vec{α}<\vec{w}} P(\vec{α}.\vec{κ})}_i) = 1 +
      \max(h(\vec{w}),h(\vec{κ}))
  $$
  We then define $⟦τ⟧^{\calM'}$ by induction
  on $h(τ)$ by first taking $\sem{τ}^{\calM'} = \sem{τ}^\calM$ for every $τ$
  such that $h(τ) < h(εⁱ_{\vec{α}<\vec{w}} P(\vec{α}.\vec{κ}))$ (including
  the elements of $\vec{w}$ and $\vec{κ}$). We then take
  $\sem{\vec{ε}_{\vec{α}<\vec{w}} P(\vec{α}.\vec{κ})}^{\calM'} = \vec{o}$
  and we complete the definition by marking arbitrary choices for other
  ordinal witnesses.
\end{proof}

We now consider an ordering relation $κ≤τ$ and a strict ordering relation
$κ<τ$ on syntactic ordinals. Both relations will be defined in terms of a
third (ternary) relation $κ ≤_i τ$ in which $i∈\mathbb{Z}$. This relation
will be specified using the deduction rule system including \emph{ordinal
contexts}, which will contain ordinals assumed to be non-zero.
\begin{defi}
  An \emph{ordinal contexts} is a finite set of syntactic ordinals
  represented using lists generated by the following BNF grammar.
  $$ γ,δ \bnfeq ∅ \bnfor γ, κ $$
  Note that it will never be useful to store syntactic ordinals of the form
  $τ+1$ or $∞$ in an ordinal context as they are necessarily non-zero.
\end{defi}
\begin{figure}
% Infinite, reflexivity, successor on both sides and definition of <
\begin{prooftree}
\AxiomC{$i ≤ 0$}
\RightLabel{$=$}
\UnaryInfC{$γ ⊢ κ ≤_i κ$}
\DisplayProof\hfill
\AxiomC{$γ ⊢ κ ≤_{i+1} \tau$}
\RightLabel{$s_l$}
\UnaryInfC{$γ ⊢ κ + 1 ≤_i \tau$}
\DisplayProof\hfill
\AxiomC{$γ ⊢ κ ≤_{i-1} \tau$}
\RightLabel{$s_r$}
\UnaryInfC{$γ ⊢ κ ≤_i \tau + 1$}
\end{prooftree}

\smallskip

% Both context rules
\begin{prooftree}
\AxiomC{$γ, κ ⊢ κ ≤_{i-1} \tau$}
\AxiomC{$w_j = κ ≠ \OMaxi$}
\RightLabel{$ε$}
\BinaryInfC{$γ, κ ⊢ {ε_{\vec{α}<\vec{w}}P(\vec{α}.\vec{υ})}_j  ≤_i \tau$}
\DisplayProof\hfill
\AxiomC{$γ ⊢ κ ≤_i \tau$}
\AxiomC{$w_j = κ ≠ \OMaxi$}
\RightLabel{$ε_w$}
\BinaryInfC{$γ ⊢ {ε_{\vec{α}<\vec{w}}P(\vec{α}.\vec{υ})}_j ≤_i \tau$}
\end{prooftree}

\caption{Rules for ordinal ordering and strict ordering.}
\label{fig:ordcomp}
\end{figure}

\begin{defi}
  The syntactic ordinals are equipped with a family of relations $(≤_i)$
  with $i \in \mathbb{Z}$. Intuitively, $κ ≤_i τ$ can be understood as
  ``$κ + i ≤ τ$'' when $i ≥ 0$ and as ``$κ ≤ τ + (-i)$'' when $i ≤ 0$.
  Given a context of positive ordinals $γ$, the relation $(≤_i)$ is
  defined using the deduction rules of Figure~\ref{fig:ordcomp}. We then
  take $κ ≤_0 τ$ as the definition of $κ ≤ τ$ and $κ ≤_1 τ$ as the
  definition of $κ < τ$.
\end{defi}

Note that the deduction rule system of Figure \ref{fig:ordcomp} can be
implemented as a deterministic and terminating procedure. Indeed, it is
easy to see that the ($s_r$) rule commutes with the ($s_l$), ($ε$) and
($ε_w$) rules. When both rules ($ε$) and ($ε_w$) may apply it is better
to use ($ε$) as it yields a lower index, and thus proves more judgments
according to Lemma \ref{lem:ordord}, \ref{it2:ordord}.
\begin{lem}\label{lem:ordord}
  For every ordinal contexts $γ$ and $δ$, every syntactic ordinals $κ₁$,
  $κ₂$ and $κ₃$, and for every integers $i$ and $j$ we have:
  \begin{enumerate}[label=(\arabic*)]
    \item \label{it1:ordord} if $γ ⊢ κ₁ ≤_i κ_2$ then $γ,δ ⊢ κ₁ ≤_i κ_2$,
    \item \label{it2:ordord} if $γ ⊢ κ₁ ≤_i κ_2$ and $j ≤ i$ then
          $γ⊢ κ₁ ≤_j κ_2$,
    \item \label{it3:ordord} if $γ ⊢ κ₁ ≤_i κ_2$ and $γ ⊢ κ_2 ≤_j κ_3$ then
          $γ ⊢ κ₁ ≤_{i+j} κ_3$.
  \end{enumerate}
\end{lem}
\begin{proof}
  The proofs of \ref{it1:ordord} and \ref{it2:ordord} are immediate by
  induction on the derivation. We prove \ref{it3:ordord} by induction
  on the sum of the sizes of the derivations of $γ ⊢ κ₁ ≤_i κ_2$ and
  $γ ⊢ κ_2 ≤_j κ_3$.
  If the last applied rule on either side is ($=$), then we have
  $κ₁ = κ_2$ and $i ≤ 0$ or $κ_2 = κ_3$ and $j ≤ 0$. In both case
  we can conclude using \ref{it2:ordord}.
  If the last rule used on the left is ($s_l$) then $κ₁ = κ + 1$. By
  induction hypothesis we have $γ ⊢ κ ≤_{i+j+1} κ_3$ and thus
  $γ ⊢ κ₁ ≤_{i+j} κ_3$. A similar argument can be used if the last
  rule used on the right is ($s_r$).
  If the last used rule on the left is ($ε$) or ($ε_w$) then we have $κ₁ =
  {ε_{\vec{α}<\vec{w}}P(\vec{α}.\vec{υ})}_m$ and $w_m = κ ≠ \OMaxi$. By
  induction hypothesis, we get $γ ⊢ κ ≤_{i+j-1} κ_3$ if we applied the ($ε$)
  rule or $γ ⊢ κ ≤_{i+j} κ_3$ if we applied the ($ε_w$) rule. In both cases
  this implies $γ ⊢ κ₁ ≤_{i+j} κ_3$.
  If the last rule used on the right is the ($ε$) or ($ε_w$) then we must be
  in one of the previous cases. Indeed, the rules that can be applied on the
  left when $κ_2$ is an ordinal witness are ($=$), ($s_l$), ($ε$) and ($ε_w$).
\end{proof}

\begin{lem}\label{lem:leqicorrect}
  Let $γ$ be a closed context, $κ₁$, $κ₂$ be closed syntactic ordinals and
  $i$ be an integer such that such that $γ ⊢ κ₁ ≤_i κ₂$ is derivable. For
  any model, if $\sem{τ} ≠ 0$ for all $τ ∈ γ$ then
  $\sem{κ₁} + i ≤ \sem{κ_2}$ when $i ≥ 0$ and
  $\sem{κ₁} ≤ \sem{κ_2} + (-i)$ when $i ≤ 0$.
\end{lem}
\begin{proof}
  The proof is done by induction on the derivation of $γ ⊢ κ₁ ≤_i κ₂$. The
  cases for the ($=$), ($s_l$) and ($s_r$) rules are immediate. In the case
  of the ($ε$) rule we have $κ₁ = {ε_{\vec{α}<\vec{w}}P(\vec{α}.\vec{υ})}_j$
  with $w_j = κ ≠ \OMaxi$. As a consequence, $\sem{κ₁}$ is either equal to
  some ordinal $o_j < \sem{κ}$ or to $0$.
  Since $\sem{κ} ≠ 0$, we have $\sem{κ₁} < \sem{τ_j}$ in both cases and we
  can thus conclude by induction hypothesis. In the case of the ($ε_w$) rule
  the proof is similar, but it is possible that $\sem{κ} = 0$ so we only
  have $\sem{κ₁} ≤ \sem{κ}$.
\end{proof}

%%% Local Variables:
%%% ispell-personal-dictionary: "~/.hunspell-en"
%%% ispell-local-dictionary: "british"
%%% End:

\section{Size change matrices}\label{matrices}
We will now consider the formalism that will be used to relate our syntactic
ordinals to the size-change principle \cite{scp} in the following sections.
The main idea will be to represent the size informations contained in the
circular structure of our proofs using matrices. We will then be able to
easily compose size informations using matrix product.
\begin{defi}
  We consider the set $\{-1,0,∞\}$ ordered as $-1 < 0 < ∞$. It is equipped
  with a semi-ring structure using the minimum operator $(\min)$ as its
  addition, and the composition operator $(\circ)$ defined below as its
  product. Note that the neutral element of $(\min)$ is $-1$ and that the
  neutral element of $(\circ)$ is $∞$.
  \begin{align*}
    x \circ ∞ &= ∞ & -1 \circ x = -1 \hspace{1em} \hbox{ if } x ≠ ∞ \\
    ∞ \circ x &= ∞ & x \circ -1 = -1 \hspace{1em} \hbox{ if } x ≠ ∞ \\
    0 \circ 0 &= 0
  \end{align*}
  Intuitively, $-1$ will be used to indicate that the size of some object
  decreases, $0$ will be used when the size does not increase and $∞$ will
  be used when there is no size information.
\end{defi}

\begin{defi}
  A \emph{size-change matrix} is simply a matrix with coefficient in
  $\{-1,0,∞\}$. Given an $n × m$ matrix $A$ and an $m × p$ matrix $B$,
  the product of $A$ and $B$, denoted $A B$, is an $n × p$ matrix $C$
  defined as follows.
  $$ C_{i,j} = \min_{1 ≤ k ≤ m} A_{i,k} \circ B_{k,j} $$
  Note that this exactly corresponds to the usual matrix product expressed
  with the operations of our semi-ring $(\{-1,0,∞\},min,\circ)$.
\end{defi}

\begin{lem}\label{lem:matassoc}
The size-change matrix product is associative.
\end{lem}
\begin{proof}
  We consider an $n × m$ matrix $A$, an $m × p$ matrix $B$ and a $p × q$
  matrix $C$. The products $L = A B$ and $R = B C$ are well-defined, and
  we have $L_{i,j} = \min_{1 ≤ k ≤ m} A_{i,k} \circ B_{k,j}$ and
  ${R_{i,j} = \min_{1 ≤ k ≤ p} B_{i,k} \circ C_{k,j}}$. As $L$ is an
  $n × p$ matrix and $R$ is an $m × q$ matrix, the products $L C$ and
  $A R$ are well-defined and both produce an $n × q$ matrix. We thus
  need to show that  $ \min_{1 ≤ k ≤ p} L_{i,k} \circ C_{k,j} =
  \min_{1 ≤ k ≤ m} A_{i,k} \circ R_{k,j} $.
  \begin{align*}
    \min_{1 ≤ k ≤ p} L_{i,k} \circ C_{k,j}
    &= \min_{1 ≤ k ≤ p} (\min_{1 ≤ l ≤ m} A_{i,l} \circ B_{l,k})
       \circ C_{k,j} \\
    &= \min_{1 ≤ k ≤ p} \min_{1 ≤ l ≤ m}
       (A_{i,l} \circ B_{l,k}) \circ C_{k,j} \\
    &= \min_{1 ≤ k ≤ m} \min_{1 ≤ l ≤ p}
       A_{i,k} \circ (B_{k,l} \circ C_{l,j}) \\
    &= \min_{1 ≤ k ≤ m} A_{i,k} \circ
       (\min_{1 ≤ l ≤ p} B_{k,l} \circ C_{l,j})
    = \min_{1 ≤ k ≤ m} A_{i,k} \circ R_{k,j}
  \end{align*}
\end{proof}

To conclude this section, we will now link the notion of size-change matrix
to an order relation. In particular, we will show that the matrix product
indeed corresponds to the composition of size informations. In other words,
the product corresponds to the application of the transitivity of the order
relation on vectors.
\begin{defi}
  Let $A$ be an $n × m$ size-change matrix, $(X,≤)$ be an ordered set and
  $\vec{x}$, $\vec{y}$ be two vectors of $X$ with $|\vec{x}| = n$ and
  $|\vec{y}| = m$. We write $\vec{y} <_A \vec{x}$ if for all $1 ≤ i ≤ n$
  and for all $1 ≤ j ≤ m$ we have $y_j < x_i$ when $A_{i,j} = -1$, and
  $y_j ≤ x_i$ when $A_{i,j} = 0$.
\end{defi}

\begin{lem}\label{lem:matcompat}
  Let $(X,≤)$ be an ordered set and $\vec{x}$, $\vec{y}$ and $\vec{z}$
  be three vectors of $X$  with $|\vec{x}| = n$, $|\vec{y}| = m$ and
  $|\vec{z}| = p$. If $A$ is an $n × m$ size-change matrix such that
  $\vec{y} <_A \vec{x}$ and if $B$ is an $m × p$ size-change matrix
  such that $\vec{z} <_B \vec{y}$ then $\vec{z} <_{AB} \vec{x}$.
\end{lem}
\begin{proof}
  Let us take $C = A B$. By definition, if $C_{i,j} = -1$ there must
  be $k$ such that ${A_{i,k} ∘ B_{k,j} = -1}$. This can only happen if
  $A_{i,k} = B_{k,j} = -1$, if $A_{i,k} = -1$ and $B_{k,j} = 0$, or if
  $A_{i,k} = 0$ and $B_{k,j} = -1$. In these three cases we respectively
  have $z_j < y_k < x_i$, $z_j < y_k ≤ x_i$ and $z_j ≤ y_k < x_i$, which
  all imply $z_j < x_i$.
  Now, if $C_{i,j} = 0$ then there must be $k$ such that
  $A_{i,k} ∘ B_{k,j} = 0$, which implies $z_j ≤ y_k ≤ x_i$.
\end{proof}

\section{Circular proofs and size change principle}\label{sct}
We will now introduce an abstract notion of circular proof, with a related
notion of \emph{well-foundedness}. The idea is to represent proofs as
directed acyclic graphs, and to label their edges with size relations
between syntactic ordinals. These size relations (expressed using
size-change matrices) are then processed using the size change principle
\cite{scp}.
In this paper, it will first allow us to build circular
subtyping proofs to handle inductive and coinductive types in
Section~\ref{language}. It will then be used to build circular typing
proofs in Section~\ref{fixpoint} to ensure the termination of recursive
programs.

Our notion of circular proof is parametrised by a notion of
\emph{abstract judgments}, their deduction rules and their semantics.
They will correspond, for example, to typing judgments or to local
subtyping judgments, with their respective deduction rules and
interpretations. We believe that the framework presented here
could be applied to other type systems involving a notion of size.

\begin{defi}\label{def:abstract}
  A language of \emph{abstract judgments} is given by a set $\mathcal{J}$
  of symbolic judgments, and an associated set $Λ$ of \emph{individuals}.
  Every symbol $J ∈ \mathcal{J}$ should depends on exactly one element of
  $Λ$ and on $|J|$ syntactic ordinals (possibly $0$).
  Optionally, for some $J ∈ \mathcal{J}$ and for all $\vec{κ} ∈ \ord^{|J|}$
  there may be a choice operator $ε_x ¬J(x,\vec{κ})$ in $Λ$, where $x$ is
  a bound variable. It will be used as a counter-example to ``for all
  $t∈Λ$, the judgment  $J(t,\vec{κ})$ is valid''.
  We denote $⟦Λ⟧ ⊂ Λ$ the set of all the individuals that do not contain
  choice operators.
\end{defi}
  
Intuitively, an abstract judgments can be seen as a predicate, which
validity depends on the truth of the denoted judgment. In the following,
such predicates will be used to build syntactic ordinal witnesses
according to Section~\ref{ordinals}. We will thus work with syntactic
ordinals of the form ${ε_{\vec{α} < \vec{κ}} J(t,\vec{α}.\vec{υ})}_i$ or
${ε_{\vec{α} < \vec{κ}} ∀x J(x,\vec{α}.\vec{υ})}_i$, for example. However,
note that we will only be able to quantify over all the individuals when
a corresponding choice choice operator $ε_x ¬J(x,\vec{κ})$ is provided.
\begin{defi}\label{def:ordjudg}
  Given a language of abstract judgments $(\mathcal{J}, Λ)$, we can build
  a language of predicates $\mathcal{P}$ using the following BNF grammar,
  where $J ∈ \mathcal{J}$ and $t ∈ Λ$.
  \begin{align*}
    P, Q \bnfeq & J(t,\vec{α})
         \bnfor ¬J(t,\vec{α})
         \bnfor ∀x J(x,\vec{α})
         \bnfor ∀x ¬J(x,\vec{α})
  \end{align*}
  We then obtain a fixed language of (parametric) syntactic ordinals by
  instantiating Definitions~\ref{def:synt_ord} and \ref{def:par_synt_ord}
  using $\mathcal{P}$.
\end{defi}

We will now consider the interpretation of individuals and abstract
judgments. Intuitively, an individual (potentially containing choice
operators) will be interpreted by a \emph{pure} individual (i.e., one
that does not contain choice operators). An abstract judgment is then
interpreted as predicates over a pure individual and (actual) ordinals.
\begin{defi}\label{def:abstract_sem}
  Let $(\mathcal{J}, Λ)$ be a language of abstract judgments. Every individual
  $t∈Λ$ is interpreted by a pure individual $⟦t⟧ ∈ ⟦Λ⟧$, and every abstract
  judgment $J ∈ \mathcal{J}$ of arity $n$ is interpreted by a function
  $⟦J⟧ : \sem{Λ} × \ordsem^n → \{0,1\}$.
  The predicates over ordinals built according to the previous definition are
  then interpreted as follows.
  \begin{align*}
    ⟦J(t,\vec{α})⟧      &= \vec{α} \mapsto ⟦J⟧(⟦t⟧,\vec{α}) \
    &⟦¬J(t,\vec{α})⟧    &= \vec{α} \mapsto 1 - ⟦J⟧(⟦t⟧,\vec{α}) \\
    ⟦∀x J(x,\vec{α})⟧   &= \vec{α}
      \mapsto \textstyle{\min_{t∈⟦Λ⟧} ⟦J⟧(t,\vec{α})} \
    &⟦∀x ¬J(x,\vec{α})⟧ &= \vec{α}
      \mapsto \textstyle{1 - \max_{t∈⟦Λ⟧} ⟦J⟧(t,\vec{α})}
  \end{align*}
  Moreover, we require that for every individual of the form
  $ε_x ¬J(x,\vec{κ})$ with $J ∈ \mathcal{J}$ and $\vec{κ} ∈ \ord$,
  we have $⟦ε_x ¬J(x,\vec{κ})⟧ = u ∈ \pure$ such that
  $⟦J⟧(u,⟦\vec{κ}⟧) = 0$ if such a $u$ exists, otherwise $u$ is
  chosen to be an arbitrary element of $\pure$.
\end{defi}
\begin{defi}\label{def:abstseq}
  An \emph{abstract sequent} $γ ⊢ J(t,\vec{κ})$ is built using an ordinal
  context $γ ⊆ \ord$, an abstract judgment $J ∈ \mathcal{J}$, an individual
  $t ∈ Λ$ and syntactic ordinals $\vec{κ} ∈ \ord^{\left| J \right|}$. We
  say that the abstract sequent $γ ⊢ J(t,\vec{κ})$ is true if we have
  $\sem{J}(\sem{t},\sem{\vec{κ}}) = 1$ whenever $\sem{τ} ≠ 0$ for all
  $τ ∈ γ$.
\end{defi}

To relate the notion of size-change matrices to abstract sequents, we
introduce \emph{ordinal constraints}. They will allow us to concisely
represent, in the form of a sequence of index, a conjunction of strict
relations between the ordinals of a given vector.
\begin{defi}
  A list of \emph{ordinal constraints} $C$ of arity $n$ is given by a
  function $C$ from $\{1,\dots,n\}$ to $\{0,1,\dots,n\}$. Given a vector
  of ordinals $\vec{o} ∈ \ordsem^n$, we denote $C(\vec{o})$ the vector
  of size $n$ defined as $C(\vec{o})_i = ⟦\OMaxi⟧$ if $C(i) = 0$ and as
  $C(\vec{o})_i = o_j$ if $C(i) = j ≠ 0$.
  We say that $C$ is satisfied by $\vec{o} ∈ \ordsem^n$ when
  $o_i < {C(\vec{o})}_i$ for all $1 ≤ i ≤ n$.
\end{defi}

Building circular proofs will require the generalisation of abstract
sequents. In other words, we will sometimes need to prove that an
abstract sequent is true for any ordinal parameters (satisfying some
constraints) and for any individual. To this aim, we introduce the
notion of \emph{general abstract sequent}.
\begin{defi}
  A \emph{general abstract sequent} is an abstract sequent that is
  quantified over. It may be of the form
  $∀ \vec{α} \left(γ ⊢ C(\vec{α}) ⇒ J(t,\vec{α}.\vec{υ})\right)$ or
  $∀ \vec{α} ∀ x \left(γ ⊢ C(\vec{α}) ⇒ J(x,\vec{α}.\vec{υ})\right)$,
  where $γ$ is an ordinal context only containing variables of $\vec{α}$,
  $C$ is a list of ordinal constraints of arity $\left|\vec{α}\right|$,
  $J$ is an abstract judgement and $t$ is an individual.
  We say that the general abstract sequent
  $∀\vec{α} \left(γ ⊢ C(\vec{α}) ⇒ J(t,\vec{α}.\vec{υ})\right)$
  (resp. $∀ \vec{α} ∀ x \left(γ ⊢ C(\vec{α}) ⇒ J(x,\vec{α}.\vec{υ})\right)$)
  is true if $\sem{J}(\sem{t},\vec{o}.\sem{\vec{υ}}) = 1$ (resp.
  $\min_{t∈\pure} \sem{J}(t,\vec{o}.\sem{\vec{υ}}) = 1$) for all $\vec{o} ∈
  \ordsem^n$ such that $o_i ≠ 0$ if $α_i ∈ γ$ and such
  that $C$ is satisfied by $\vec{o}$.
\end{defi}
Note that in a general abstract sequent, a judgement $J$ may use ordinals
$\vec{υ}$ that are not quantified over. In the following, we will often
omit to mention them explicitly. In particular, our definition implies that
the ordinal context $γ$ and the ordinals of $C(\vec{α})$ cannot use ordinals
of $\vec{υ}$ (therefore, they can only use ordinals of $\vec{α}$). This
restriction is not essential, but simplifies the definitions to come.

\begin{figure}
  \begin{prooftree}
    \AxiomC{$∀ \vec{α} (γ ⊢ C(\vec{α}) ⇒ J(t,\vec{α}))$}
    \AxiomC{$(γ[\vec{α} := \vec{κ}],δ ⊢
      κ_i < C(\vec{κ})_i)_{1 ≤ i ≤ |\vec{α}|}$}
    \RightLabel{$\G$}
    \BinaryInfC{$γ[\vec{α} := \vec{κ}], δ ⊢ J(t,\vec{κ})$}
  \end{prooftree}

  \bigskip

  \begin{prooftree}
    \AxiomC{$\H{∀ \vec{α} (γ ⊢ C(\vec{α}) ⇒ J(t,\vec{α}))}{k}$}
    \noLine
    \UnaryInfC{$\smash{\raisebox{-0.5mm}{\vdots}}\vphantom{0mm}$}
    \noLine
    \UnaryInfC{$γ[\vec{α}:=\vec{κ}] ⊢ J(t,\vec{κ})$}
    \AxiomC{where $\vec{κ} =  \vec{ε}_{\vec{α}<C(\vec{α})} ¬J(t,\vec{α})$}
    \RightLabel{$\I{k}$}
    \BinaryInfC{$∀ \vec{α} (γ ⊢ C(\vec{α}) ⇒ J(t,\vec{α}))$}
  \end{prooftree}

  \bigskip

  \begin{prooftree}
    \AxiomC{$∀ \vec{α} ∀ x (γ ⊢ C(\vec{α}) ⇒ J(x,\vec{α}))$}
    \AxiomC{$(γ[\vec{α} := \vec{κ}],δ ⊢
      κ_i < C(\vec{κ})_i)_{1 ≤ i ≤ |\vec{α}|}$}
    \RightLabel{$\GP$}
    \BinaryInfC{$γ[\vec{α} := \vec{κ}], δ ⊢ J(t,\vec{κ})$}
  \end{prooftree}

  \bigskip

  \begin{prooftree}
    \AxiomC{$\H{∀ \vec{α} ∀ x (γ ⊢ C(\vec{α}) ⇒ J(x,\vec{α}))}{k}$}
    \noLine
    \UnaryInfC{$\smash{\raisebox{-0.5mm}{\vdots}}\vphantom{0mm}$}
    \noLine
    \UnaryInfC{$γ[\vec{α}:=\vec{κ}] ⊢ J(ε_x ¬J(x,\vec{κ}),\vec{κ})$}
    \RightLabel{$\IP{k}$}
    \AxiomC{where $\vec{κ} =  \vec{ε}_{\vec{α}<C(\vec{α})} ¬∀x J(x,\vec{α})$}
    \BinaryInfC{$∀ \vec{α} ∀ x (γ ⊢ C(\vec{α}) ⇒ J(x,\vec{α}))$}
  \end{prooftree}

  \caption{Generalisation and induction rules for general abstract sequents.}
  \label{fig:gen_abs_seq}
\end{figure}

\begin{defi}
  A \emph{circular deduction system} is given by a set of deduction
  rules defined over abstract sequents (i.e., their conclusions and premises
  are abstract sequents), together with the rules of
  Figure~\ref{fig:gen_abs_seq}.
  The aim of the \emph{generalisation rules} ($\G$) and ($\GP$) is to prove
  an abstract sequent using a general abstract sequent. In particular, the
  ordinal constraints used in their first premise should be satisfied in the
  conclusion (see their second premise).
  The \emph{induction rules} ($\I{k}$) and ($\IP{k}$) may be used to prove
  a general abstract sequent using itself as an hypothesis (this is the
  meaning of the square brackets).\footnote{Note that the ($\IP{k}$)
  rule relies on the individuals of the form $ε_x ¬J(x,\vec{κ})$ required by
  Definition~\ref{def:abstract}.} Note that a natural number $k$ (unique
  in a proof) is used to keep track of the originating induction rule.
\end{defi}

\begin{figure}
  \begin{prooftree}
    \AxiomC{$\left[∀{()} (⊢ () ⇒ J(t,\vec{υ}))\right]_0$}
    \AxiomC{$\varnothing$}
    \RightLabel{$\G$}
    \BinaryInfC{$⊢ J(t,\vec{υ})$}
    \RightLabel{$\I{0}$}
    \UnaryInfC{$∀{()} (⊢ {()} ⇒ J(t,\vec{υ}))$}
    \AxiomC{$\varnothing$}
    \RightLabel{$\G$}
    \BinaryInfC{$γ ⊢ J(t,\vec{υ})$}
  \end{prooftree}
  \caption{Example of a circular proof that is not well-founded.}
  \label{fig:badex}
\end{figure}
The rules ($\I{k}$), ($\G$), ($\IP{k}$) and ($\GP$) are the only ones
allowed to manipulate general abstract sequents. The induction rules alone
are responsible for the circular structure of the proofs in a circular
deduction system. In particular, they allow for clearly invalid proofs such
as the on of Figure~\ref{fig:badex}, which can be used to prove an arbitrary
abstract sequent. After applying a generalisation rule over the empty vector
of ordinals $()$, the invalid proof is constructed by using an induction rule
and by applying the new hypothesis directly.

As a circular deduction system can be used to build incorrect circular
proofs, we will need to rely on a well-foundedness criterion. In other
words, a derivable (general) abstract sequent will only be considered
correct if its derivation is well-founded.
In this paper, we will rely on the size-change principle \cite{scp} to
obtain a sufficient condition for a given proof to be well-founded. To
this aim, circular proofs first need to be decomposed into \emph{blocks}.
\begin{defi}
  Given a proof $Π$ expressed in a circular proof system, a \emph{block}
  is a subproof $B$ of $Π$ such that its conclusion is either the conclusion
  of $Π$ or some general abstract sequent, and its premises (if any) are
  also general abstract sequents.
  We require blocks to be minimal, which means that they should not
  contain general abstract sequents (except in their conclusions and
  premises). This condition implies that a proof admits a unique
  decomposition into blocks. A block $B$ has an arity $|B|$ which is
  $0$ if the conclusion of the block is also the conclusion of $Π$,
  and it is the size of the quantified vector of ordinals $\vec{α}$
  in the conclusion of $B$ otherwise.
\end{defi}

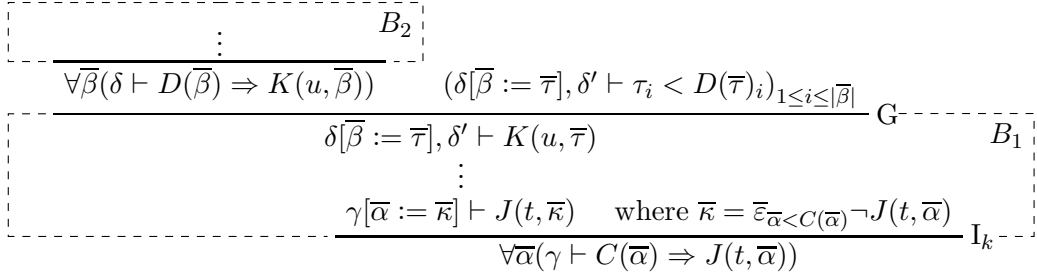
\begin{figure}
  \begin{prooftree}
    \AxiomC{\tikzmark{fff}}
    \noLine
    \UnaryInfC{$\vdots$}
    \LeftLabel{\raisebox{-1.4mm}{\tikzmark{ggg}}}
    \RightLabel{\raisebox{-1.4mm}{\tikzmark{eee}}}
    \UnaryInfC{$∀ \vec{β} (δ ⊢ D(\vec{β}) ⇒ K(u,\vec{β}))$}
    \AxiomC{$(δ[\vec{β} := \vec{τ}], δ' ⊢
      τ_i < D(\vec{τ})_i)_{1 ≤ i ≤ |\vec{β}|}$}
    \LeftLabel{\raisebox{-1.4mm}{\tikzmark{ccc}}}
    \RightLabel{$\G$\raisebox{0.65mm}{\tikzmark{ddd}}}
    \BinaryInfC{$δ[\vec{β} := \vec{τ}], δ' ⊢ K(u,\vec{τ})$}
    \noLine
    \UnaryInfC{$\smash{\raisebox{-0.5mm}{\vdots}}\vphantom{0mm}$}
    \noLine
    \UnaryInfC{$γ[\vec{α}:=\vec{κ}] ⊢ J(t,\vec{κ})$}
    \AxiomC{\hspace{-11.5em}where $\vec{κ} =
      \vec{ε}_{\vec{α}<C(\vec{α})} ¬J(t,\vec{α})$}
    \LeftLabel{\raisebox{-1.4mm}{\tikzmark{aaa}}}
    \RightLabel{$\I{k}$\raisebox{0.3mm}{\tikzmark{bbb}}\hspace{3mm}}
    \BinaryInfC{$∀ \vec{α} (γ ⊢ C(\vec{α}) ⇒ J(t,\vec{α}))$}
  \end{prooftree}
  \begin{tikzpicture}[remember picture, overlay]%
    \draw[dashed] ([xshift=0pt,yshift=2pt]{pic cs:ddd}) --+
                  (1.8,0) node [below left] {$B₁$} |-
                  ([xshift=0pt,yshift=2pt]{pic cs:bbb});
    \draw[dashed] ([xshift=0pt,yshift=2pt]{pic cs:ccc}) --+
                  (-0.5,0) |-
                  ([xshift=0pt,yshift=2pt]{pic cs:aaa});
    \draw[dashed] ([xshift=0pt,yshift=2pt]{pic cs:ggg}) --+
                  (-0.5,0) |-
                  ({pic cs:fff});
    \draw[dashed] ({pic cs:fff})--+
                  (2.7,0) node [below left] {$B₂$} |-
                  ([xshift=0pt,yshift=2pt]{pic cs:eee});
  \end{tikzpicture}%
  \caption{Construction of an edge of the call graph
    (see Definition~\ref{def:edgeconstr}).}
  \label{fig:defdef}
\end{figure}
\begin{defi}\label{def:edgeconstr}
  Let $Π$ be a proof expressed in a circular proof system. The \emph{call
  graph} of $Π$ is the graph induced by the block structure of $Π$. Its
  vertices are the blocks of $Π$, and every block $B$ has one outgoing
  edge for each of its premises. It is directed toward the block proving
  the considered premise, which may be directly above $B$ in $Π$, $B$
  itself, or even a block below $B$. In the latter two cases, the premise
  must correspond to an hypothesis in square brackets introduced by an
  instance of the ($\I{k}$) or ($\IP{k}$) rules.

  Every edge $(B₁,B₂)$ of a call graph is labeled by a size-change matrix
  $M$. To give its definition, we need to remark that a premise of a block
  necessarily uses the ($\G$) or ($\GP$) rules. Indeed, they are the
  only available rules having a general abstract sequent as a premise. As
  a consequence, we can represent the block $B₁$ as in
  Figure~\ref{fig:defdef}, if we only include the premises involved in the
  definition of the edge $(B₁,B₂)$.\footnote{The structure is the same for
  the three other cases: ($\GP$) with ($\IP{k}$), ($\GP$) with ($\I{k}$) or
  ($\G$) with ($\IP{k}$).}
  The $|\vec{α}| × |\vec{β}|$ matrix $M$ attached to the edge $(B₁,B₂)$ is
  then defined as $M_{i,j} = -1$ when $δ[\vec{β} := \vec{τ}], δ' ⊢ τ_j < κ_i$
  is derivable, $M_{i,j} = 0$ when only $δ[\vec{β} := \vec{τ}], δ' ⊢ τ_j
  ≤ κ_i$ is derivable, and $M_{i,j} = ∞$ otherwise.
\end{defi}

As the edges of a call graph are labeled with matrices, any path in its
transitive closure can also be assigned a label using the matrix product
of the labels along the path.
In particular, if there is a path from $B₁$ to $B₂$ with label $M$ and
a path from $B₂$ to $B₃$ with label $N$, then there is a path from $B₁$
to $B₃$ with label $M N$.
Since a call graph has finitely many vertices and edges, the number of
possible labels for a path in the transitive closure of the graph is also
finite. If we consider two paths with the same label to be equal, then
there can only be finitely many distinct paths in the transitive closure
of a call graph. It can hence be computed in finite time by composing
edges until saturation.
\begin{defi}\label{def:wellfounded}
  We say that a proof is \emph{well-founded} if every idempotent loop
  in the transitive closure of its call graph (i.e. closed path with
  label $M$ such that $M M = M$) has at least one $-1$ on the diagonal
  of its label. Note that such loops are necessarily labeled with square
  matrices.
\end{defi}

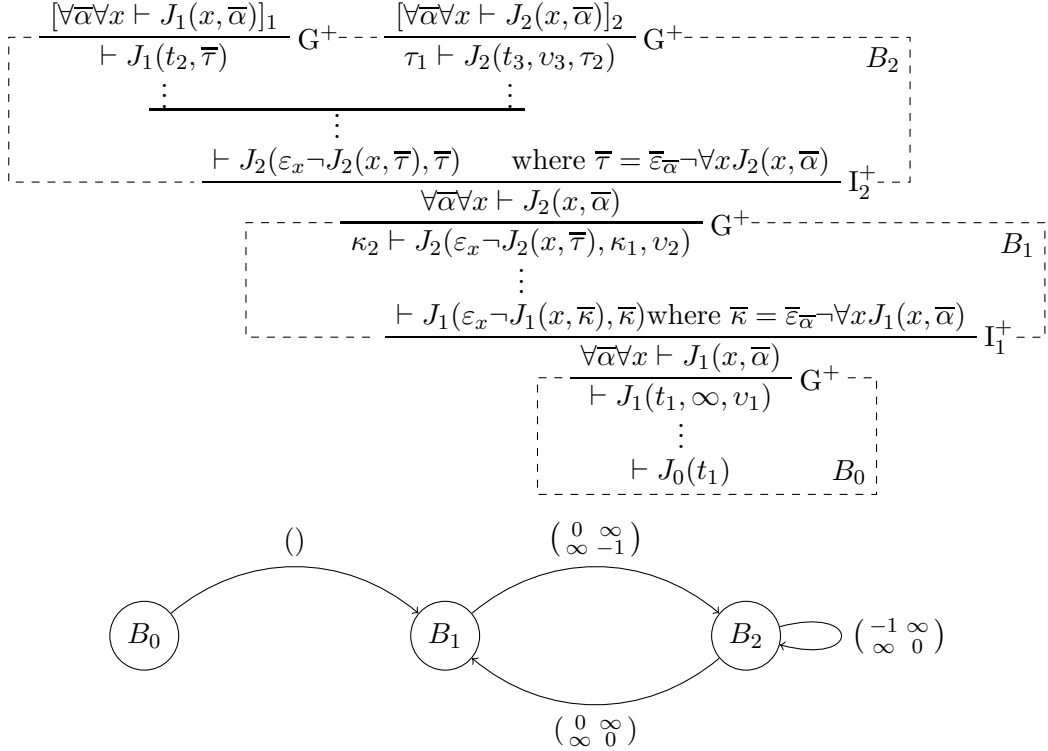
\begin{figure}
  \centering
  \begin{prooftree}
    \AxiomC{$\H{∀\vec{α} ∀x ⊢ J₁(x,\vec{α})}{1}$}
    \LeftLabel{\raisebox{-1.4mm}{\tikzmark{a}}}
    \RightLabel{$\GP$\raisebox{0.9mm}{\tikzmark{b}}}
    \UnaryInfC{$⊢ J₁(t₂,\vec{τ})$}
    \noLine
    \UnaryInfC{$\smash{\raisebox{-0.5mm}{\vdots}}\vphantom{0mm}$}
    \AxiomC{$\H{∀\vec{α} ∀x ⊢ J₂(x,\vec{α})}{2}$}
    \LeftLabel{\raisebox{-1.4mm}{\tikzmark{c}}}
    \RightLabel{$\GP$\raisebox{0.9mm}{\tikzmark{d}}}
    \UnaryInfC{$τ₁ ⊢ J₂(t₃,υ₃,τ₂)$}
    \noLine
    \UnaryInfC{$\smash{\raisebox{-0.5mm}{\vdots}}\vphantom{0mm}$}
    \BinaryInfC{$\smash{\raisebox{-0.5mm}{\vdots}}\vphantom{0mm}$}
    \noLine
    \UnaryInfC{$⊢ J₂(εₓ ¬ J₂(x,\vec{τ}),\vec{τ})$}
    \AxiomC{\hspace{-8em} where $\vec{τ} = \vec{ε}_{\vec{α}} ¬∀x J₂(x,\vec{α})$}
    \LeftLabel{\raisebox{-1.4mm}{\tikzmark{e}}}
    \RightLabel{$\IP{2}$\raisebox{0.4mm}{\tikzmark{f}}}
    \BinaryInfC{$∀\vec{α} ∀x ⊢ J₂(x,\vec{α})$}
    \LeftLabel{\raisebox{-1.4mm}{\tikzmark{ee}}}
    \RightLabel{$\GP$\raisebox{0.9mm}{\tikzmark{ff}}\hspace{1cm}}
    \UnaryInfC{$κ₂ ⊢ J₂(εₓ ¬ J₂(x,\vec{τ}),κ₁,υ₂)$}
    \noLine
    \UnaryInfC{$\smash{\raisebox{-0.5mm}{\vdots}}\vphantom{0mm}$}
    \noLine
    \UnaryInfC{$⊢ J₁(εₓ ¬ J₁(x,\vec{κ}),\vec{κ})$}
    \AxiomC{\hspace{-10em} where $\vec{κ} = \vec{ε}_{\vec{α}}¬∀x J₁(x,\vec{α})$}
    \LeftLabel{\raisebox{-1.4mm}{\tikzmark{g}}}
    \RightLabel{$\IP{1}$\raisebox{0.4mm}{\tikzmark{h}}\hspace{4mm}}
    \BinaryInfC{$∀\vec{α} ∀x ⊢ J₁(x,\vec{α})$}
    \LeftLabel{\raisebox{-1.4mm}{\tikzmark{gg}}}
    \RightLabel{$\GP$\raisebox{0.9mm}{\tikzmark{hh}}}
    \UnaryInfC{$⊢ J₁(t₁,∞,υ₁)$}
    \noLine
    \UnaryInfC{$\smash{\raisebox{-0.5mm}{\vdots}}\vphantom{0mm}$}
    \noLine
    \UnaryInfC{$⊢ J₀(t₁)$}
    \noLine
    \UnaryInfC{\tikzmark{i}\tikzmark{j}}
  \end{prooftree}
  \begin{tikzpicture}[remember picture, overlay]
    \draw[dashed] ([xshift=0pt,yshift=2pt]{pic cs:b}) --
                  ([xshift=0pt,yshift=2pt]{pic cs:c});
    \draw[dashed] ([xshift=0pt,yshift=2pt]{pic cs:d}) --+
                  (3.0,0) node [below left] {$B₂$} |-
                  ([xshift=0pt,yshift=2pt]{pic cs:f});
    \draw[dashed] ([xshift=0pt,yshift=2pt]{pic cs:a}) --+
                  (-0.3,0) |-
                  ([xshift=0pt,yshift=2pt]{pic cs:e});

    \draw[dashed] ([xshift=0pt,yshift=2pt]{pic cs:ff}) --+
                  (3.9,0) node [below left] {$B₁$} |-
                  ([xshift=0pt,yshift=2pt]{pic cs:h});
    \draw[dashed] ([xshift=0pt,yshift=2pt]{pic cs:ee}) --+
                  (-1.15,0) |-
                  ([xshift=0pt,yshift=2pt]{pic cs:g});

    \draw[dashed] ([xshift=0pt,yshift=2pt]{pic cs:j}) --+
                  (2.6,0) node [above left] {$B₀$} |-
                  ([xshift=0pt,yshift=2pt]{pic cs:hh});
    \draw[dashed] ([xshift=0pt,yshift=2pt]{pic cs:gg}) --+
                  (-0.33,0) |-
                  ([xshift=0pt,yshift=2pt]{pic cs:i});
  \end{tikzpicture}
  \begin{tikzpicture}
    \node[circle,draw] (B0) at (0,0) {$B₀$};
    \node[circle,draw] (B1) at (4,0) {$B₁$};
    \node[circle,draw] (B2) at (8,0) {$B₂$};
    \path [->] (B0) edge[bend left=40] node[above]
          {$\left(\right)$} (B1);
    \path [->] (B1) edge[bend left=40] node[above]
          {$\left(\begin{smallmatrix}
            0 & ∞\\ ∞ & -1\end{smallmatrix}\right)$} (B2);
    \path [->] (B2) edge[bend left=40] node[below]
          {$\left(\begin{smallmatrix}
            0 & ∞\\ ∞ &  0\end{smallmatrix}\right)$} (B1);
    \path [->,every loop/.style={looseness=12}]
          (B2) edge[loop right] node[right]
          {$\left(\begin{smallmatrix}
            -1& ∞\\ ∞ &  0\end{smallmatrix}\right)$} (B2);
  \end{tikzpicture}
  \caption{Example of a circular proof and the corresponding call graph.}
  \label{fig:circularex}
\end{figure}

\begin{exa}\label{exa:sctcirc}
  We now consider the example of circular proof given in the upper part of
  Figure~\ref{fig:circularex}. For simplicity, the individuals and ordinals
  are not given explicitly. We will however assume that (besides reflexivity)
  it is possible to derive $κ₂ ⊢ υ₂ < κ₂$ and $τ₁ ⊢ υ₃ < τ₁$. The proof can
  be decomposed into three blocks $B₀$, $B₁$ and $B₂$. The corresponding call
  graph is given in the lower part of Figure~\ref{fig:circularex}.
  Its transitive closure contains five idempotent loops. There are none on
  the block $B₀$, two on the block $B₁$ with labels
  {\tiny$\left(\begin{smallmatrix}0 & ∞\\ ∞ & -1\end{smallmatrix}\right)$} and
  {\tiny$\left(\begin{smallmatrix}-1 & ∞\\ ∞ & -1\end{smallmatrix}\right)$},
  and three on block $B₂$ with labels
  {\tiny$\left(\begin{smallmatrix}-1 & ∞\\ ∞ & 0\end{smallmatrix}\right)$},
  {\tiny$\left(\begin{smallmatrix}-1 & ∞\\ ∞ & -1\end{smallmatrix}\right)$} and
  {\tiny$\left(\begin{smallmatrix}0 & ∞\\ ∞ & -1\end{smallmatrix}\right)$}.
  We can thus conclude that our proof example is indeed well-founded since
  every idempotent loop is labeled with a matrix having at least one
  $-1$ on its diagonal\footnote{The proof example of
  Figure~\ref{fig:circularex} corresponds to the block decomposition of
  Figure~\ref{fig:munu} page \pageref{fig:munu}.}.
\end{exa}
\begin{exa}
  The circular proof of Figure~\ref{fig:badex} is built using only two blocks.
  The upper block has one loop labelled with the empty matrix. It is therefore
  not well-founded.
\end{exa}

\begin{lem}\label{lem:Gcorrect}
  The four deduction rules of Figure~\ref{fig:gen_abs_seq} are correct. In
  other words, if the premises of such a rule are semantically valid, then
  so is its conclusion.
\end{lem}
\begin{proof}
  The ($\G$) and ($\GP$) rules can be seen as the composition of standard
  elimination rules for the universal quantifier, followed by a weakening
  of the ordinal context. They are therefore correct. For the ($\I{k}$) and
  ($\IP{k}$) rules, we consider the semantics of the choice operators over
  ordinals (and the choice operator over individuals for the ($\IP{k}$)
  rule). By definition, if the conclusion of the sequent is false, then
  there is a counterexample that the choice operator can use. However in
  this case, the premise of the rule is false as well, which implies the
  correctness by contraposition.
\end{proof}
Note that the correctness of the ($\I{k}$) and ($\IP{k}$) rules rely on
the fact that we ignore the hypothesis they introduce. They justification
for such hypotheses are handled globally by our notion of well-founded
proof (Definition~\ref{def:wellfounded}).

\begin{thm}\label{th:wellfounded}
  Let us assume that all the deduction rules for abstract sequents are
  correct with respect to the semantics. If an abstract sequent admits
  a well-founded circular proof then it is true in any model.
\end{thm}
\begin{proof}
  Let us consider an abstract sequent that is derivable using a
  well-founded circular proof. We will assume, by contradiction,
  that there is a model $\calM$ such that the considered abstract
  sequent is false. As all the deduction rules are supposed correct
  (by hypothesis and by Lemma \ref{lem:Gcorrect}), the call-graph
  of our proof necessarily contains cycles. We will thus unroll the
  proof to exhibit an infinite branch that will imply the existence
  of an infinite, decreasing sequence of ordinals (which is a
  contradiction).

  We will now build an infinite sequence $(B_i,\vec{o}_i,\calM_i)_{i∈\N}$
  of triples of a block, a vector of ordinals and a model. We will take
  $B₀$ to be the block at the root of our proof, $\vec{o}₀$ to be the empty
  vector and $\calM₀$ to be $\calM$. By construction, we will enforce that
  for all $i$ the conclusion of $B_i$ is false in $\calM_i$ and that
  $\vec{o}_i$ is a counterexample (thus $|\vec{o}_i| = |B_i|$). We will
  also require that the call-graph contains an edge linking $B_i$ to
  $B_{i+1}$ labeled with a matrix $M_i$ such that $\vec{o}_{i+1} <_{M_i}
  \vec{o}_i$ (the conclusion of $B_{i+1}$ is thus a general abstract
  sequent).

  Note that the first element of the sequence $(B₀,\vec{o}₀,\calM₀)$
  satisfies the above conditions. In particular, the conclusion of $B₀$
  has been assumed to be false (independently of any ordinal). Moreover,
  the matrix labeling the edge between $B_0$ and $B_1$ will be empty.

  Let us now suppose that the sequence has been constructed for all
  $0≤j≤i$, and define $(B_{i+1},\vec{o}_{i+1},\calM_{i+1})$. If $i ≠ 0$
  then the conclusion of $B_i$ must be a general sequent, which means
  that the last rule in $B_i$ is either $\I{k}$ or $\IP{k}$. Without
  loss of generality we can assume that it is $\IP{k}$, and thus $B_i$
  ends with the following rule.
  \begin{prooftree}
    \AxiomC{$γ[\vec{α}:=\vec{κ}] ⊢ J(ε_x ¬J(x,\vec{κ}),\vec{κ})$}
    \AxiomC{where $\vec{κ} = \vec{ε}_{\vec{α}<C(\vec{α})} ¬∀x J(x,\vec{α})$}
    \RightLabel{$\IP{k}$}
    \BinaryInfC{$∀ \vec{α} ∀ x (γ ⊢ C(\vec{α}) ⇒ J(x,\vec{α}))$}
  \end{prooftree}
  By construction, we know that $∀\vec{α}∀x(γ ⊢ C(\vec{α})⇒J(x,\vec{α}))$
  is false in the model $\calM_i$ and that $\vec{o}_i$ is a counterexample.
  This means that
  $\sem{γ[\vec{α} := \vec{o}_i]}^{\calM_i}$ contains only positive
  ordinals, $C$ is satisfied by $\vec{o}$ and
  $\sem{J}^{\calM_i}(t,\vec{o}_i) = 0$ for all $t ∈ Λ$.
  Thus, using Lemma \ref{lem:changemodel} we can define $\calM_{i+1}$ to be
  a model such that $\sem{\vec{κ}}^{\calM_{i+1}} = \vec{o}_i$.
  By definition \ref{def:abstract_sem} the individual
  $t = \sem{ε_x ¬J(x,\vec{κ})}^{\calM_{i+1}}$ satisfies
  $\sem{J(t,\vec{o}_i)} = 0$. This establishes that the
  premise of our $\IP{k}$ rule is a false abstract sequent in the model
  $\calM_{i+1}$.

  As all the deduction rules for abstract sequents are supposed correct,
  at least one premise of the block $B_i$ must be false in the model
  $\calM_{i+1}$. The first rule of such a leaf must be either $\G$ or
  $\GP$ as they are the only deduction rules having a general abstract
  sequent as premise. Without loss of generality we can assume a $\GP$ rule.
  \begin{prooftree}
    \AxiomC{$∀ \vec{β} ∀ y (δ ⊢ D(\vec{β}) ⇒ K(y,\vec{β}))$}
    \AxiomC{$(δ[\vec{β} := \vec{τ}],δ' ⊢ τ_i < D(\vec{τ})_i)_{1≤i≤|\vec{β}|}$}
    \RightLabel{$\GP$}
    \BinaryInfC{$δ[\vec{β} := \vec{τ}], δ' ⊢ K(u,\vec{τ})$}
  \end{prooftree}
  As the conclusion of this rule is false is the model $\calM_{i+1}$, we
  know that $\sem{δ[\vec{β} := \vec{τ}], δ'}^{\calM_{i+1}}$ only contains
  positive ordinals and that $\sem{K}^{\calM_{i+1}}(\sem{u}^{\calM_{i+1}},
  \sem{\vec{τ}}^{\calM_{i+1}}) = 0$. By Proposition \ref{lem:leqicorrect},
  the right premises of our $\GP$ rule cannot be false. Therefore,
  $∀\vec{β}∀y (δ ⊢ D(\vec{β}) ⇒ K(y,\vec{β}))$
  must be false in the model $\calM_{i+1}$.
  Therefore, we can define $B_{i+1}$ to be the block proving this sequent
  and $\vec{o}_{i+1}$ to be $\sem{\vec{τ}}^{\calM_{i+1}}$, which is indeed
  a counterexample for this sequent.

  By definition, there is an edge linking the block $B_i$ to the block
  $B_{i+1}$ in the call-graph. It is labeled with a matrix $M_i$ and we
  will show $\vec{o}_{i+1} <_{M_i} \vec{o}_i$ to conclude the construction
  of our sequence.
  Let us take $1 ≤ m ≤ |\vec{o}_i|$ and $1 ≤ n ≤ |\vec{o}_{i+1}|$ and
  consider $(M_i)_{m,n}$. If it is equal to $-1$ then there is a proof
  of $\smash{δ[\vec{β} := \vec{τ}],δ' ⊢ τ_n<κ_m}$ and hence proposition
  \ref{lem:leqicorrect} gives us $\sem{τ_n}^{\calM_{i+1}} <
  \sem{κ_m}^{\calM_{i+1}}$.
  We can hence conclude that $o_{i+1,n} < o_{i,m}$ since we have
  $\sem{κ_m}^{\calM_{i+1}} = o_{i,m}$ and
  $ o_{i+1,n} = \sem{τ_n}^{\calM_{i+1}}$ by definition of
  $\calM_{i+1}$ and $\vec{o}_{i+1}$ respectively.
  If it is $0$ then a similar reasoning can be applied to get
  $o_{i+1,n} ≤ o_{i,m}$ and if it is $∞$ then there is nothing to prove.

  To conclude, we will now use the same argument as in the proof of
  \cite[Theorem~4]{scp}. For all $0 ≤ i < j$, we define $M_{i,j}$ to
  be the matrix $M_i M_{i+1} \dots M_{j-1}$. The number of possible
  different tuples of the form $(B_i,B_j,M_{i,j})$ being finite, we
  can apply Ramsey's theorem for pairs to find an infinite, increasing
  sequence of natural numbers $(u_n)_{n∈\N}$ such that the tuples of
  the form $(B_{u_i},B_{u_j},M_{u_i,u_j})$ with $0 ≤ i < j$ are all
  equal. We will call $M$ the matrix contained in all of these tuples.
  Thanks to the associativity of the matrix product and to the
  definition of $M_{i,j}$, this implies that
  $M M = M_{u_0,u_1} M_{u_1,u_2} = M_{u_0,u_2} = M$.

  Finally, we can use Lemma \ref{lem:matcompat} to obtain
  $\vec{o}_j <_{M_{i,j}} \vec{o}_i$ for all $0 ≤ i < j$. Our circular
  proof being well-founded, the matrix $M$ must have a $-1$ on the
  diagonal at some index $k$. Therefore, $\vec{o}_{u_{i+1}} <_M
  \vec{o}_{u_i}$ implies that $o_{u_{i+1},k} < o_{u_i,k}$ for all
  $i ∈ \N$, which gives an infinite, decreasing sequence of ordinals
  $(o_{u_i,k})_{i ∈ \N}$ and thus a contradiction.
\end{proof}
%%% Local Variables:
%%% ispell-personal-dictionary: "~/.hunspell-en"
%%% ispell-local-dictionary: "british"
%%% End:

\section{Language and type system}\label{language}
In this section, we consider a first (restricted) version of our
language and type system. It does not provide general recursion
and is shown strongly normalising in Section~\ref{semantics}.
Surprisingly, recursion is still possible (for specific algebraic
data types) using $λ$-calculus recursors that are typable thanks to
subtyping (see Section~\ref{scottrec}).
The language is formed using three syntactic entities: terms,
types and syntactic ordinals (see Section~\ref{ordinals}).
Syntactic ordinals are used to annotate types with a size
information that is used to show the well-foundedness of
subtyping proofs. They are only introduced internally and they
are not accessible to the user.
However, we will see in Section~\ref{fixpoint} that the type
system can be naturally extended to allow the user to express
size invariants using ordinals.
Although the system is Curry-style (or implicitly typed), terms,
types and ordinals are defined mutually inductively due the choice
operators that are contained in their syntax.

\begin{defi}\label{def:syntax}
  Let $\mathcal{V}_{Λ} = \{x,y,z,\dots\}$, $\mathcal{V}_\mathcal{F} =
  \{X,Y,Z,\dots\}$ be two disjoint and countable sets of $λ$-variables
  and propositional variables respectively.
  The set of terms (or individuals) $Λ$, the set of types (or formulas)
  $\mathcal{F}$ and the set of syntactic ordinals $\mathcal{O}$ are
  defined mutually inductively. The terms and types are defined using
  the following two BNF grammars.
  \begin{align*}
    t, u
    \bnfeq &x
    \bnfor λ x.t
    \bnfor t\;u
    \bnfor \{(l_i = t_i)_{i ∈ I}\}
    \bnfor t.l_k
    \bnfor C_k\,t
    \bnfor \case{t}{(C_i → t_i)_{i ∈ I}}
    \bnfor ε_{x∈A}(t∉B)\\[1ex]
    A, B
    \bnfeq&X
    \bnfor \{(l_i : A_i)_{i ∈ I}\}
    \bnfor \{(l_i : A_i)_{i ∈ I}; \dots\}
    \bnfor [(C_i \of A_i)_{i ∈ I}]
    \bnfor A → B
    \bnfor \\
          &∀X. A
    \bnfor ∃X. A
    \bnfor μ_κ X. A
    \bnfor ν_{κ} X. A
    \bnfor ε_{X}(t ∈ A)
    \bnfor ε_{X}(t ∉ A)
  \end{align*}
  The syntactic ordinals are build according to Definitions~\ref{def:synt_ord}
  and \ref{def:ordjudg} using abstract judgments of the form $J(t,\vec{τ}) =
  t:A$ and $J(t,\vec{τ}) = t∈A⊂B$, where the ordinals of $\vec{τ}$ may appear
  in the formulas $A$ and $B$. Note that choice operators for individuals are
  provided for all the abstract judgments of the second form. Formally,
  Definition~\ref{def:abstract_sem} requires terms of the form
  $ε_x ¬(x:A)$ and $ε_x ¬(x∈A⊂B)$. The former will not be provided in the
  syntax since we will never use the ($\GP$) and ($\IP{k}$) rules on typing
  judgments. The latter will be syntactically encoded as $ε_{x∈A}(x∉B)$,
  which will have the intended semantics. Note that in general, we require
  terms of the form $ε_{x∈A}(t∉B)$ not to contain any free $λ$-variable
  (e.g., ${λ y. ε_{x∈A}(y\,x ∉ B)}$ is not valid).
\end{defi}

The term language contains the usual syntax of the $λ$-calculus extended with
records, projections, constructors and pattern matching (see the reduction
rules of Figure~\ref{fig:reduction}).
A term of the form ${ε_{x∈A}(t ∉ B)}$ corresponds to a choice operator
denoting a closed term $u$ of type $A$ such that $t[x := u]$ does not have
type $B$.\footnote{Note that in a choice operator like $ε_{x∈A}(t∉B)$,
the variable $x$ is bound in the term $t$.} The restriction to closed choice
is absolutely necessary for their interpretation in the semantics.
\begin{nota}
  In our meta-language, we use the notation $\{(l_i = t_i)_{i ∈ I}\}$ (where
  $I$ is a finite subset of $\mathbb{N}$) to denote a record. For example, if
  $I = \{1, 2\}$ then $\{(l_i=t_i)_{i∈I}\}$ corresponds to $\{l₁=t₁;l₂=t₂\}$.
  Similar notations are used for pattern matchings, product types and sum
  types. In particular, if $i ∈ \mathbb{N}$ then $l_i$ is a record field
  label and $C_i$ is a constructor (or variant).
\end{nota}

In addition to the usual types of System F, our system provides sums and
products (corresponding to variants and records), existential types,
inductive types and coinductive types.
Note that our product types may be either \emph{strict} or \emph{extensible}.
A record having an extensible product type (marked with an ellipsis) will be
allowed to contain more fields than those explicitly specified, while records
with a strict product type will only contain the specified fields. From a
subtyping point of view, extensible records are obviously more interesting.
However, strict product types will allow us to express a stronger type safety
result based on a semantic proof (Theorem~\ref{th:safety}).
Our inductive and coinductive types carry size information in the form of a
syntactic ordinals $κ$. The ordinal $∞$ is supposed to be large enough so
that the construction of $μ_{∞} F$ and $ν_{∞} F$ converges. In particular,
when $F$ is covariant then correspond to the least and greatest fixpoints of
$F$.
Choice operators $ε_{X}(t ∈ A)$ and $ε_{X}(t ∉ A)$ are also provided for
types.\footnote{In the choice operators $ε_X(t∈A)$ and $ε_X(t∉A)$ for types,
the variable $X$ is bound in $A$ only.} As for our term choice operators,
they correspond to witnesses of the property they denote, and they will be
interpreted as such in the semantics. However, contrary to term choice
operators, they do not need to be closed to be given a semantical
interpretation.

\begin{nota}
  To lighten the syntax and reduce the need for parentheses we will use some
  syntactic sugars. We will sometimes group binders and write $λ x\,y.t$ for
  $λ x.λ y.t$, and $∀ X\,Y.A$ for $∀ X.∀ Y.A$. Morever, we will consider that
  binders have the lowest priority, which means that $λ x.x\,x$ is to be read
  as $λ x.(x\,x)$, and $∀ X. A ⇒ B$ as $∀ X.(A ⇒ B)$. We will write $μ X. A$
  for $μ_{∞} X. A$ and $ν X. A$ for $ν_{∞} X. A$, and we will sometimes use
  the letter $F$ to denote a type with one parameter $X \mapsto A$ so that we
  can write $F(μ_{κ} F)$ for $A[X := μ_{κ} X.A]$. In pattern matchings, we
  will use the notation $C_k x → t$ to denote $C_k → λ x.t$. Finally, we will
  will write $t.C_k$ for the term $\case{t}{C_k → λ x.x}$, also written
  $\case{t}{C_k\,x → x}$.
\end{nota}

\begin{figure}
  \begin{multicols}{2}
  \noindent
  \raggedcolumns
  \begin{align*}
    (λ x. t) u                        &\succ t[x := u] \\
    \{(l_i : t_i)_{i∈ I}\}.l_j        &\succ
      \begin{cases}
        t_j &\hbox{ if } j ∈ I\\
        Ω &\hbox{ otherwise }
      \end{cases}\\
    \case{C_j\;u}{(C_i → t_i)_{i∈ I}} &\succ
      \begin{cases}
        t_j\;u &\hbox{ if } j ∈ I\\
        Ω &\hbox{ otherwise }
      \end{cases}
  \end{align*}
  \columnbreak
  \begin{align*}
    \case{(λ x. t)}{(C_i → t_i)_{i ∈ I}}            &\succ Ω\\
    (λ x. t).l_k                                    &\succ Ω\\
    \{(l_i = t_i)_{i ∈ I}\}\;u                      &\succ Ω\\
    (C_k\;t)\;u                                     &\succ Ω\\
    \case{\{(l_i = t_i)_{i∈I}\}}{(C_i → t_i)_{i∈I}} &\succ Ω\\
    (C_k\;t).l_i                                    &\succ Ω
  \end{align*}
  \end{multicols}
  \caption{Reduction rules of the language (without general recursion).}
  \label{fig:reduction}
\end{figure}
We now define the reduction relation of our language, which contains
$β$-reduction and rules for pattern matching and record projection.
The terms corresponding to runtime errors are also reduced to a
diverging term $Ω$ for termination to subsume type safety. 
\begin{defi}\label{def:reduction}
  The reduction relation $(\succ) \subseteq Λ \times Λ$ is defined as the
  contextual closure of the rules given in Figure~\ref{fig:reduction}. Its
  reflexive, transitive closure is denoted $(≻^*)$.
\end{defi}

As our system relies on choice operators, usual typing contexts assigning
a type to free variables are not required. In particular, open terms will
never appear in typing and subtyping rules.
\begin{figure}

% Arrow introduction
\begin{prooftree}
\AxiomC{$\subtype{}{\lambda x.t}{A \to B}{C}$}
\AxiomC{$\type{t[x := \epsilon_{x\in A}(t \notin B)]}{B}$}
\RightLabel{$\to_i$}
\BinaryInfC{$\type{\lambda x.t}{C}$}
\end{prooftree}

\smallskip

% Arrow elimination and epsilon
\begin{prooftree}
\AxiomC{$\type{t}{A \to B}$}
\AxiomC{$\type{u}{A}$}
\RightLabel{$\to_e$}
\BinaryInfC{$\type{t\;u}{B}$}
\DisplayProof\hfill
\AxiomC{$\subtype{}{\epsilon_{x\in A}(t \notin B)}{A}{C}$}
\RightLabel{$\epsilon$}
\UnaryInfC{$\type{\epsilon_{x\in A}(t \notin B)}{C}$}
\end{prooftree}

\smallskip

% Product introduction and elimination
\begin{prooftree}
\AxiomC{$\subtype{}{\{(l_i = t_i)_{i \in I}\}}{
  \{(l_i : A_i)_{i \in I}\}}{B}$}
\AxiomC{$(\type{t_i}{A_i})_{i \in I}$}
\RightLabel{$\times_i$}
\BinaryInfC{$\type{\{(l_i = t_i)_{i \in I}\}}{B}$}
\DisplayProof\hfill
\AxiomC{$\type{t}{\{l_k : A; \dots\}}$}
\RightLabel{$\times_e$}
\UnaryInfC{$\type{t.l_k}{A}$}
\end{prooftree}

\smallskip

% Sum introduction and elimination
\begin{prooftree}
\AxiomC{$\subtype{}{C_k\,t}{[C_k \of A]}{B}$}
\AxiomC{$\type{t}{A}$}
\RightLabel{$+_i$}
\BinaryInfC{$\type{C_k\,t}{B}$}
\DisplayProof\hfill
\AxiomC{$\type{t}{[(C_i \of A_i)_{i \in I}]}$}
\AxiomC{$(\type{t_i}{A_i \to B})_{i \in I}$}
\RightLabel{$+_e$}
\BinaryInfC{$\type{\case{t}{(C_i \to t_i)_{i \in I}}}{B}$}
\end{prooftree}

\caption{Typing rules for the system without general recursion.}
\label{sntypingrules}
\end{figure}

\begin{figure}

% Arrow
\begin{prooftree}
\AxiomC{$\subtype{\gamma}{\epsilon_{x \in A_2}(t\;x \notin B_2)}{A_2}{A_1}$}
\AxiomC{$\subtype{\gamma}{t\;\epsilon_{x \in A_2}(t\;x \notin B_2)}{B_1}{B_2}$}
\RightLabel{$\to$}
\BinaryInfC{$\subtype{\gamma}{t}{A_1 \to B_1}{A_2 \to B_2}$}
\end{prooftree}

\smallskip

% Equal, forall left and forall right
\begin{prooftree}
\AxiomC{\phantom{$\subtype{\gamma}{t}{A[]}{A}$}}
\RightLabel{$=$}
\UnaryInfC{$\subtype{\gamma}{t}{A}{A}$}
\DisplayProof\hfill
\AxiomC{$\subtype{\gamma}{t}{A[X:= U]}{B}$}
\RightLabel{$\forall_l$}
\UnaryInfC{$\subtype{\gamma}{t}{\forall X. A}{B}$}
\DisplayProof\hfill
\AxiomC{$\subtype{\gamma}{t}{A}{B[X := \epsilon_X(t \notin B)]}$}
\RightLabel{$\forall_r$}
\UnaryInfC{$\subtype{\gamma}{t}{A}{\forall X. B}$}
\end{prooftree}

\smallskip

% Exists left and exists right
\begin{prooftree}
\AxiomC{$\subtype{\gamma}{t}{B}{A[X:= U]}$}
\RightLabel{$\exists_r$}
\UnaryInfC{$\subtype{\gamma}{t}{B}{\exists X. A}$}
\DisplayProof\hfill
\AxiomC{$\subtype{\gamma}{t}{B[X := \epsilon_X(t \in B)]}{A}$}
\RightLabel{$\exists_l$}
\UnaryInfC{$\subtype{\gamma}{t}{\exists X. B}{A}$}
\end{prooftree}

\smallskip

% Product and sum
\begin{prooftree}
\AxiomC{$(\subtype{\gamma}{t.l_i}{A_i}{B_i})_{i \in I}$}
\RightLabel{$\times_s$}
\UnaryInfC{$\subtype{\gamma}{t}{
  \{(l_i : A_i)_{i \in I}\}}{\{(l_i : B_i)_{i \in I}\}}$}
\DisplayProof\hfill
\AxiomC{$I_1 \subseteq I_2$}
\AxiomC{$(\subtype{\gamma}{t.C_i}{A_i}{B_i})_{i \in I_1}$}
\RightLabel{$+$}
\BinaryInfC{$\subtype{\gamma}{t}{
  [(C_i : A_i)_{i \in I_1}]}{[(C_i : B_i)_{i \in I_2}]}$}
\end{prooftree}

\smallskip

\begin{prooftree}
\AxiomC{$I_2 \subseteq I_1$}
\AxiomC{$(\subtype{\gamma}{t.l_i}{A_i}{B_i})_{i \in I_2}$}
\RightLabel{$\times_{se}$}
\BinaryInfC{$\subtype{\gamma}{t}{
  \{(l_i : A_i)_{i \in I_1}\}}{\{(l_i : B_i)_{i \in I_2}; \dots\}}$}
\end{prooftree}

\smallskip

\begin{prooftree}
\AxiomC{$I_2 \subseteq I_1$}
\AxiomC{$(\subtype{\gamma}{t.l_i}{A_i}{B_i})_{i \in I_2}$}
\RightLabel{$\times_e$}
\BinaryInfC{$\subtype{\gamma}{t}{
  \{(l_i : A_i)_{i \in I_1}; \dots\}}{\{(l_i : B_i)_{i \in I_2}; \dots\}}$}
\end{prooftree}

\smallskip

% Mu right and nu left
\begin{prooftree}
\AxiomC{$\subtype{\gamma}{t}{A}{F(μ_\tau F)}$}
\AxiomC{$\gamma \vdash \tau < \kappa$}
\RightLabel{$μ_r$}
\BinaryInfC{$\subtype{\gamma}{t}{A}{μ_\kappa F}$}
\DisplayProof\hfill
\AxiomC{$\subtype{\gamma}{t}{F(ν_\tau F)}{B}$}
\AxiomC{$\gamma \vdash \tau < \kappa$}
\RightLabel{$ν_l$}
\BinaryInfC{$\subtype{\gamma}{t}{ν_\kappa F}{B}$}
\end{prooftree}

\smallskip

\begin{prooftree}
\AxiomC{$\subtype{\gamma}{t}{A}{F(μ F)}$}
\RightLabel{$μ_r^\infty$}
\UnaryInfC{$\subtype{\gamma}{t}{A}{μ F}$}
\DisplayProof\hfill
\AxiomC{$\subtype{\gamma, \kappa}{t}{
  F(μ_{τ} F)}{B}$}
\AxiomC{\hspace{-4ex}with $τ = ε_{α < κ}(t ∈ F(μ_{τ} F))$}
\RightLabel{$μ_l$}
\BinaryInfC{$\subtype{\gamma}{t}{μ_{κ} F}{B}$}
\end{prooftree}

\smallskip

\begin{prooftree}
\AxiomC{$\subtype{\gamma, \kappa}{t}{A}{
  F(ν_{τ} F)}$}
\AxiomC{\hspace{-4ex}with $τ = ε_{α < κ} (t ∉ F(ν_{τ} F))$}
\RightLabel{$ν_r$}
\BinaryInfC{$\subtype{\gamma}{t}{A}{ν_{κ} F}$}
\DisplayProof\hfill
\AxiomC{$\subtype{\gamma}{t}{F(ν F)}{B}$}
\RightLabel{$ν_l^\infty$}
\UnaryInfC{$\subtype{\gamma}{t}{ν F}{B}$}
\end{prooftree}
\caption{Subtyping rules for the system without general recursion.}
\label{subtypingrules}
\end{figure}

\begin{defi}
  In addition to rather usual typing judgments of the form $\type{t}{A}$,
  we introduce local subtyping judgements of the form $\subtype{γ}{t}{A}{B}$
  meaning ``if $t$ has type $A$, then it also has type $B$'' (in the
  positivity context $γ$). Usual subtyping judgments of the form
  $\nsubt{γ}{A}{B}$ are then encoded as $\subtype{γ}{ε_{x∈A}(x∉B)}{A}{B}$.
  The typing and subtyping rules of the system are given in
  Figures~\ref{sntypingrules} and \ref{subtypingrules} respectively.
  Both forms of judgments can be used as abstract sequents (in the sense
  of Definition~\ref{def:abstseq}) to build well founded circular proofs
  (see Section~\ref{sct}). In fact, we will only use the ($\GP$) and
  ($\IP{k}$) rules\footnote{The ($\G$) and ($\I{k}$) will be used in
  Section~\ref{fixpoint} to handle general recursion.}, and only allow
  circularity on subtyping proofs.
\end{defi}

Thanks to local subtyping judgements, quantifiers are exclusively handled in
the subtyping part of the system. The use of choice operators enables many
valid permutations of quantifiers with other connectives, while preserving
the syntax-directed nature of the system.
Let aside the ($\GP$) and ($\IP{k}$) rules, only one typing rule applies for
every term constructor, and essentially one local subtyping rule applies for
every two type constructors (see the beginning of Section~\ref{algorithm}).
In the context of our type system, the ($\GP$) and ($\IP{k}$) rules can be
written as follows.
\begin{prooftree}
  \AxiomC{$∀ \vec{α} ∀ x (γ ⊢ C(\vec{α}) ⇒ x ∈ A ⊂ B)$}
  \AxiomC{$(γ[\vec{α} := \vec{κ}],δ ⊢
    κ_i < C(\vec{κ})_i)_{1 ≤ i ≤ |\vec{α}|}$}
  \RightLabel{$\GP$}
  \BinaryInfC{$\subtype{γ[\vec{α}:=\vec{κ}],δ}
    {t}{A[\vec{α}:=\vec{κ}]}{B[\vec{α}:=\vec{κ}]}$}
\end{prooftree}
\smallskip
\begin{prooftree}
  \AxiomC{$\H{∀ \vec{α} ∀ x (γ ⊢ C(\vec{α}) ⇒ x ∈ A ⊂ B)}{k}$}
  \noLine
  \UnaryInfC{$\smash{\raisebox{-0.5mm}{\vdots}}\vphantom{0mm}$}
  \noLine
  \UnaryInfC{$\nsubt{γ[\vec{α}:=\vec{κ}]}
    {A[\vec{α}:=\vec{κ}]}{B[\vec{α}:=\vec{κ}]}$}
  \RightLabel{$\IP{k}$}
  \AxiomC{where $\vec{κ} = \vec{ε}_{\vec{α}<C(\vec{α})} (A \not\subset B)$}
  \BinaryInfC{$∀ \vec{α} ∀ x (γ ⊢ C(\vec{α}) ⇒ x ∈ A ⊂ B)$}
\end{prooftree}

Overall, our rules use syntactic ordinals of the forms
$ε_{α < κ} (t ∈ F(ν_{τ} F))$, $ε_{α < κ} (t ∉ F(ν_{τ} F))$
and $\vec{ε}_{\vec{α}<C(\vec{α})} (A \not\subset B)$. They are all
built from our two forms of abstract judgments according to
Definition~\ref{def:abstract} (up to notations). We respectively
write $t ∈ A$ and $t ∉ A$ for $t : A$ and $¬(t : A)$, and we
also write $A \not\subset B$ for $¬∀x(x ∈ A ⊂ B)$.

\begin{exa}\textbf{Mitchell's containment axiom.}
  In our system, it is possible to derive Mitchell's containment
axiom \cite{mitchell2}, as well as one of its variations.
\begin{align*}
∀X.\mathrm{F}(X) → \mathrm{G}(X) &\subset ∀X.\mathrm{F}(X) → ∀X.\mathrm{G}(X) \\
∀X.\mathrm{F}(X) → \mathrm{G}(X) &\subset ∃X.\mathrm{F}(X) → ∃X.\mathrm{G}(X)
\end{align*}
The derivation of the former is given in Figure~\ref{fig:mitchell} (it
is not circular). Note that the choice operators for terms and types
are all well defined (their definitions are not cyclic).

\begin{figure}
\centering

\begin{prooftree}
  \AxiomC{}
  \RightLabel{$$=$$}
  \UnaryInfC{$ ⊢ x_{1} ∈ \mathrm{F}(X_{0}) ⊂ \mathrm{F}(X_{0})$}
  \RightLabel{$∀_l$}
  \UnaryInfC{$ ⊢ x_{1} ∈ ∀X.\mathrm{F}(X) ⊂ \mathrm{F}(X_{0})$}
  \AxiomC{}
  \RightLabel{$$=$$}
  \UnaryInfC{$ ⊢ x_{0} \; x_{1} ∈ \mathrm{G}(X_{0}) ⊂ \mathrm{G}(X_{0})$}
  \RightLabel{$∀_r$}
  \UnaryInfC{$ ⊢ x_{0} \; x_{1} ∈ \mathrm{G}(X_{0}) ⊂ ∀X.\mathrm{G}(X)$}
  \RightLabel{$→$}
  \BinaryInfC{$ ⊢ x_{0} ∈ \mathrm{F}(X_{0}) → \mathrm{G}(X_{0}) ⊂ ∀X.\mathrm{F}(X) → ∀X.\mathrm{G}(X)$}
  \RightLabel{$∀_l$}
  \UnaryInfC{$ ⊢ x_{0} ∈ ∀X.\mathrm{F}(X) → \mathrm{G}(X) ⊂ ∀X.\mathrm{F}(X) → ∀X.\mathrm{G}(X)$}
\end{prooftree}
\begin{center}
\begin{dot2tex}[dot,options=-tmath]
  digraph G {
  }
\end{dot2tex}
\end{center}

where
\begin{align*} x_{0} &= ε_{x ∈ ∀X.\mathrm{F}(X) → \mathrm{G}(X)}(x ∉ ∀X.\mathrm{F}(X) → ∀X.\mathrm{G}(X))\\
x_{1} &= ε_{x ∈ ∀X.\mathrm{F}(X)}(x_{0} \; x ∉ ∀X.\mathrm{G}(X))\\
X_{0} &= ε_{X}(x_{0} \; x_{1} ∉ \mathrm{G}(X)) \end{align*}

\caption{Derivation of Mitchell's containment axiom.}\label{fig:mitchell}
\end{figure}

%%% Local Variables:
%%% ispell-personal-dictionary: "~/.hunspell-en"
%%% ispell-local-dictionary: "british"
%%% End:

\end{exa}

\begin{exa}\textbf{Mixed inductive and coinductive types.}
  Our system is suitable for handling types containing alternations
of inductive and coinductive types. Let us consider the following two
types $S$ and $L$ where $F(X,Y)$ is a predicate covariant in $X$ and in
$Y$.
$$ \mathrm{S} = μX.νY.[\mathrm{A} \of X \st \mathrm{B} \of Y] \hspace{2em} \mathrm{L} = νY.μX.[\mathrm{A} \of X \st \mathrm{B} \of Y] $$
The elements of $S$ can be thought of as streams of $A$'s and
$B$'s that only contain finitely many $A$'s. The elements of $\mathrm{L}$ are
streams that do not contain infinitely many consecutive $A$'s. In our
system, it is possible to prove $\mathrm{S} \subset \mathrm{L}$ using the circular
proof displayed in Figure~\ref{fig:munu}. Note that the block decomposition
of the proof is given in Example~\ref{exa:sctcirc}. We can thus conclude
that it is well-founded (and thus valid).

\begin{figure*}
  \centering
  \begin{prooftree}
  \AxiomC{$\H{∀α_{0},α_{1} ( ⊢  S_{α_{1}} ⊂ \mathrm{G}(L_{α_{0}}))}{1}$}
  \RightLabel{$\GP$}
  \UnaryInfC{$ ⊢ x_{2}.A ∈ S_{κ_{5}} ⊂ \mathrm{G}(L_{κ_{4}})$}
  \AxiomC{$\H{∀α_{0},α_{1} ( ⊢  \mathrm{F}(S_{α_{1}}) ⊂ \mathrm{G}(L_{α_{0}}))}{2}$}
  \RightLabel{$\GP$}
  \UnaryInfC{$κ_{4} ⊢ x_{2}.B ∈ \mathrm{F}(S_{κ_{5}}) ⊂ \mathrm{G}(L_{κ_{6}})$}
  \RightLabel{$ν_r$}
  \UnaryInfC{$ ⊢ x_{2}.B ∈ \mathrm{F}(S_{κ_{5}}) ⊂ L_{κ_{4}}$}
  \RightLabel{$+$}
  \BinaryInfC{$ ⊢ x_{2} ∈ [\mathrm{A} \of S_{κ_{5}} \st \mathrm{B} \of \mathrm{F}(S_{κ_{5}})] ⊂ [\mathrm{A} \of \mathrm{G}(L_{κ_{4}}) \st \mathrm{B} \of L_{κ_{4}}]$}
  \RightLabel{$μ_r$}
  \UnaryInfC{$ ⊢ x_{2} ∈ [\mathrm{A} \of S_{κ_{5}} \st \mathrm{B} \of \mathrm{F}(S_{κ_{5}})] ⊂ \mathrm{G}(L_{κ_{4}})$}
  \RightLabel{$ν_l$}
  \UnaryInfC{$ ⊢ x_{2} ∈ \mathrm{F}(S_{κ_{5}}) ⊂ \mathrm{G}(L_{κ_{4}})$}
  \RightLabel{$\IP{2}$}
  \UnaryInfC{$∀α_{0},α_{1} ( ⊢  \mathrm{F}(S_{α_{1}}) ⊂ \mathrm{G}(L_{α_{0}}))$}
  \RightLabel{$\GP$}
  \UnaryInfC{$κ_{2} ⊢ x_{1} ∈ \mathrm{F}(S_{κ_{3}}) ⊂ \mathrm{G}(L_{κ_{1}})$}
  \RightLabel{$μ_l$}
  \UnaryInfC{$ ⊢ x_{1} ∈ S_{κ_{2}} ⊂ \mathrm{G}(L_{κ_{1}})$}
  \RightLabel{$\IP{1}$}
  \UnaryInfC{$∀α_{0},α_{1} ( ⊢  S_{α_{1}} ⊂ \mathrm{G}(L_{α_{0}}))$}
  \RightLabel{$\GP$}
  \UnaryInfC{$∞ ⊢ x_{0} ∈ \mathrm{S} ⊂ \mathrm{G}(L_{κ_{0}})$}
  \RightLabel{$ν_r$}
  \UnaryInfC{$ ⊢ x_{0} ∈ \mathrm{S} ⊂ \mathrm{L}$}
\end{prooftree}
\begin{center}
\end{center}

  \bigskip
  \begin{multicols}{2}
    \noindent
    \small
    \begin{align*}%
      \mathrm{F}(X) &= νY.[\mathrm{A} \of X \st \mathrm{B} \of Y] \\
      S_\alpha &= \mu_\alpha X \nu Y [ A \of X | B \of Y ]= \mu_\alpha X F(X)\\
      \mathrm{G}(Y) &= μX.[\mathrm{A} \of X \st \mathrm{B} \of Y] \\
      L_\alpha &= \nu_\alpha Y \mu X [ A \of X | B \of Y ]= \nu_\alpha Y G(Y)\\
      κ_{0} &= ε_{α<∞}(x_{0}∉\mathrm{G}(L_{α}))\\
κ_{1} &= {\vec{ε}_{α_{1},α_{2}<\OMaxi,\OMaxi}(S_{α_{2}} \not\subset \mathrm{G}(L_{α_{1}}))}_1\\
κ_{2} &= {\vec{ε}_{α_{1},α_{2}<\OMaxi,\OMaxi}(S_{α_{2}} \not\subset \mathrm{G}(L_{α_{1}}))}_2\\
κ_{3} &= ε_{α<κ_{2}}(x_{1}∈\mathrm{F}(S_{α}))\\
κ_{4} &= {\vec{ε}_{α_{1},α_{2}<\OMaxi,\OMaxi}(\mathrm{F}(S_{α_{2}}) \not\subset \mathrm{G}(L_{α_{1}}))}_1\\
κ_{5} &= {\vec{ε}_{α_{1},α_{2}<\OMaxi,\OMaxi}(\mathrm{F}(S_{α_{2}}) \not\subset \mathrm{G}(L_{α_{1}}))}_2\\
κ_{6} &= ε_{α<κ_{4}}(x_{2}.B∉\mathrm{G}(L_{α}))\\
x_{0} &= ε_{x ∈ \mathrm{S}}(x ∉ \mathrm{L})\\
x_{1} &= ε_{x ∈ S_{κ_{2}}}(x ∉ \mathrm{G}(L_{κ_{1}}))\\
x_{2} &= ε_{x ∈ \mathrm{F}(S_{κ_{5}})}(x ∉ \mathrm{G}(L_{κ_{4}}))%
    \end{align*}
  \end{multicols}
  \caption{Example of circular proof involving inductive and coinductive types.}
  \label{fig:munu}
\end{figure*}

%%% Local Variables:
%%% ispell-personal-dictionary: "~/.hunspell-en"
%%% ispell-local-dictionary: "british"
%%% End:

\end{exa}

%%% Local Variables:
%%% ispell-personal-dictionary: "~/.hunspell-en"
%%% ispell-local-dictionary: "british"
%%% End:

\section{Fixpoint-less recursion for Scott encoding}\label{scottrec}

In this section, we are going to demonstrate the expressivity of our
system by exhibiting typable, pure $λ$-calculus recursors for Scott
encoded data types. Scott encoding is similar to Church encoding, but
it relies on (co-)inductive types as well as polymorphism. As first
examples, we are going to consider the Church and Scott encodings of
natural numbers. Although they have little (if any) practical interest,
they demonstrate well the use of polymorphism and fixpoints. The type
of Church numerals $\mathbb{N}_C$ and the type of Scott numerals
$\mathbb{N}_S$ are defined below, together with their respective zero
and successor functions.
\begin{align*}
\mathbb{N}_C &= ∀X.(X → X) → X → X\\
0_C &: \mathbb{N}_C = λf\, x.x \\
S_C &: \mathbb{N}_C → \mathbb{N}_C = λn\, f\, x.f \; (n \; f \; x) \\
\mathbb{N}_S &= μN.∀X.(N → X) → X → X\\
0_S &: \mathbb{N}_S = λf\, x.x \\
S_S &: \mathbb{N}_S → \mathbb{N}_S = λn\, f\, x.f \; n
\end{align*}

Using Church encoding, we are able to define (and of course type-check
using our implementation) the usual terms for predecessor $P_C$,
recursor $R_C$, but also the less well-known Maurey infimum
$ (\leq) $, which requires inductive type \cite{krivine}.
The latter requires some type annotations for our implementation to
guess the correct instantiation of unifications variables. In particular,
the type $N_T = (\mathrm{T} → \mathrm{T}) → \mathrm{T} → \mathrm{T}$ where $\mathrm{T} = μX.((X → \mathbb{B}) → \mathbb{B})$ must be used for natural
numbers. Note that $T,F :  \mathbb{B} $ denote booleans.
\begin{align*}
P_C     &: \mathbb{N}_C → \mathbb{N}_C = λn.n \; (λp\, x{:}\mathbb{N}_C\, y{:}\mathbb{N}_C.p \; (S_C \; x) \; x) \; (λx\, y.y) \; 0_C \; 0_C \\
R_C &: ∀P.(P → \mathbb{N}_C → P) → P → \mathbb{N}_C → P = λf\, a\, n.n \; (λx\, p{:}\mathbb{N}_C.f \; (x \; (S_C \; p)) \; p) \; (λp.a) \; 0_C \\
 (\leq)      &: \mathbb{N}_C → \mathbb{N}_C →  \mathbb{B}  = λn\, m.(n : N_T) \; (λf\, g.g \; f) \; (λi.T) \; ((m : N_T) \; (λf\, g.g \; f) \; (λi.F))
\end{align*}

Scott numerals were initially introduced because they admit a constant
time predecessor, whereas Church numerals do not. Usually, programming
using Scott numerals requires the use of a recursor similar to that of
Gödel's System T. Such a recursor can be easily programmed using
general recursion, however this would require introducing typable
terms that are not strongly normalising. In our type system, we can
typecheck a strongly normalisable recursor due to Michel
Parigot\cite{parigotX}. It is displayed below together with several
terms and types involved in its definition.
\begin{align*}
\mathrm{pred} &: \mathbb{N}_S → \mathbb{N}_S = λn.n \; (λp.p) \; 0_S  \\
\mathrm{U}(P) &= ∀Y.Y → \mathbb{N}_S → P \\
\mathrm{T}(P) &= ∀Y.(Y → \mathrm{U}(P) → Y → \mathbb{N}_S → P) → Y → \mathbb{N}_S → P \\
 \mathbb{N}'  &= ∀P.\mathrm{T}(P) → \mathrm{U}(P) → \mathrm{T}(P) → \mathbb{N}_S → P \\
 \zeta  &\,: ∀P.P → \mathrm{U}(P) = λa\, r\, q.a\\
 \delta  &\,: ∀P.P → (\mathbb{N}_S → P → P) → \mathrm{T}(P) = λa\, f\, p\, r\, q.f \; (\mathrm{pred} \; q) \; (p \; r \; ( \zeta  \; a) \; r \; q) \\
 R_S  &\,: ∀P.P → (\mathbb{N}_S → P → P) → \mathbb{N}_S → P = λa\, f\, n.(n :  \mathbb{N}' ) \; ( \delta  \; a \; f) \; ( \zeta  \; a) \; ( \delta  \; a \; f) \; n
\end{align*}
It is easy to check that the term $ R_S $ is indeed a recursor for
Scott numerals. It is similar to a $λ$-calculus fixpoint combinator
but it only allows a limited number of unfoldings. As the recuror is
typable, Theorem~\ref{th:strongnorm} implies that it is strongly
normalising.
The crutial point for typing the recursor is the subtyping relation
$ \mathbb{N}_S \subset  \mathbb{N}'  $, which is derivable in our system. It is
however not clear what are the terms of type $ \mathbb{N}' $ (that are not
in $\mathbb{N}_S$). Note that the type annotation $n :  \mathbb{N}' $ is required for
type-checking $ R_S $ using our implementation, but we do not
need to give the type of $ \zeta $ or $ \delta $.

The recursor for Scott numerals can be adapted to other algebraic data
types like lists or trees. Surprisingly, it can also be adapted to some
coinductive data types. For instance, it is possible to encode streams
using the following definitions.
\begin{align*}
 \mathbb{S}(A) &= νK.∃S.\{\mathrm{hd} : S → A; \mathrm{tl} : S → K; \mathrm{st} : S\} \\
\mathrm{hd} &: ∀A. \mathbb{S}(A) → A = λs.s.hd \; s.st\\
\mathrm{tl} &: ∀A. \mathbb{S}(A) →  \mathbb{S}(A) = λs.s.tl \; s.st\\
\mathrm{cons} &: ∀A.A →  \mathbb{S}(A) →  \mathbb{S}(A) = λa\, l.\{\mathrm{hd}  = λu.a; \mathrm{tl}  = λu.l; \mathrm{st}  = ()\}
\end{align*}
Here, the existentially quantified type can be seen as the representation
of the internal state of the stream. In particular, it must be provided to
compute the first element or the tail of the stream. The order in which the
fixpoint and the existential type is essential to allow the typing
of ``$\mathrm{cons}$''. Note that the internal state is also used to keep strong
normalisation by introducing some laziness into the data type. In other
words, a function call is required to compute the head or the tail of the
stream. The definition of our stongly normalising coiterator $\mathrm{coiter}$ for
streams is given bellow.
\begin{align*}
\mathrm{T}(A,P) &= ∀Y.(P × Y) → \{\mathrm{hd} : (P × Y) → A; \mathrm{tl} : Y; \mathrm{st} : P × Y\}\\
\mathbb{S}'(A,P) &= \{\mathrm{hd} : (P × \mathrm{T}(A, P)) → A; \mathrm{tl} : \mathrm{T}(A, P); \mathrm{st} : P × \mathrm{T}(A, P)\}\\
 \zeta   &\,: ∀A.∀P.(P → A) → ∀X.(P × X) → A = λf\, s.f \; s.1\\
 \delta  &\,: ∀A.∀P.(P → A) → (P → P) → \mathrm{T}(A, P) \\ &= λf\, n\, s.\{\mathrm{hd}  =  \zeta   \; f; \mathrm{tl}  = s.2; \mathrm{st}  = (n \; s.1, s.2)\}\\
\mathrm{coiter} &\,: ∀A.∀P.P → (P → A) → (P → P) →  \mathbb{S}(A)\\ &= λs\, f\, n.\begin{array}[t]{l}\LET A,P \ST f:P → A \IN\\ \left\{\setlength{\arraycolsep}{0.2em}\begin{array}{ll}\mathrm{hd}  &=  \zeta   \; f;\\
\mathrm{tl}  &=  \delta  \; f \; n;\\
\mathrm{st}  &= (s,  \delta  \; f \; n)\end{array}\right\} : \mathbb{S}'(A, P)\end{array}
\end{align*}
Note that in the definition of $\mathrm{coiter}$ we deliberately used
the same names as in the definitions of $ R_S $ to highlight their
similarities. The minimum type annotation for our implementation to
type-check $\mathrm{coiter}$ involve the subtyping relation
$\mathbb{S}'(\mathrm{A}, \mathrm{P}) \subset  \mathbb{S}(\mathrm{A})$. The let-binding syntax
in $\mathrm{coiter}$ is used to name universally quantified types (see
Section~\ref{algorithm}). It is only used in the implementation and it
is not part of the theoretical type system.
In particular, the types of $ \zeta  $ and $ \delta $ are not required. As
for Scott numeral, a question arise about the inhabitants of the type
$∃P.\mathbb{S}'(\mathrm{A}, P)$.

The main difference between the encoding of Scott numerals and the
encoding of streams is the use of native records. It is in fact possible
to use native sums for encoding Scott numerals, but a function type is
still required to program a strongly normalising recursor. We were not
able to program a strongly normalisable recursor for the usual type of
unary natural numbers $μN.[\mathrm{Z} \st \mathrm{S} \of N]$, and we conjecture that this
is not possible. However, if we encode the sum type using a record type,
a recursor can be given. The type of unary natural numbers then becomes
$\mathbb{N} = μX.∀Y.\{\mathrm{z} : Y; \mathrm{s} : X → Y\} → Y$, which is very similar to
the type of Scott numeral.

%%% Local Variables:
%%% ispell-personal-dictionary: "~/.hunspell-en"
%%% ispell-local-dictionary: "british"
%%% End:

\section{Realisability semantics}\label{semantics}
In this section, we build a realisability model that is shown adequate
with our type system. In particular, a formula $A$ is interpreted as
a set of strongly normalising \emph{pure terms} $\sem{A}$. Consequently,
if $⊢ t : A$ is derivable then we will have $\sem{t} ∈ \sem{A}$, where
$\sem{t}$ is the interpretation of $t$ as a pure term.
\begin{defi}
  A term is said to be \emph{pure} if it does not contain subterms of
  the form $ε_{x∈ A}(t ∉ B)$. We denote $\pure ⊂ Λ$ the set of pure
  terms (or pure individuals according to the terminology of
  Definition~\ref{def:abstract_sem}). A pure term $t ∈ \pure$ is said
  to be strongly normalising if there is no infinite sequence of
  reduction starting from $t$ using the rules of Figure~\ref{fig:reduction}.
  We denote $\calN ⊂ \pure$ the set of strongly normalising pure terms.
\end{defi}

\begin{defi}
  The set $\mathcal{H}$ of head contexts (i.e. terms with a hole in head
  position) is generated by the following grammar.
  \begin{align*}
    H
    \bnfeq [\,]
    \bnfor H\;t
    \bnfor H.l
    \bnfor \case{H}{(C_i → t_i)_{i ∈ I}}
  \end{align*}
  Given a term $t∈Λ$ and a context $H∈\mathcal{H}$, we denote $H[t]$ the
  term formed by plugging $t$ into the hole of $H$.
  We extend naturally the notion of reduction to context by writing
  $H ≻ H'$ when $H[t] ≻ H'[t]$ for any term $t ∈ Λ$ (including, for
  instance, $λ$-variables).
  We denote $(≻_H)$ the \emph{head reduction} relation defined as the
  contextual closure of the rules of Figure~\ref{fig:reduction},
  restricted to contexts of $\mathcal{H}$.
  We say that a term is in \emph{head normal form} if it cannot be
  reduced using $(≻_H)$.
\end{defi}

\begin{defi}\label{def:sat}
  We say that a set of pure terms $Φ ⊂ \pure$ is \emph{saturated} if it
  is closed by head reduction\footnote{Requiring closure under head
  reduction is unusual, but necessary for subtyping on sum types.}
  and if the following conditions hold.
  \begin{enumerate}
  \item If $H[t[x := u]] ∈ Φ$ and $u∈\calN$ then $H[(λ x.t)\;u] ∈Φ$.
  \item If $H[t\;u] ∈ Φ$, then $H[\case{D\;u}{D → t}] ∈ Φ$.
  \item If $H[t] ∈ Φ$ then $H[\{ l = t; (l_i = t_i)_{i ∈ I} \}.l] ∈ Φ$
        provided that $t_i ∈ \calN$ for all $i ∈ I$.
  \item If $H[t ∣ (C_i → t_i)_{i ∈ I}] ∈ Φ$ and $I ⊂ J$ then
        $H[t ∣ (C_i → t_i)_{i ∈ J}] ∈ Φ$ provided that $t_j ∈ \calN$
        for all $j ∈ J ∖ I$.
  \end{enumerate}
\end{defi}
\begin{lem}\label{lem:simplesat}
  If $Φ ⊂ \pure$ is saturated, then for all $t ∈ \calN$ and $u ∈ Φ$,
  $t ≻_H u$  implies $t ∈ Φ$.
\end{lem}
\begin{proof}
  Immediate by the definitions of saturated sets and head reduction.
\end{proof}

\begin{lem}\label{lem:nsat}
  The set $\calN$ is saturated.
\end{lem}
\begin{proof}
  The set $\calN$ is obviously closed under head reduction, so it remains
  to show that it satisfies the four conditions of Definition~\ref{def:sat}.
  \begin{enumerate}
    \item Let us take $H[t[x:=u]] ∈ \calN$ and suppose, by contradiction,
          that $H[(λ x.t)\;u] ∉ \calN$. There cannot be an infinite
          reduction of $H$, $t$ or $u$. Hence, an infinite reduction of
          $H[(λx.t)\;u]$ must start with $H[(λx.t)\;u] ≻^* H'[(λx.t')\;u']
          ≻ H'[t'[x:=u']]$, where $H ≻^* H'$, $t ≻^* t'$ and $u ≻^* u'$.
          We then contradict $H[t[x:=u]]∈\calN$ by transforming this
          reduction into $H[(λ x.t)\;u] ≻ H[t[x:=u]] ≻^* H'[t'[x:=u']]$.

    \item Let us take $H[t\,u] ∈ \calN$ and suppose, by contradiction,
          that $H[\case{D\;u}{D → t}] ∉ \calN$. As in the previous case,
          there cannot be an infinite reduction of $H$, $u$ or $t$. As
          a consequence, an infinite reduction of $H[\case{D\;u}{D→t}]$
          necessarily starts with $H[\case{D\;u}{D → t}] ≻^*
          H'[\case{D\;u'}{D→t'}] ≻ H'[t'\,u']$, where $H ≻^* H'$,
          $t ≻^* t'$ and $u ≻^* u'$. This can be transformed into
          $H[\case{D\;u}{D → t}] ≻ H[t\,u] ≻^* H'[t'\,u']$, which
          contradicts $H[t\,u] ∈ \calN$.

    \item Let us take $H[t] ∈ \calN$ and $t_i ∈ \calN$ for all $i ∈ I$,
          and suppose, by contradiction, that $H[\{l=t; (l_i=t_i)_{i∈I}\}
          .l] ∉ \calN$. There cannot be an infinite reduction of $H$, $t$
          nor of any of the $t_i$. Consequently, an infinite reduction of
          $H[\{ l = t; (l_i = t_i)_{i ∈ I}\}.l]$ must start with
          $H[\{ l = t; (l_i = t_i)_{i ∈ I}\}.l] ≻^* H[\{ l = t'; (l_i =
          t'_i)_{i ∈ I}\}.l]$, where $H ≻^* H'$, $t ≻^* t'$ and $t_i ≻^*
          t'_i$ for all $i ∈ I$. We then obtain a contradiction with
          $H[t] ∈ \calN$ by transforming this reduction into
          $H[\{ l = t; (l_i = t_i)_{i ∈ I}\}.l] ≻^* H[t] ≻ H'[t']$.

    \item Let us take $H[\case{t}{(C_i → t_i)_{i ∈ I}}] ∈ \calN$, a set
          of index $J$ with $I ⊂ J$ and for all $j ∈ J \setminus I$ a term
          $t_j ∈ \calN$. We suppose, by contradiction, that $H[\case{t}{(C_i
          → t_i)_{i ∈ J}}] ∉ \calN$. There cannot be an infinite sequence of
          reduction of $H$, $t$ nor any of the $t_j$ for $j ∈ J$. As a
          consequence, an infinite reduction of $H[\case{t}{(C_i →
          t_i)_{i∈J}}]$ necessarily starts with $H[\case{t}{(C_i→t_i)_{i∈J}}]
          ≻^* H'[\case{C_k u}{(C_i → t'_i)_{i ∈ J}}] ≻ H'[t'_k\,u]$, where
          $H ≻^* H'$, $t ≻^* C_k u$ for some $k ∈ J$ and $t_i ≻^* t'_i$ for
          all $i ∈ J$. We can then obtain a contradiction using
          $H[\case{t}{(C_i → t_i)_{i ∈ I}}] ≻^* H'[\case{C_k u}{(C_i→
          t'_i)_{i ∈ I}}] ≻ H'[t'_k\,u]$ if $k∈I$, and
          $H[\case{t}{(C_i→t_i)_{i∈I}}] ≻^*H'[\case{C_k u}{(C_i →
          t'_i)_{i ∈ I}}] ≻ H'[Ω]$ otherwise.
  \end{enumerate}
  \vspace{-4.5mm}% HACK FIXME
\end{proof}

\begin{defi}\label{def:nz}
  The set of \emph{neutral terms} $\calNZ ⊂ \pure$ is the smallest set
  such that:
  \begin{enumerate}
    \item for every $λ$-variable $x$ we have $x ∈ \calNZ$,
    \item for every $u ∈ \calN$ and $t ∈ \calNZ$ we have $t\,u ∈ \calNZ$,
    \item for every $i∈\N$ and $t∈\calNZ$ we have $t.l_i ∈ \calNZ$,
    \item for every $(C_i, t_i)_{i ∈ I} ∈ (\mathcal{C} \times \calN)^I$ and
          $t ∈ \calNZ$ we have $\case{t}{(C_i → t_i)_{i ∈ I}} ∈ \calNZ$.
  \end{enumerate}
  Note that $\calNZ$ is not saturated.
\end{defi}
\begin{defi}
  Given a set of pure valus $Φ ⊆ \pure$, we denote $\overline{Φ} ⊆ \pure$
  the smallest saturated set containing $Φ$.
\end{defi}

\begin{lem}\label{nzinn}
  We have $\calNZ ⊂ \calNZS ⊂ \calN$.
\end{lem}
\begin{proof}
  We obviously have $\calNZ ⊂ \calNZS$ and $\calNZ ⊂ \calN$. Moreover, it
  is clear that the saturation operation is covariant. As a consequence,
  we have $\calNZS ⊂ \calNS = \calN$.
\end{proof}

\begin{defi}
  Given two sets $Φ₁$, $Φ₂ ⊂ \pure$ we define $(Φ₁ ⇒ Φ₂) ⊂ \pure$ as
  follows. $$ (Φ₁⇒Φ₂) = \{t∈⟦Λ⟧ \st ∀ u∈Φ₁, \;t\;u ∈ Φ₂\}$$
\end{defi}
\begin{lem}\label{lem:inclusion}
  Let $Φ₁$, $Φ₂$, $Ψ₁$, $Ψ₂ ⊆ \pure$ be sets of pure terms such that
  $Φ₂ ⊆ Φ₁$ and $Ψ₁ ⊆ Ψ₂$. We have $(Φ₁ ⇒ Ψ₁) ⊆ (Φ₂ ⇒ Ψ₂)$.
\end{lem}
\begin{proof}
  Immediate by definition.
  % We take $t ∈ (Φ₁ ⇒ Ψ₁)$ and show $t ∈ (Φ₂ ⇒ Ψ₂)$. By definition we
  % know that if $u ∈ Φ₁$ then $t\;u ∈ Ψ₁$. Let us now take $u ∈ Φ₂$.
  % By hypothesis $u ∈ Φ₁$ and hence $t\;u ∈ Ψ₁$. We finally obtain
  % $t\;u ∈ Ψ₂$ by hypothesis.
\end{proof}

\begin{lem}\label{lem:nnzero}
  We have $\calNZ ⊆ (\calN ⇒ \calNZ) ⊆ (\calNZ ⇒ \calN) ⊆ \calN$.
\end{lem}
\begin{proof}
  By Lemma \ref{nzinn} we know that $\calNZ ⊆ \calN$ and hence we obtain
  $(\calN ⇒ \calNZ) ⊆ (\calNZ ⇒ \calN)$ using Lemma~\ref{lem:inclusion}. If we
  take $t∈\calNZ$, then by definition $t\,u ∈ \calNZ$ for all $u∈\calN$.
  Therefore we obtain $\calNZ ⊆ (\calN ⇒ \calNZ)$. Finally, if we take
  $t ∈ (\calNZ⇒\calN)$ then by definition $t\;x ∈ \calN$ since $x∈\calNZ$.
  Hence $t ∈ \calN$, which gives $(\calNZ ⇒ \calN) ⊆ \calN$.
\end{proof}

In the semantics, a closed term $t ∈ Λ$ will be interpreted as a pure term
$\sem{t} ∈ \pure$ with the same structure. The choice operators $t$ will be
replaced by (possibly open) pure terms in $\sem{t}$. A formula $A ∈ \calF$
will be interpreted by a saturated set of pure terms $\sem{A}$ such that
$\calNZS ⊆ \sem{A} ⊆ \calN$.
Note that a syntactic ordinals $κ ∈ \mathcal{O}$ will be interpreted by
an actual ordinal $\sem{κ} ∈ \ordsem$ according to Section~\ref{ordinals}.
Of course, the interpretation of syntactic ordinals will involve the
interpretation of terms and formulas through abstract judgments. The
interpretation of our three syntactic entities is thus defined mutually
inductively, as was their syntax.
\begin{defi}
  The set of every type interpretations $\formsem$ is defined as follows,
  its elements will be called \emph{reducibility candidates} (or simply
  \emph{candidates}).
  $$\formsem = \{Φ ⊆ \pure \st Φ\;\text{saturated},\; \calNZS⊆Φ⊆\calN\}$$
\end{defi}

To simplify the definition of the semantics, we will extend the syntax of
formulas with the elements of their domain of interpretation. We already
used this technique in Section~\ref{ordinals} for syntactic ordinals,
and it will allow us to work only with closed syntactic elements. Most
notably, we will use substitutions with elements of the semantics instead
of relying on a semantical map for interpreting free variables.
\begin{defi}\label{paramdef}
  The sets of \emph{parametric terms} $\parterm$ and the set of
  \emph{parametric formulas} $\parform$ are formed by extending the
  syntax of formulas with the elements of $\formsem$. Terms do not need
  to be extended directly, however, the definition of $\parform$
  impacts the definition of $\parterm$ since terms and formulas are
  defined mutually inductively.\footnote{The set of parametric syntactic
  ordinals $\parord$ of Definition~\ref{def:par_synt_ord} should also
  be impacted.}
  A closed parametric term (resp. formula, resp. syntactic ordinal) is
  a parametric term (resp. formula, resp. syntactic ordinal) that does
  not contain free propositional variables nor free ordinal variables.
  Note however that $λ$-variables are allowed. This is due to the
  definition of $\calNZ$.
\end{defi}

\begin{figure}
\centering

\begin{align*}
  \sem{t}
    &= t \;\text{ if } t ∈ \pure\\
  \sem{x}
    &= x \\
  \sem{t\,u}
    &= \sem{t}\,\sem{u} \\
  \sem{\lambda x.t}
    &= \lambda x.\sem{t} \\
  \sem{C\,u}
    &= C\,\sem{u} \\
  \sem{\case{u}{(C_i → t_i)_{i∈ I}}}
    &= \case{\sem u}{(C_i → \sem{t_i})_{i∈ I}}\\
  \sem{\{(l_i = t_i)_{i ∈ I}\}}
    &= \{(l_i = \sem{t_i})_{i ∈ I}\}\\
  \sem{\eps_{x∈ A}(t ∉ B)}
    &= \left\{
      \begin{array}{l}
        \!\!\! u ∈ \sem{A} \text{ s.t. } \sem{t[x:=u]} ∉ \sem{B} \\
        \!\!\! \text{any } t ∈ \calNZ \text{ if there is no such } u
      \end{array} \right. \\
  \sem{Φ}
    &= Φ \;\text{ if } Φ ∈ \formsem\\
  \sem{A → B}
    &= (\sem{A} ⇒ \sem{B})\\
  \sem{\{(l_i : A_i)_{i ∈ I}; \dots\}}
    &= \{t ∈ \calN \st ∀i∈I, t.l_i ∈ \sem{A_i}\}\\
  \sem{\{(l_i : A_i)_{i ∈ I≠∅}\}}
    &= \{t ∈ \calN \st ∀i∈I, t.l_i ∈ \sem{A_i} \text{ and }
       ∀i∉I, t.l_i ≻_H^* Ω\}\\
  \sem{\{\}}
    &= \overline{\calNZ ∪ \{\{\}\}}\\
  \sem{[(C_i : A_i)_{i ∈ I}]}
    &= \{t ∈ \calN \st ∀Φ ∈ \formsem, ∀ (t_i ∈ (\sem{A_i} ⇒ Φ))_{i ∈ I},
       \case{t}{(C_i → t_i)_{i ∈ I}} ∈ Φ\}\\
  \sem{∀X.A}
    &= \cap_{Φ ∈ \formsem} \sem{A[X := Φ]}\\
  \sem{∃X.A}
    &= \cup_{Φ ∈ \formsem} \sem{A[X := Φ]}\\
  \sem{μ_κ X.A}
    &= \left(\cup_{o < \sem{\kappa}} \sem{A[X := \mu_o X A]}\right)
       ∪ \calNZS\\
  \sem{ν_κ X.A}
    &= \left(\cap_{o < \sem{\kappa}} \sem{A[X := \nu_o X A]}\right)
       ∩ \calN\\
  \sem{ε_X(t ∈ A)}
    &= \left\{
      \begin{array}{l}
        \!\!\!Φ ∈ \formsem \text{ s.t. } \sem{t} ∈ \sem{A[X := Φ]}\\
        \!\!\!\calN \text{ if there is no such } Φ
      \end{array} \right. \\
  \sem{ε_X(t ∉ A)}
    &= \left\{
      \begin{array}{l}
        \!\!\!Φ ∈ \formsem \text{ s.t. } \sem{t} ∉ \sem{A[X := Φ]}\\
        \!\!\!\calN \text{ if there is no such } Φ
      \end{array} \right.
\end{align*}

\caption{Semantical interpretation of closed parametric terms and types.}
\label{interpr}
\end{figure}

\begin{defi}
  The interpretation of a closed parametric term $t ∈ \parterm$ (resp.
  closed parametric formula $A ∈ \parform$) is defined to be a pure
  term $\sem{t} ∈ \pure$ (resp. a set of pure terms $\sem{A} ∈ \formsem$)
  defined inductively according to Figure~\ref{interpr} and
  Definition~\ref{def:ordsem}. Note that the semantics of terms, types and
  syntactic ordinals should be defined mutually inductively due to choice
  operators (or witnesses). In particular, the abstract judgments used in
  the definition of choice operators for ordinals are interpreted in the
  obvious way according to Definition~\ref{def:abstract_sem}.
\end{defi}

In the interpretation of choice operators of the form $ε_{x∈A}(t∉B)$, it
is important that no $λ$-variable other that $x$ is bound in $t$. This is
enforced by a syntactic restriction given in Definition~\ref{def:syntax}.
Without this restriction, a term $\sem{λ y.ε_{x ∈ A}(t ∉ B)}$ with $y$
free in $t$ would correspond to a function that is not always definable
using a pure term. Thus, our model would have circular (and hence invalid)
definitions. Note that the axiom of choice is required to interpret the
choice operators.

It is also worth noting that the interpretation of the types of the form
$μ_{κ} F$ (resp. $ν_{κ} F$) involves a union with $\calNZS$ (resp. an
intersection with $\calN$). It is required as otherwise we would obtain
$\sem{μ₀ F} = ∅$ (resp. $\sem{ν₀ F} = Λ$) for the zero ordinal, and these
sets are not proper candidates for the interpretation of formulas.

\begin{lem}\label{lem:substitution}
  The semantical interpretation of terms, formulas and syntactic ordinals
  commutes with the substitution of the three kinds of variables. We thus
  have, for example, $\sem{t[X:=A]} = \sem{t[X:=\sem{A}]}$ or
  $\sem{A[α:=κ]} = \sem{A[α:=\sem{κ}]}$.
\end{lem}
\begin{proof}
  Immediate by induction on the definition of the semantics.
\end{proof}

\begin{lem}\label{lem:arrowsem}
  For all candidates $Φ$, $Ψ ∈ \formsem$, we have $(Φ⇒Ψ) ∈ \formsem$.
\end{lem}
\begin{proof}
  Since $Φ$ and $Ψ$ are candidates, we know $\calNZS⊆Φ⊆\calN$ and
  $\calNZS⊆Ψ⊆\calN$. As a consequence, we can use Lemma~\ref{lem:inclusion}
  to obtain $(\calN⇒\calNZS) ⊆ (Φ⇒Ψ)$ (using $Φ⊆\calN$ and $\calNZS⊆Ψ$) and
  $(Φ⇒Ψ) ⊆ (\calNZS⇒\calN)$ (using $\calNZS⊆Φ$ and $Ψ⊆\calN$). We then
  obtain $\calNZS ⊆ (Φ⇒Ψ) ⊆ \calN$ with Lemma~\ref{lem:nnzero}.
  It remains to show that $(Φ⇒Ψ)$ is saturated, so we will first show that
  it is closed under head reduction. Let us take $t ∈ (Φ⇒Ψ)$ such that
  $t ≻_H t'$ and show that $t' ∈ (Φ⇒Ψ)$. We take $u ∈ Φ$ and show
  $t'\;u ∈ Ψ$. Since $Ψ$ is closed under head reduction and
  $t\;u ≻_H t'\;u$ it is enough to show $t\;u ∈ Ψ$, which follows from
  the definition of $t ∈ (Φ⇒Ψ)$. It remains to prove the four saturation
  conditions.
  \begin{enumerate}
    \item Let us suppose that $H[t[x:=u]] ∈ (Φ⇒Ψ)$ and that $u∈\calN$. We
      need to show that $H[(λx.t)\;u] ∈ (Φ⇒Ψ)$ so we take $v∈Φ⊆\calN$ and
      we prove $H[(λx.t)\;u]\;v ∈ Ψ$. As we have $H[t[x:=u]] ∈ (Φ⇒Ψ)$ we
      know that $H[t[x:=u]]\;v ∈ Ψ$. We can thus conclude using the
      saturation condition (1) on $Ψ$ with the context $H\;v$.
    \item We now suppose $H[t\;u] ∈ (Φ⇒Ψ)$ and show
      $H[\case{D\;u}{D → t}] ∈ (Φ⇒Ψ)$. We thus take $v∈Φ⊆\calN$ and we
      prove $H[\case{D\;u}{D → t}]\;v ∈ Ψ$. As $H[t\;u] ∈ (Φ⇒Ψ)$ we know
      that $H[t\;u]\;v ∈ Ψ$ and thus we can conclude using the saturation
      condition (2) of $Ψ$ with the context $H\;v$.
    \item Let us now suppose that $H[t] ∈ (Φ⇒Ψ)$ and that $t_i ∈ \calN$ for
      all $i∈I$. We need to show that $H[\{l=t; (l_i=t_i)_{i∈I}\}.l]∈(Φ⇒Ψ)$
      so we take $v∈Φ⊆\calN$ and we prove $H[\{l=t; (l_i=t_i)_{i∈I}\}.l]\;v
      ∈ Ψ$. As $H[t] ∈ (Φ⇒Ψ)$ we know that $H[t]\;v ∈ Ψ$ and thus conclude
      with the saturation condition (3) of $Ψ$ with the context $H\;v$.
    \item We now suppose $H[t ∣ (C_i → t_i)_{i ∈ I}] ∈ (Φ⇒Ψ)$ and $I ⊂ J$
      with $t_j ∈ \calN$ for all $j ∈ J ∖ I$. We need to show
      $H[t ∣ (C_i → t_i)_{i ∈ J}] ∈ (Φ⇒Ψ)$ so we take $v ∈ Φ ⊆ \calN$ and
      we prove $H[t ∣ (C_i → t_i)_{i ∈ J}]\;v ∈ Ψ$. As we have
      $H[t ∣ (C_i → t_i)_{i ∈ I}] ∈ (Φ⇒Ψ)$ we know that
      $H[t ∣ (C_i → t_i)_{i ∈ I}]\;v ∈ Ψ$ and thus we can conclude using
      the saturation condition (4) of $Ψ$ with the context $H\;v$.
  \end{enumerate}
  \vspace{-4.5mm}% HACK FIXME
\end{proof}

\begin{lem}\label{lem:sumsem}
  If for all $i∈I$ we have $Φ_i ∈ \formsem$
  then $\sem{[(C_i \of Φ_i)_{i∈I}]} ∈ \formsem$.
\end{lem}
\begin{proof}
  By definition, it is easy to see that $\sem{[(C_i \of Φ_i)_{i∈I}]} ⊆ \calN$.
  Let us take $t ∈ \calNZS$ and show that $t ∈ \sem{[(C_i \of Φ_i)_{i∈I}]}$.
  We thus take $Φ ∈ \formsem$ and $t_i ∈ (Φ_i ⇒ Φ)$ for all $i∈I$, and we
  show $\case{t}{(C_i → t_i)_{i ∈ I}} ∈ Φ$. This is immediate as we have
  $\case{t}{(C_i → t_i)_{i ∈ I}} ∈ \calNZS$ and $\calNZS ⊆ Φ$.
  It remains to show that $\sem{[(C_i \of Φ_i)_{i∈I}]}$ is saturated, so we
  will first show that it is closed under head reduction. Let us take
  $t ∈ \sem{[(C_i \of Φ_i)_{i∈I}]}$ such that $t ≻_H t'$ and show that
  $t' ∈ \sem{[(C_i \of Φ_i)_{i∈I}]}$. We thus take $Φ ∈ \formsem$ and
  $t_i ∈ (Φ_i⇒Φ)$ for all $i∈I$, and we show $\case{t'}{(C_i→t_i)_{i∈I}}
  ∈ Φ$. Since $t ∈ \sem{[(C_i \of Φ_i)_{i∈I}]}$ we know that
  $\case{t}{(C_i→t_i)_{i∈I}} ∈ Φ$. We thus conclude since $Φ$ is
  saturated and $\case{t}{(C_i→t_i)_{i∈I}} ≻_H \case{t'}{(C_i→t_i)_{i∈I}}$.
  It remains to prove the four saturation conditions.
  \begin{enumerate}
    \item Let us suppose that we have $H[t[x:=u]] ∈ \sem{[(C_i \of
      Φ_i)_{i∈I}]}$ and $u ∈ \calN$, and show $H[(λx.t)\;u] ∈
      \sem{[(C_i \of Φ_i)_{i∈I}]}$. We take $Φ ∈ \formsem$ and
      $t_i ∈ (Φ_i⇒Φ)$ for all $i∈I$, and show
      $\case{H[(λx.t)\;u]}{(C_i→t_i)_{i∈I}} ∈ Φ$. Since
      $H[t[x:=u]] ∈ \sem{[(C_i \of Φ_i)_{i∈I}]}$ we have
      $\case{H[t[x:=u]]}{(C_i→t_i)_{i∈I}} ∈ Φ$. We can thus conclude
      using the saturation condition (1) on $Φ$ with the context
      $\case{H}{(C_i→t_i)_{i∈I}}$.
    \item We suppose $H[t\;u] ∈ \sem{[(C_i \of Φ_i)_{i∈I}]}$ and show
      $H[\case{D\;u}{D → t}] ∈ \sem{[(C_i \of Φ_i)_{i∈I}]}$. We thus take
      $Φ ∈ \formsem$ and $t_i ∈ (Φ_i⇒Φ)$ for all $i∈I$, and show that we
      have $\case{H[\case{D\;u}{D → t}]}{(C_i→t_i)_{i∈I}} ∈ Φ$.
      As $H[t\;u] ∈ \sem{[(C_i \of Φ_i)_{i∈I}]}$ we know that
      $\case{H[t\;u]}{(C_i→t_i)_{i∈I}} ∈ Φ$ and thus we can conclude
      using the saturation condition (2) of $Φ$ with the context
      $\case{H}{(C_i→t_i)_{i∈I}}$.
    \item Let us now suppose that $H[t] ∈ \sem{[(C_i \of Φ_i)_{i∈I}]}$ and
      that $t_i ∈ \calN$ for  all $i∈I$. We need to show that
      $H[\{l=t; (l_i=t_i)_{i∈I}\}.l] ∈ \sem{[(C_i \of Φ_i)_{i∈I}]}$
      so we take $Φ ∈ \formsem$ and $t_i ∈ (Φ_i⇒Φ)$ for all $i∈I$, and
      show $\case{H[\{l=t; (l_i=t_i)_{i∈I}\}.l]}{(C_i→t_i)_{i∈I}} ∈ Φ$.
      As $H[t] ∈ \sem{[(C_i \of Φ_i)_{i∈I}]}$ we know that
      $\case{H[t]}{(C_i→t_i)_{i∈I}} ∈ Φ$ and thus conclude with the
      saturation condition (3) of $Φ$ with the context
      $\case{H}{(C_i→t_i)_{i∈I}}$.
    \item We now suppose $H[t ∣ (C_i→t_i)_{i∈I}] ∈ \sem{[(C_i \of
      Φ_i)_{i∈I}]}$ and $I ⊂ J$ with $t_j ∈ \calN$ for all $j ∈ J ∖ I$.
      We need to show $H[t ∣ (C_i→t_i)_{i∈J}] ∈ \sem{[(C_i \of
      Φ_i)_{i∈I}]}$ so we take $Φ ∈ \formsem$ and $t_i ∈ (Φ_i⇒Φ)$ for
      all $i∈I$, and show $\case{H[t ∣ (C_i → t_i)_{i ∈ J}]}{(C_i→
      t_i)_{i∈I}} ∈ Φ$. As $H[t ∣ (C_i→t_i)_{i∈I}] ∈ \sem{[(C_i
      \of Φ_i)_{i∈I}]}$ we get $\case{H[t ∣ (C_i→t_i)_{i∈I}]}{(C_i
      →t_i)_{i∈I}} ∈ Φ$ and we can then use the saturation
      condition (4) of $Φ$ with the context
      $\case{H}{(C_i→t_i)_{i∈I}}$.
  \end{enumerate}
  \vspace{-4.5mm}% HACK FIXME
\end{proof}

\begin{lem}\label{lem:prodsem}
  If for all $i∈I$ we have $Φ_i ∈ \formsem$
  then $\sem{\{(l_i : Φ_i)_{i∈I};\dots\}} ∈ \formsem$.
\end{lem}
\begin{proof}
  Similar to the proofs of Lemmas~\ref{lem:arrowsem} and
  \ref{lem:sumsem}.
\end{proof}

\begin{lem}\label{lem:sprodsem}
  If for all $i∈I$ we have $Φ_i ∈ \formsem$
  then $\sem{\{(l_i : Φ_i)_{i∈I}\}} ∈ \formsem$.
\end{lem}
\begin{proof}
  Immediate if $I = ∅$ and similar to the proofs of
  Lemmas~\ref{lem:arrowsem} and \ref{lem:sumsem}
  otherwise.
\end{proof}

\begin{thm}\label{th:snsemantics}
  For every closed parametric term $t ∈ \parterm$ (resp. ordinal
  $κ ∈ \parord$, resp. type $A ∈ \parform$) we have $\sem{t} ∈ \pure$
  (resp. $\sem{κ} ∈ \ordsem$, resp. $\sem{A} ∈ \formsem$).
\end{thm}
\begin{proof}
  We do a proof by induction. For terms, all the cases are immediate by
  induction hypothesis. For instance, if $u = \sem{ε_{x ∈ A}(t ∉ B)}$
  then we have $u ∈ \sem{A} ⊆ \calN ⊆ \pure$ by induction hypothesis,
  or $u ∈ \calNZS ⊆ \pure$.
  For ordinals, the proof is immediate by Definition~\ref{def:ordsem}
  and using the induction hypothesis to interpret predicates in ordinal
  witnesses.
  For types of the form $Φ ⊆ \formsem$, $ε_X(t ∈ A)$ or $ε_X(t ∉ A)$ the
  proof is immediate. For types of the form  $A⇒B$, $[(C_i \of A_i)_{i∈I}]$,
  $\{(l_i : A_i)_{i∈I};\dots\}$ or $\{(l_i : A_i)_{i∈I}\}$ then we
  respectively use Lemma~\ref{lem:arrowsem}, \ref{lem:sumsem},
  \ref{lem:prodsem} or \ref{lem:sprodsem} with the
  induction hypotheses and Lemma~\ref{lem:substitution}. The remaining four
  possible forms of types are treated bellow.
  \begin{itemize}
    \item For types of the form $∀X.A$, the induction hypothesis gives
      $\sem{A[X:=Φ]} ∈ \formsem$ for all $Φ ∈ \formsem$. We can then
      conclude using the fact that an intersection of candidates is
      itself a candidate.
    \item For types of the form $∃X.A$, the proof is similar to the
      previous case, using the fact that a union of candidates is itself
      a candidate.
    \item For types of the form $μ_κX.A$, we show $\sem{μ_o X.A} ∈ \formsem$
      for all $o ≤ \sem{κ}$ by induction on the ordinal $o$. This is enough
      as we can then conclude using Lemma~\ref{lem:substitution} to show
      $\sem{μ_κ X.A} = \sem{μ_{\sem{κ}} X.A} = \sem{μ_o X.A} ∈ \formsem$.
      If $o = 0$ then we have $\sem{μ_0 X.A} = \calNZS$ and the proof is
      thus immediate.
      Otherwise, we have $\sem{μ_o X.A} = \cup_{o'<o} \sem{A[X:=μ_{o'}X.A]}$.
      Using the local induction hypothesis we get $\sem{μ_{o'} X. A} ∈
      \formsem$ for all $o' < o$. Using Lemma~\ref{lem:substitution}, we then
      obtain $\sem{A[X:=μ_{o'} X. A]} = \sem{A[X:=\sem{μ_{o'} X. A}]}$ for
      all $o' < o$, which gives $\sem{A[X:=\sem{μ_{o'} X. A}]} ∈ \formsem$
      for all $o' < o$ using the global induction hypothesis. We can then
      conclude using again the fact that a union of candidates is itself a
      candidate.
    \item For types of the form $ν_κX.A$, we proceed in a similar way as in
      the previous case, using again the fact that an intersection of
      candidates is itself a candidate. Note that we have $\sem{ν_0 X.A} =
      \calN ∈ \formsem$ in the case of the zero ordinal.
  \end{itemize}
  \vspace{-4.5mm}% HACK FIXME
\end{proof}

Before going into our main soundness theorem, we need to show that the
elements of sum types behave in the expected way. In other words, such
a term sould reduce to either a neutral term (i.e., a term in $\calNZS$)
or to a constructor. Although the semantics of our sum types involve
arrows, we still obtain this result thanks to parametricity. This is why
the codomain of the arrows is quantified over universally in the
interpretation of sum types.
\begin{lem}\label{lem:possibi}
  Every strongly normalising pure term $t ∈ \calN$ has a head normal
  form that is either a $λ$-abstraction, a record, a constructor or
  a term in $\calNZ$.
\end{lem}
\begin{proof}
  The head normal form of a pure term can be written $H[u]$ where $u$
  is either a $λ$-abstraction, a record, a constructor or a $λ$-variable.
  If $H = [\,]$ then we can conclude immediately. If $H ≠ [\,]$ then we
  must have $u = x$, which implies $H[u] ∈ \calNZ$, as in every other
  cases $H[u]$ can be reduced.
\end{proof}
\begin{lem}\label{lem:sumsemantics}
  If $\sem{A_i} ∈ \formsem$ for all $i ∈ I$, then we have
  $t ∈ \sem{[(C_i \of A_i)_{i∈I}]}$ if and only if $t ∈ \calN$
  and either $t ≻_H^* v$ with $v ∈ \calNZ$ or $t ≻_H^* C_k\,v$
  with $k∈I$ and $v∈\sem{A_k}$.
\end{lem}
\begin{proof}
  ($⇒$) Let us suppose that $t ∈ \sem{[(C_i \of A_i)_{i∈I}]}$. By definition,
  we immediately have $t ∈ \calN$, so according to  Lemma~\ref{lem:possibi}
  there is a head normal form $v$ such that $t ≻_H^* v$, and we only need
  to show that $v$ cannot be a $λ$-abstraction, a record, a term of the form
  $C_k\,u$ with $k∉I$, or a term of the form $C_k\,u$ with $k∈I$ and
  $u ∉ \sem{A_k}$.
  To rule out the first three possibilities, we apply the definition of
  $\sem{[(C_i \of A_i)_{i∈I}]}$ using the fact that $λx.x ∈ (\sem{A_i} ⇒
  \calN)$ for all $i ∈ I$ to obtain $[t ∣ (C_i→λx.x)_{i∈I}] ∈ \calN$.
  We thus have $[v ∣ (C_i→λx.x)_{i∈I}] ∈ \calN$ since $t ≻^* v$, but
  this term diverges if $v$ has one of the first three forms.
  Let us now suppose that there is $k ∈ I$ such that $v = C_k\,u$. We
  consider the term $u_k = [t ∣ C_k → λx.x ∣ (C_i → λx.y)_{i∈I∖\{k\}}]$
  where $y$ is a fresh variable. Obviously, we have $λx.x ∈ (\sem{A_k} ⇒
  \sem{A_k})$ and $λx.y ∈ (\sem{A_i}⇒\calNZ) ⊆ (\sem{A_i} ⇒ \sem{A_k})$
  for all $i ∈ I ∖ \{k\}$. Therefore, we can use the definition of
  $\sem{[(C_i \of A_i)_{i∈I}]}$ to obtain $u_k ∈ \sem{A_k}$. We can then
  conclude that $C_k\,u ∈ \sem{A_k}$ as $\sem{A_k}$ is saturated and
  $u_k ≻_H^* C_k\,u$.

  ($⇐$) Let us now suppose that $t ∈ \calN$ and that $t ≻_H^* v$
  with either $v ∈ \calNZ$ or $v = C_k\,u$ with $k∈I$ and $u∈\sem{A_k}$.
  We need to show $t ∈ \sem{[(C_i \of A_i)_{i∈I}]}$, so we take a set
  $Φ ∈ \formsem$, terms $t_i ∈ (\sem{A_i} ⇒ Φ)$ for all $i ∈ I$, and we
  show $\case{t}{(C_i \to t_i)_{i \in I}} ∈ Φ$. Since $t ≻_H^* v$ we also
  have $\case{t}{(C_i → t_i)_{i∈I}} ≻_H^* \case{v}{(C_i → t_i)_{i∈I}}$
  and thus it is enough to show $\case{v}{(C_i \to t_i)_{i \in I}} ∈ Φ$
  according to Lemma~\ref{lem:simplesat}. Now, if $v ∈ \calNZ$ then we
  have $\case{v}{(C_i \to t_i)_{i \in I}} ∈ \calNZ$ and we can conclude
  immediately. If $v = C_k\,u$ with $k∈I$ and $u∈\sem{A_k}$, then we need
  to show $t_k\;u ∈ \sem{A_k}$, which follows from $t_k ∈ (\sem{A_k} ⇒ Φ)$.
\end{proof}

We will now prove our main soundness theorem, the so-called \emph{adequacy
lemma}. Note that the definition of saturation and the previous lemmas
give exactly the properties required for the proof of this theorem. In fact,
it is possible to gather the required properties by attempting to construct
the proof.
\begin{thm}\label{th:snadequacy}
  Let $γ$ be an ordinal context such that $\sem{τ} > 0$ for all $τ ∈ γ$. 
  \begin{enumerate}
    \item If $\subtype{γ}{t}{A}{B}$ is derivable by a well-founded proof
      and $\sem{t} ∈ \sem{A}$ then $\sem{t} ∈ \sem{B}$.
    \item If $\type{t}{A}$ is derivable by a well-founded proof then
      $\sem{t} ∈ \sem{A}$.
  \end{enumerate}
\end{thm}
\begin{proof}
  According to Theorem~\ref{th:wellfounded} we only have to prove that
  our typing and subtyping rules are correct. Note that the truth of our
  abstract judgments $⟦t : A⟧ = 1$ and $⟦t ∈ A ⊂ B⟧ = 1$ is defined
  according to the statement of the current theorem.
  We thus consider all the rules of Figure~\ref{typingrules} and
  \ref{subtypingrules}.
  \begin{itemize}
    \item[($→_i$)] We need to show $\sem{λx.t} ∈ \sem{C}$. However,
      according to the second induction hypothesis, it is enough to show
      $\sem{λx.t} ∈ \sem{A → B} = (\sem{A} ⇒ \sem{B})$. Using the second
      induction hypothesis we have $\sem{t[x:=ε_{x∈ A}(t∉B)]} ∈ \sem{B}$.
      By definition of the choice operator, this means that we have
      $\sem{t[x:=u]} = \sem{t}[x:=u] ∈ \sem{B}$ for all $u∈\sem{A}$. By
      Theorem~\ref{th:snsemantics} we know that $\sem{B}$ is saturated
      and that $\sem{A} ⊆ \calN$. We then use the saturation condition
      (1) to get $\sem{λx.t}\; u ∈ \sem{B}$ for all $u ∈ \sem{A}$.

    \item[($→_e$)] We need to show $\sem{t}\;\sem{u} ∈ \sem{B}$.
      By induction hypothesis we have $\sem{t} ∈ \sem{A→B}$ and 
      $\sem{u} ∈ \sem{A}$, so we can conclude by definition of
      $\sem{A→B} = (\sem{A}⇒\sem{B})$.

    \item[($ε$)] We need to show $\sem{ε_{x∈A}(t∉B)} ∈ \sem{C}$. However,
      according to the induction hypothesis, it is enough to show
      $\sem{ε_{x∈A}(t∉B)} ∈ \sem{A}$. This follows immediately from
      the definition of $\sem{ε_{x∈A}(t∉B)}$. In particular,
      $\calNZS ⊆ \sem{A}$ by Theorem~\ref{th:snsemantics}.

    \item[($×_i$)] We need to show that $\sem{\{(l_i=t_i)_{i∈I}\}} ∈
      \sem{B}$. According to the first induction hypothesis, it is
      enough to show $\sem{\{(l_i=t_i)_{i∈I}\}} ∈ \sem{\{(l_i:
      A_i)_{i∈I}\}}$. By definition, we need to take $k∈I$ and show
      $\{(l_i=\sem{t_i})_{i∈I}\}.l_k ∈ \sem{A_k}$. By induction
      hypothesis we know that $\sem{t_k} ∈ \sem{A_k}$, hence we can
      use the saturation condition (3) on $\sem{A_k}$ since it is
      saturated by Theorem~\ref{th:snsemantics}. Note that if $k∉I$
      then we immediately have $\{(l_i=\sem{t_i})_{i∈I}\}.l_k ≻_H^* Ω$.

    \item[($×_e$)] We need to show $\sem{t.l_k} = \sem{t}.l_k ∈ \sem{A}$.
      As we have $\sem{t} ∈ \sem{\{l_k:A;...\}}$ by induction hypothesis,
      we can conclude by definition of $\sem{\{l_k:A;...\}}$.

    \item[($+_i$)] We need to show $\sem{C_k\,t} ∈ \sem{B}$. According to
      the first induction hypothesis, it is enough to show $\sem{C_k\,t}
      = C_k\,\sem{t} ∈ \sem{[C_k \of A]}$. By definition, we need to take
      $Φ ∈ \formsem$, $t_k : (\sem{A} ⇒ Φ)$ and show
      $\case{C_k\,\sem{t}}{C_k → t_k} ∈ Φ$. Using the saturation condition
      (2) on $Φ$, it is enough to show $t_k\;\sem{t} ∈ Φ$. This follows by
      definition of $(\sem{A} ⇒ Φ)$ since $\sem{t} ∈ \sem{A}$ according to
      the second induction hypothesis.

    \item[($+_e$)] We need to show $\sem{\case{t}{(C_i→t_i)_{i∈I}}} ∈
      \sem{B}$. By the first induction hypothesis, we know that
      $\sem{t} ∈ \sem{[(C_i \of A_i)_{i∈I}]}$. We can thus conclude by
      definition of $\sem{[(C_i \of A_i)_{i∈I}]}$, using the remaining
      induction hypotheses.

    \item[($→$)] Let us suppose that $\sem{t} ∈ \sem{A₁→ B₁}$, and assume
      that $\sem{t} ∉ \sem{A₂→B₂}$ by contradiction. By definition of
      $\sem{A₂→B₂} = (\sem{A₂}⇒\sem{B₂})$ there must be $u ∈ \sem{A₂}$
      such that $\sem{t}\;u ∉ \sem{B₂}$. As a consequence, the term
      $v = \sem{ε_{x∈A₂}(t\;x ∉ B₂)}$ must satisfy $v∈\sem{A₂}$ and
      $\sem{t}\;v ∉ \sem{B₂}$ by definition of the choice operator. By
      the first induction hypothesis we have $v ∈ \sem{A₁}$, and hence
      $\sem{t}\;v ∈ \sem{B₁}$ by definition of $t ∈ \sem{A₁→B₁}$.
      Using the second induction hypothesis this gives
      $\sem{t}\;v ∈ \sem{B₂}$, which is a contradiction.

    \item[($=$)] This is a trivial implication.

    \item[($∀_l$)] We assume $\sem{t} ∈ \sem{∀X.A}$, and we show $\sem{t}
      ∈ \sem{B}$. Using the induction hypothesis, it is enough to show
      $\sem{t} ∈ \sem{A[X:=U]}$, which is equivalently to $\sem{t} ∈
      \sem{A[X:=\sem{U}]}$ according to Lemma~\ref{lem:substitution}.
      By definition of $\sem{∀X.A}$, we have $\sem{t} ∈ \sem{A[X:=Φ]}$
      for all $Φ∈\formsem$. We can thus conclude as $\sem{U} ∈ \formsem$
      by Theorem~\ref{th:snsemantics}.
      
    \item[($∀_r$)] We assume $\sem{t} ∈ \sem{A}$, and we show $\sem{t} ∈
      \sem{∀X.B}$. Using the induction hypothesis we obtain $\sem{t} ∈
      \sem{B[X := ε_X(t∉B)]}$. Consequently we have $\sem{t} ∈
      \sem{B[X := Φ]}$ for all $Φ ∈ \formsem$ by definition of the
      choice operator, and thus $\sem{t} ∈ \sem{∀ X. B}$.

    \item[($\exists_r$)] Similar to the ($∀_l$) case.

    \item[($\exists_l$)] Similar to the ($∀_r$) case.

    \item[($×_s$)] We assume $\sem{t} ∈ \sem{\{(l_i:A_i)_{i∈I}\}}$ and we
      show $\sem{t} ∈ \sem{\{(l_i:B_i)_{i∈I}\}}$. We can assume that $I≠∅$
      as otherwise the proof is trivial. By definition of
      $\sem{\{(l_i:A_i)_{i∈I}\}}$, we know that $\sem{t}.l_i ≻_H^* Ω$ for
      all $i∉I$. Thus, by definition of $\sem{\{(l_i:B_i)_{i∈I}\}}$, it
      only remains to take $k∈I$ and show $\sem{t}.l_k ∈ \sem{B_k}$. This
      follows from the induction hypothesis since $\sem{t}.l_k ∈ \sem{A_k}$
      by definition of $\sem{\{(l_i:A_i)_{i∈I}\}}$.
 
    \item[($×_{se}$)] Similar to the ($×_s$) case.

    \item[($×_e$)] Also similar to the ($×_s$) case.

    \item[($+$)] We assume $\sem{t} ∈ \sem{[(C_i \of A_i)_{i∈I₁}]}$ and
      we show $\sem{t} ∈ \sem{[(C_i \of B_i)_{i∈I₂}]}$. According to
      Lemma~\ref{lem:sumsemantics}, we know that $t ≻_H^* v$ with only
      two possibilities for $v$. In the case where $v ∈ \calNZS$ then we
      can conclude directly using Lemma~\ref{lem:sumsemantics} in the
      other direction.
      Otherwise, we know that $t ≻_H^* C_k\,u$ with $k∈I₁⊆I₂$ and
      $u∈\sem{A_k}$. We now consider the term $t.C_k$, which reduces as
      $t.C_k ≻_H^* \case{C_k\,u}{C_k → λx.x} ≻_H u$. Since $t∈\calN$, we
      can use Lemma~\ref{lem:simplesat} to deduce that $t.C_k ∈ \sem{A_k}$.
      Hence, we obtain $t.C_k ∈ \sem{B_k}$ by induction hypothesis. By
      Theorem~\ref{th:snsemantics} we know that $\sem{B_k}$ is saturated
      (and thus closed under head reduction). As a consequence, we can
      deduce $u ∈ \sem{B_k}$. We can then conclude using (the right to
      left direction of) Lemma~\ref{lem:sumsemantics}.
     
    \item[($μ_r$)] We assume $\sem{t} ∈ \sem{A}$ and we show $\sem{t} ∈
      \sem{μ_κ F}$. By the first induction hypothesis we obtain that
      $\sem{t} ∈ \sem{F(μ_τ F)}$, so it only remains to show
      $\sem{F(μ_τ F)} ⊆ \sem{μ_κ F}$. According to the second induction
      hypothesis, using Lemma~\ref{lem:leqicorrect}, we know that
      $\sem{τ} < \sem{κ}$. We thus obtain $\sem{F(μ_{\sem{τ}} F)} ⊆
      \sem{μ_{\sem{κ}} F}$ by definition of $\sem{μ_{\sem{κ}} F}$. We
      then obtain $\sem{F(μ_τ F)} = \sem{F(μ_{\sem{τ}} F)} ⊆
      \sem{μ_{\sem{κ}} F} = \sem{μ_κ F}$ using
      Lemma~\ref{lem:substitution} twice.

    \item[($ν_l$)] Similar to the ($μ_r$) case.

    \item[($μ_r^\infty$)] We assume $\sem{t} ∈ \sem{A}$ and we show
      $\sem{t} ∈ \sem{μ_∞ F}$. By induction hypothesis, we obtain
      $\sem{t} ∈ \sem{F (μ_∞ F)}$ so we only need to show
      $\sem{F (μ_∞ F)} ⊆ \sem{μ_∞ F}$. Since the cardinal of the
      ordinal $\sem{∞}$ is $2^{2^ω}$, it is larger than the cardinal
      of $\formsem$ which is $2^ω$. Hence the inductive definition of
      $\sem{μ_{\sem{∞}} F}$ must reach its stationary point strictly
      before $\sem{∞}$.\footnote{This stationary point is not a fixpoint
      if $F$ is not covariant, but we do not need this information.} As
      a consequence, we have $\sem{μ_{\sem{∞}} F} = \sem{μ_{\sem{∞}+1} F}
      ⊇ \sem{F (μ_{\sem{∞}} F)}$ by definition. We can thus conclude
      using Lemma~\ref{lem:substitution} on both sides.

    \item[($ν_l^\infty$)] Similar to the ($μ_r^\infty$) case.

    \item[($μ_l$)] Let us suppose that $\sem{t} ∈ \sem{μ_κ F}$ and show
      that $\sem{t} ∈ \sem{B}$. If $\sem{κ} = 0$ then this is immediate
      since in this case we have $\sem{μ_κ F} = \calNZS$, and thus
      $\sem{t} ∈ \sem{B}$ since $\calNZS ⊆ \sem{B}$ according to
      Theorem~\ref{th:snsemantics}. If $\sem{κ} ≠ 0$ then by definition
      there must be $o < \sem{κ}$ such that $\sem{t} ∈ \sem{F(μ_o F)}$.
      By definition of the choice operator, this means that
      $o = \sem{ε_{α < κ} (t ∈ F(μ_{α} F))}$ does verify $o < \sem{κ}$ and
      $\sem{t} ∈ \sem{F(μ_o F)}$. We can thus conclude using the
      induction hypothesis.

    \item[($ν_r$)] Similar to the ($μ_l$) case.
  \end{itemize}
  \vspace{-4.5mm}% HACK FIXME
\end{proof}

Intuitively, the adequacy lemma establishes the compatibility of our
semantics with our type system. We will now rely on this theorem to
obtain results such as consistency, strong normalisation or weak forms
of type safety.
\begin{thm}\label{th:consistency}
  There is no closed, pure term $t$ such that $\type{t}{∀X.X}$ or
  $\type{t}{[\,]}$ is derivable.
\end{thm}
\begin{proof}
  Let us assume that there is such a term $t$. According to the adequacy
  lemma (Theorem~\ref{th:snadequacy}), we have $\sem{t} ∈ \calNZS$ since
  $\sem{∀X.X} = \sem{[\;]} = \calNZS$ by definition. This is a
  contradiction since $\calNZS$ only contains open terms.
\end{proof}
\begin{thm}\label{th:strongnorm}
  Given a closed, pure term $t ∈ \pure$ and a closed type $A ∈ \mathcal{F}$,
  if $\type{t}{A}$ is derivable then $t$ is strongly normalising.
\end{thm}
\begin{proof}
  Using the adequacy lemma (Theorem~\ref{th:snadequacy}), we know that
  $\sem{t} ∈ \sem{A}$. However, since $t$ is pure we have $\sem{t} = t$.
  Moreover, according to Theorem~\ref{th:snsemantics} we have $\sem{A}
  ⊆ \calN$, and thus we obtain $t ∈ \calN$.
\end{proof}

Note that, as a direct consequence of strong normalisation, we know that
a well-typed term cannot produce a runtime error. Indeed, the reduction
rules of Figure~\ref{fig:reduction} introduce a non-terminating term in
case of an error (e.g., the projection of a $λ$-abstraction). We will
now consider a stronger safety result, which will apply to so-called
\emph{simple} data types. They will cover most of the common inductive
datatypes such as lists or binary trees.
\begin{defi}
  We say that a type $A ∈ \mathcal{F}$ is \emph{simple} if it is closed, and
  if it only contains sums, strict products and least fixpoints carrying the
  $∞$ ordinal. Moreover, we will assume that a simple type $A$ does not
  have two consecutive least fixpoints, and that the body of fixpoints is
  not limited to a variable (like in $μX.Y$ or $μX.X$).
\end{defi}
\begin{thm}\label{th:safety}
  If $\type{t}{A}$ is derivable for a closed, pure term $t$ and a simple type
  $A$, then $t$ reduces to a normal form $u$ such that $\type{u}{A}$ is
  derivable.
\end{thm}
\begin{proof}
  According to Theorem \ref{th:strongnorm}, we know that $t$ must reduce to
  a normal form $u$. Moreover, $u$ is closed since no free variables are
  introduced by our reduction rules.
  We proceed by induction on the size of $u$. In the case where $A = μX.B$
  we know that $\sem{B[X  := A]} ⊆ \sem{A}$. Let us define $A' = B[X := A]$
  if $A = μX.B$ and $A' = A$ otherwise. The hypotheses on least fixpoints
  are still true in $A'$ since $B$ cannot be equal to $X$ by hypothesis.
  Moreover, since pure types may not contain two consecutive fixpoints and
  $A$ cannot be $μX.X$, $A'$ is either a sum type or a strict product type.

  If $A' = [(C_i : A_i)_{i \in I}]$ then, by Lemma~\ref{lem:sumsemantics}
  we know that $u = C_k\,v$ with $k ∈ I$ and $v ∈ \sem{A_k}$. In particular,
  $u$ is in normal form (and thus in head normal form) and it cannot be
  open, which means that $u ∉ \calNZS$. Since $u$ is in normal form, we
  know that $v$ is also in normal form. The induction hypothesis provides us
  with a derivation of $\type{v}{A_k}$. In the case where $A' = A
  = [(C_i : A_i)_{i∈I}]$ then we can conclude using the follwing derivation.
  \begin{prooftree}
    \AxiomC{$\{k\} ⊆ I$}
    \RightLabel{$=$}
    \AxiomC{$\subtype{}{t.C_k}{A_k}{A_k}$}
    \RightLabel{$+$}
    \BinaryInfC{$\subtype{}{C_k\,v}{[C_k : A_k]}{[(C_i : A_i)_{i∈I}]}$}
    \AxiomC{$\type{v}{A_k}$}
    \RightLabel{$+_i$}
    \BinaryInfC{$\type{C_k\,v}{[(C_i : A_i)_{i∈I}]}$}
  \end{prooftree}
  Otherwise, if we have $A = μX.[(C_i : A_i)_{i∈I}]$ then $A' = [(C_i :
  A_i[X:=A])_{i∈I}]$ and we can construct the following derivation. Note
  that in this case, $A_k$ is rather of the form $A_k[X:=A]$, so we in
  fact have a proof of $\type{v}{A_k[X:=A]}$.
  \begin{prooftree}
    \AxiomC{$\{k\} ⊆ I$}
    \RightLabel{$=$}
    \AxiomC{$\subtype{}{t.C_k}{A_k[X:=A]}{A_k[X:=A]}$}
    \RightLabel{$+$}
    \BinaryInfC{$\subtype{}{C_k\,v}{[C \of A_k]}{[(C_i : A_i[X:=A])_{i∈I}]}$}
    \RightLabel{$\mu_r^\infty$}
    \UnaryInfC{$\subtype{}{C_k\,v}{[C \of A_k]}{μX.[(C_i : A_i)_{i∈I}]}$}
    \AxiomC{$\type{v}{A_k[X:=A]}$}
    \RightLabel{$+_i$}
    \BinaryInfC{$\type{C_k\,v}{μX.[(C_i : A_i)_{i∈I}]}$}
  \end{prooftree}
 
  Now, if $A' = \{(l_i:A_i)_{i∈I}\}$ is a strict product type then the
  proof is similar. However, we first need to remark that $v = \{(l_i=
  v_i)_{i∈I}\}$ with $v_i ∈ \sem{A_i}$ for all $i∈I$. Note that all the
  other possible forms of normal forms can be ruled our using similar
  techniques as in the proof of Lemma~\ref{lem:sumsemantics}. By
  induction hypothesis, we can obtain a proof of $\type{v_i}{A_i}$ for
  all $i∈I$ and the reconstruct proofs as in the case of the sum
  types.
\end{proof}

To conclude this section, we will discuss the closure by head reduction
imposed in our definition of saturation. This condition is not usually
required, but it is needed here for a subtle reason. Although it is
used in the proof of Theorem~\ref{th:safety}, the main aim of this
condition is to allow for the correctness of the subtyping rule for
sums recalled bellow.
\begin{prooftree}
  \AxiomC{$I₁ ⊆ I₂$}
  \AxiomC{$(\subtype{\gamma}{t.C_i}{A_i}{B_i})_{i∈I₁}$}
  \RightLabel{$+$}
  \BinaryInfC{$\subtype{γ}{t}{[(C_i : A_i)_{i∈I₁}]}{[(C_i : B_i)_{i∈I₂}]}$}
\end{prooftree}
Indeed, closure under head reduction is necessary to accommodate the simple
witnesses of the form $t.C_i$. It would be possible
to use more complex witnesses similar to those introduced by the
following encoding of sums as products.
$$ [(Cᵢ \of Aᵢ)_{i∈I}] = ∀X \{(Cᵢ : Aᵢ ⇒ X)_{i∈I}\} ⇒ X $$
However, there is a fundamental problem with this encoding as the witnesses
would mention all the types $A_i$ and $B_i$ due to
subtyping on the arrow types.
As a consequence, such witnesses would prevents the derivation of subtyping
relations like $∀X [C \of A] ⊂ [C \of ∀X A]$ or $[C \of ∃X A] ⊂ ∃X [C \of A]$.
The simple witnesses mention none of these types, and thus give a
workaround to this problem.
%%% Local Variables:
%%% ispell-personal-dictionary: "~/.hunspell-en"
%%% ispell-local-dictionary: "british"
%%% End:

\section{Fixpoint and termination}\label{fixpoint}
We will now extend the system with general recursion using a fixpoint
combinator $Y x.t$, while preserving a termination property on
programs. Obviously, strong normalisation is compromised by the
reduction rule $Y x.t ≻ t[x := Y x.t]$ of the fixpoint. Nonetheless,
we will prove normalisation for all the weak reduction strategies,
(i.e., those that do not reduce under $λ$-abstractions, and hence
under the right members of case analyses).

Moreover, to prove the termination of certain programs, we will need
to express the fact that some functions are size-preserving. For
example, proving the termination of quicksort will require the partition
function to return two lists that are no bigger than the input list. To
this aim, we provide quantification over ordinals in types. We will thus
be able to write $∀A.∀B.∀α.(A ⇒ B) ⇒ L_α(A) ⇒ L_α(B)$ for the type of
the map function on lists, where $L_α(A) = μ_α X. [ \mathrm{Nil} \of \{\},
\mathrm{Cons} \of \{car : A, cdr : X\}]$. It is important to note that
this is a subtype of $∀A.∀B.(A ⇒ B) ⇒ L_∞(A) ⇒ L_∞(B)$.

Finally, proving the termination of recursive programs will generally
require us to extend our typing judgments with ordinal contexts. We
will then be able to assume that certain ordinals are positive while
building typing proofs. For example, if we know that $l : L_α(A)$ and
we want to type the case analysis $[l \st \mathrm{Nil} → u \st \mathrm{Cons}
→ v]$, then we can assume that $α > 0$ when typing $u$ and $v$.
Indeed, if $α = 0$ then we know that $l$ is a neutral term and the
typing proof is trivial. Without this technique, we would for example
not be able to use the previously size-preserving type for the map
function on lists. To transfer positivity hypotheses from subtyping
judgments to typing judgements, we will rely on new connectives $A ∧ α$
and $A ∨ α$. The former will be interpreted as $A$ if $α ≠ 0$ and as
$∀X.X$ otherwise, and the latter will be interpreted as $A$ id $α ≠ 0$
and as $∃X.X$ otherwise. They will appear in the premises of our typing
rules, and they will be handled using new subtyping rules.

\begin{defi}
  We extend the syntax of terms and types given in
  Definition~\ref{def:syntax} with a fixpoint combinator and
  new connectives as follows.
  \begin{align*}
    t, u \bnfeq &\dots \bnfor Y x.t \\[1ex]
    A, B \bnfeq &\dots \bnfor ∀α.A \bnfor ∃α.A \bnfor A∧α \bnfor A∨α
  \end{align*}
  Note that this new definition also impacts abstract judgments and
  syntactic ordinals. However, we will still work with abstract
  judgments of the form $t : A$ and $t ∈ A ⊆ B$. As for $λ$-abstractions,
  terms of the form $Yx.t$ are not allowed to bind variables through choice
  operators of the form $ε_{x∈A}(t∉B)$.
\end{defi}
\begin{nota}
  We will use the abbreviations $A ∧ γ$ and $A ∨ γ$, where
  $γ = κ₁,\dots,κₙ$ is an ordinal context, to denote $A ∧ κ₁ \dots ∧ κₙ$
  and $A ∨ κ₁ \dots ∨ κₙ$ respectively. In particular, if $γ = ∅$ then
  we have $A ∧ γ = A ∨ γ = A$. We will also use the notation $γ₁, γ₂$ for
  the union of the ordinals contexts $γ₁$ and $γ₂$.
\end{nota}

Before going into the typing and subtyping rules of the extended system,
we first need to consider a syntactic condition on terms. It will be used
to strengthen several typing rules by allowing us to assume the positivity
of syntactic ordinals in some cases.
\begin{defi}
  We say that a term $t ∈ Λ$ is \emph{weakly normal} and we write $t↓$
  if either $t = ε_{x∈A} u ∉ B$, $t = λx.u$, $t = C u$ and $u↓$, or
  $t = \{(l_i = u_i)_{i∈I}\}$ and $u_i ↓$ for all $i ∈ I$.
\end{defi}

\begin{figure}

% Arrow introduction
\begin{prooftree}
\AxiomC{$\subtype{γ}{λ x\;t}{(A → B) ∨ γ_0}{C}$}
\AxiomC{$\type[γ, γ_0]{t[x := ε_{x∈ A}(t ∉ B)]}{B}$}
\RightLabel{$→_i$}
\BinaryInfC{$\type[γ]{λx.t}{C}$}
\end{prooftree}

\smallskip

% Arrow elimination and epsilon
\begin{prooftree}
\AxiomC{$\type[γ]{t}{(A → B) ∧ γ_0}$}
\AxiomC{$\type[γ, γ_0]{u}{A}$}
\RightLabel{$→_e$}
\BinaryInfC{$\type[γ]{t\;u}{B}$}
\DisplayProof\hfill
\AxiomC{$\subtype{γ}{ε_{x∈ A}(t ∉ B)}{A}{C}$}
\RightLabel{$ε$}
\UnaryInfC{$\type[γ]{ε_{x∈ A}(t ∉ B)}{C}$}
\end{prooftree}

\smallskip

% Product introduction
\begin{prooftree}
\AxiomC{$\subtype{γ}{\{(l_i=t_i)_{i ∈ I}\}}{\{(l_i : A_i)_{i∈I}\} ∨ γ_0}{B}$}
\AxiomC{\hspace{-1em}$(\type[γ, γ_0]{t_i}{A_i})_{i ∈ I}$}
\AxiomC{\hspace{-1em}$γ_0 = ∅$ or $∀ i, t_i ↓$}
\RightLabel{$×_i$}
\TrinaryInfC{$\type[γ]{\{(l_i = t_i)_{i ∈ I}\}}{B}$}
\end{prooftree}

\smallskip

% Sum introduction
\begin{prooftree}
\AxiomC{$\subtype{γ}{C\,t}{[C \of A] ∨ γ_0}{B}$}
\AxiomC{$\type[γ, γ_0]{t}{A}$}
\AxiomC{$γ_0 = ∅$ or $t ↓$}
\RightLabel{$+_i$}
\TrinaryInfC{$\type[γ]{C\,t}{B}$}
\end{prooftree}

\smallskip

% Sum elimination
\begin{prooftree}
\AxiomC{$\type[γ]{t}{[(C_i \of A_i)_{i ∈ I}] ∧ γ_0}$}
\AxiomC{$(\type[γ, γ_0]{t_i}{A_i → B})_{i ∈ I}$}
\RightLabel{$+_e$}
\BinaryInfC{$\type[γ]{\case{t}{(C_i → t_i)_{i ∈ I}}}{B}$}
\end{prooftree}

\smallskip

% Product elimination and fixpoint
\begin{prooftree}
\AxiomC{$\type[γ]{t}{\{l_k : A; \dots\}}$}
\RightLabel{$×_e$}
\UnaryInfC{$\type[γ]{t.l_k}{A}$}
\DisplayProof\hfill
\AxiomC{$\type[γ]{t[x := Y x.t]}{A}$}
\RightLabel{$Y$}
\UnaryInfC{$\type[γ]{Y x.t}{A}$}
\end{prooftree}

\caption{Typing of the system extended with general recursion.}
\label{typingrules}
\end{figure}

\begin{figure}

% forall left and forall right  on ordinals
\begin{prooftree}
\AxiomC{$\subtype{γ}{t}{A[α:= κ]}{B}$}
\RightLabel{$∀^o_l$}
\UnaryInfC{$\subtype{γ}{t}{∀α. A}{B}$}
\DisplayProof\hfill
\AxiomC{$\subtype{γ}{t}{A}{B[α := ε_{α<\OMaxi}(t ∉ B)]}$}
\RightLabel{$∀^o_r$}
\UnaryInfC{$\subtype{γ}{t}{A}{∀α. B}$}
\end{prooftree}

\smallskip

% Exists left and exists right on ordinals
\begin{prooftree}
\AxiomC{$\subtype{γ}{t}{B}{A[α:= κ]}$}
\RightLabel{$∃^o_r$}
\UnaryInfC{$\subtype{γ}{t}{B}{∃ α. A}$}
\DisplayProof\hfill
\AxiomC{$\subtype{γ}{t}{B[X := ε_{α<\OMaxi}(t ∈ B)]}{A}$}
\RightLabel{$∃^o_l$}
\UnaryInfC{$\subtype{γ}{t}{∃ α. B}{A}$}
\end{prooftree}

\smallskip

% Conjunction and disjunction
\begin{prooftree}
\AxiomC{$\subtype{γ, κ}{t}{A}{B}$}
\RightLabel{$∧_l$}
\UnaryInfC{$\subtype{γ}{t}{A ∧ κ}{B}$}
\DisplayProof\hfill
\AxiomC{$\subtype{γ}{t}{A}{B}$}
\AxiomC{$κ ∈ γ$}
\RightLabel{$∧_r$}
\BinaryInfC{$\subtype{γ}{t}{A}{B ∧ κ}$}
\end{prooftree}

\smallskip

\begin{prooftree}
\AxiomC{$\subtype{γ, κ}{t}{A}{B}$}
\RightLabel{$∨_r$}
\UnaryInfC{$\subtype{γ}{t}{A}{B ∨ κ}$}
\DisplayProof\hfill
\AxiomC{$\subtype{γ}{t}{A}{B}$}
\AxiomC{$κ ∈ γ$}
\RightLabel{$∨_l$}
\BinaryInfC{$\subtype{γ}{t}{A ∨ κ}{B}$}
\end{prooftree}

\caption{Extra subtyping rules for the extended system.}
\label{subtypingrules2}
\end{figure}

\begin{defi}
  Our typing judgments now have the form $γ⊢t:A$, where $γ$ is an
  ordinal context. The typing rules of the extended system are
  given in Figure~\ref{typingrules}. Its subtyping rules still
  include those of Figure~\ref{subtypingrules}, but the rules of
  Figure~\ref{subtypingrules2} are added to handle the new
  connectives. Note that we allow circular subtyping proofs using
  the ($\GP$) and ($\IP{k}$) rules as in Section~\ref{language},
  and circular typing rules using the ($\G$) and ($\I{k}$) rules.
\end{defi}
The typing rules of the system need to be changed completely to account
for the ordinal contexts. Note that they are strongly linked to the new
connectives $A∧α$ and $A∨α$ in types. Moreover, the ($×_i$) and ($+_i$)
rules require some terms to be weakly normal to learn the positivity
of certain syntactic ordinals.
Furthermore, the system now includes circular typing proofs to handle
general recursion. Note that the typing rule of the fixpoint is very
simple as it only performs an unfolding. In practice, we will only
need to allow circularity on typing judgments of the form $γ ⊢ Yx.t:A.$.
In this context, the ($\G$) and ($\I{k}$) rules can be written as in
Figure~\ref{fig:circulartyping}, where we write ${ε_{\vec{α}<C(\vec{α})}
(Yx.t ∉ A)}_i$ for the ordinal ${ε_{\vec{α} < C(\vec{α})} ¬ (Yx.t : A)}_i$
(see Section~\ref{sct}).
\begin{figure}
  \begin{prooftree}
    \AxiomC{$∀ \vec{α} (γ ⊢ C(\vec{α}) ⇒ Yx.t : A)$}
    \AxiomC{$(γ[\vec{α} := \vec{κ}],δ ⊢
      κ_i < C(\vec{κ})_i)_{1 ≤ i ≤ |\vec{α}|}$}
    \RightLabel{$\G$}
    \BinaryInfC{$γ[\vec{α}:=\vec{κ}],δ ⊢ Yx.t : A[\vec{α}:=\vec{κ}]$}
  \end{prooftree}
  \smallskip
  \begin{prooftree}
    \AxiomC{$\H{∀ \vec{α} (γ ⊢ C(\vec{α}) ⇒ Yx.t : A)}{k}$}
    \noLine
    \UnaryInfC{$\smash{\raisebox{-0.5mm}{\vdots}}\vphantom{0mm}$}
    \noLine
    \UnaryInfC{$γ[\vec{α}:=\vec{κ}] ⊢ Yx.t : A[\vec{α}:=\vec{κ}]$}
    \RightLabel{$\I{k}$}
    \AxiomC{where $\vec{κ} = \vec{ε}_{\vec{α}<C(\vec{α})} (Yx.t ∉ A)$}
    \BinaryInfC{$∀ \vec{α} (γ ⊢ C(\vec{α}) ⇒ Yx.t : A)$}
  \end{prooftree}
  \caption{Specialised circular typing rules for general recursion.}
  \label{fig:circulartyping}
\end{figure}

\begin{figure*}
  \centering
  {\tiny
   \def\defaultHypSeparation{\hskip 0.4em}
   \begin{prooftree}
  \AxiomC{}
  \RightLabel{$$=$$}
  \UnaryInfC{$κ_{0} ⊢ n_{0} ∈ \mathrm{F}(\mathbb{N}_{κ_{1}}) ⊂ \mathrm{F}(\mathbb{N}_{κ_{1}})$}
  \RightLabel{$∧_r$}
  \UnaryInfC{$κ_{0} ⊢ n_{0} ∈ \mathrm{F}(\mathbb{N}_{κ_{1}}) ⊂ \mathrm{F}(\mathbb{N}_{κ_{1}}) ∧ {κ_{0}}$}
  \RightLabel{$μ_l$}
  \UnaryInfC{$ ⊢ n_{0} ∈ \mathbb{N}_{κ_{0}} ⊂ \mathrm{F}(\mathbb{N}_{κ_{1}}) ∧ {κ_{0}}$}
  \RightLabel{$\mathrm{Ax}$}
  \UnaryInfC{$ ⊢ n_{0} : \mathrm{F}(\mathbb{N}_{κ_{1}}) ∧ {κ_{0}}$}
  \AxiomC{}
  \RightLabel{$$}
  \UnaryInfC{$κ_{0} ⊢ \mathrm{z} : \mathbb{N}$}
  \AxiomC{}
  \RightLabel{$\mathrm{Ax}$}
  \UnaryInfC{$κ_{0} ⊢ p_{0} : \mathbb{N}_{κ_{1}}$}
  \AxiomC{$\H{∀α_{0} ( ⊢  \mathrm{id} : \mathbb{N}_{α_{0}} → \mathbb{N})}{0}$}
  \RightLabel{$\G$}
  \UnaryInfC{$κ_{0} ⊢ \mathrm{id} : \mathbb{N}_{κ_{1}} → \mathbb{N}$}
  \RightLabel{$→_e$}
  \BinaryInfC{$κ_{0} ⊢ \mathrm{id} \; p_{0} : \mathbb{N}$}
  \AxiomC{}
  \RightLabel{$$}
  \UnaryInfC{$κ_{0} ⊢ \mathrm{s} : \mathbb{N} → \mathbb{N}$}
  \RightLabel{$→_e$}
  \BinaryInfC{$κ_{0} ⊢ \mathrm{s} \; (\mathrm{id} \; p_{0}) : \mathbb{N}$}
  \RightLabel{$→_i$}
  \UnaryInfC{$κ_{0} ⊢ λp.\mathrm{s} \; (\mathrm{id} \; p) : \mathbb{N}_{κ_{1}} → \mathbb{N}$}
  \RightLabel{$+_e$}
  \TrinaryInfC{$ ⊢ \case{n_{0}}{ Z → \mathrm{z} ∣ Sp → \mathrm{s} \; (\mathrm{id} \; p)} : \mathbb{N}$}
  \RightLabel{$→_i$}
  \UnaryInfC{$ ⊢ λn.\case{n}{ Z → \mathrm{z} ∣ Sp → \mathrm{s} \; (\mathrm{id} \; p)} : \mathbb{N}_{κ_{0}} → \mathbb{N}$}
  \RightLabel{$Y$}
  \UnaryInfC{$ ⊢ \mathrm{id} : \mathbb{N}_{κ_{0}} → \mathbb{N}$}
  \RightLabel{$\I{0}$}
  \UnaryInfC{$∀α_{0} ( ⊢  \mathrm{id} : \mathbb{N}_{α_{0}} → \mathbb{N})$}
  \RightLabel{$\G$}
  \UnaryInfC{$ ⊢ \mathrm{id} : \mathbb{N} → \mathbb{N}$}
\end{prooftree}
}
  \bigskip
  \begin{multicols}{2}
    \noindent
    \small
    \begin{align*}%
      \mathrm{F}(X) &= [\mathrm{Z} \st \mathrm{S} \of X]\\
      \mathbb{N}_{α} &= μ_{α}X.\mathrm{F}(X)\\
      \mathbb{N} &= \mathbb{N}_∞\\
      %\mathbb{N} &= μX.\mathrm{F}(X)\\
      \mathrm{z} &\,: \mathbb{N} = Z\\
      \mathrm{s} &\,: \mathbb{N} → \mathbb{N} = λn.S n\\
      \mathrm{id} &= Yid.λn.\case{n}{ Z → \mathrm{z} ∣ Sp → \mathrm{s} \; (id \; p)}\\
      κ_{0} &= ε_{α_{1}<\OMaxi}(\mathrm{id}∉\mathbb{N}_{α_{1}} → \mathbb{N})\\
κ_{1} &= ε_{α<κ_{0}}(n_{0}∈\mathrm{F}(\mathbb{N}_{α}))\\
n_{0} &= ε_{n ∈ \mathbb{N}_{κ_{0}}}(\case{n}{ Z → \mathrm{z} ∣ Sp → \mathrm{s} \; (\mathrm{id} \; p)} ∉ \mathbb{N})\\
p_{0} &= ε_{p ∈ \mathbb{N}_{κ_{1}}}(\mathrm{s} \; (\mathrm{id} \; p) ∉ \mathbb{N})%
    \end{align*}
  \end{multicols}
  \caption{Typing proof of the recursive identity function on $\mathbb{N}$.}
  \label{fig:id}
\end{figure*}
\begin{exa}
  We consider the identity function on unary natural numbers. It can be
  typed using the derivation given in Figure~\ref{fig:id}, which is the
  simplest possible example of a circular typing proof.
  Follownig the terminology of Section~\ref{sct}), the proof is formed
  using two blocks. The former, that we will call $B_0$, starts at the
  root of the proof and only contains one typing rule. The latter, that
  we will call $B_1$, contains all the rest of the proof. The call graph
  corresponding to the proof contains one edge from $B_0$ to $B_1$,
  labelled with the empty matrix, and one edge from $B_1$ to itself,
  labelled with the $1×1$ matrix $(-1)$ since we can prove $κ₀⊢κ₁<κ₀$.

  It is important to note that the positivity of $κ₀$ must be known to obtain
  $κ₁<κ₀$. It is thus essential to use the type $F(\mathbb{N}_{κ₁}) ∧ κ₀$ (and
  not $F(\mathbb{N}_{κ₁})$) for the first premisse of the ($+_e$) rule. This
  allows us to assume that $κ₀$ is positive when typing its other premisses.
  There would be no way of building a typing proof withoug doing so.
\end{exa}

%%% Local Variables:
%%% ispell-personal-dictionary: "~/.hunspell-en"
%%% ispell-local-dictionary: "british"
%%% End:

We will now modify the semantics that was given in Section~\ref{semantics}
to account for the fixpoint combinator and the new connectives. Here, we
will not be able to interpret types as subsets of $\calN$, since the
reduction rule of the fixpoint will break strong normalisation. We will
however be able to preserve normalisation for all the weak reduction
strategies (i.e., those that do not reduce under $λ$-abstractions, and
thus case analyses as well).
\begin{defi}
  We denote $(≻_w) ⊆ Λ×Λ$ the one step weak reduction relation.
  It is defined as the least relation containing the rules of
  Figure~\ref{fig:reduction} and $Yx.t ≻_w t[x:=Yx.t]$, and that
  is contextually closed for weak contexts (i.e., contexts formed
  without a $λ$-abstraction constructor). Its reflexive, transitive
  closure is denoted $(≻_w^*)$.
\end{defi}
\begin{defi}\label{nzdef2}
  We denote $\calW ⊆ \pure$ the set of all the pure terms that are
  strongly normalising for the $(≻_w)$ reduction relation. In other words,
  we have $t ∈ \calW$ if and only if there is no infinite sequence of
  reduction of $t$ using $(≻_w)$.
\end{defi}

Using the set $\calW$ we can define a notion of saturated set, as well
as a set $\calWZ$ like in Section~\ref{semantics}. We are then able to
prove corresponding lemmas using the same techniques.
\begin{defi}\label{def:sat2}
  A set of pure terms $Φ ⊆ \pure$ is said to be \emph{weakly saturated}
  if it satisfies the conditions of Definition~\ref{def:sat}, where every
  occurrence of $\calN$ is replaced by $\calW$, plus the following
  condition.
  \begin{enumerate}
    \item[(5)] If $H[t[x:=Yx.t]] ∈ Φ$, then $H[Yx.t] ∈ Φ$.
  \end{enumerate}
\end{defi}
\begin{defi}
  The set $\calWZ$ is defined as $\calNZ$ (see Definition~\ref{def:nz}),
  but using $\calW$ instead of $\calN$. We denote by $\calWZS$ the least
  weakly saturated set containing $\calWZ$.
\end{defi}
\begin{exa}
  The term $y\;(Yr.λx.r)$ is in $\calWZ$, but not in $\calNZ$.
\end{exa}

Using the above definitions, we can obtain similar properties as in
Section~\ref{semantics}. This is mainly due to the fact that the
proof of theses lemmas do not considers reductions which are allowed
for $(≻)$ but forbidden for $(≻_w)$. We will first show that $\calW$
is weakly saturated, but this requires a small lemma that was
immediate in Section~\ref{semantics}.
\begin{lem}\label{nsat2:aux}
  For any terms $t ∈ Λ$ and $u ∈ \calW$ such that $u ≻_w^* u'$, if
  $t[x:=u'] ∈ \calW$ has an infinite weak reduction then $t[x:=u]$
  also has one.
\end{lem}
\begin{proof}
  We reason coinductively. We first distinguish the occurrences of
  $x$ in $t$ that appear under an abstraction by denoting them $x₀$,
  while denoting the others $x₁$. We hence obtain $t[x:=u] =
  (t[x₁:=u])[x₀:=u]$ and $t[x:=u'] = (t[x₁:=u'])[x₀:=u']$ up to a
  renaming of $x$.
  Let us now consider the first step of an infinite reduction of
  $t[x:=u']  ≻_w  t'[x₀:=u']$, with $t[x₁:=u'] ≻_w t'$ (there
  cannot be any weak reduction for the occurrences of $u'$ replacing
  $x₀$). We thus have $t[x:=u] = (t[x₁:=u])[x₀:=u] ≻_w^*
  (t[x₁:=u'])[x₀:=u] ≻_w t'[x₀:=u]$.
  This step being productive, we can apply the coinduction hypothesis
  with $t'$ to get an infinite weak reduction of $t'[x₀:=u]$ from the
  infinite weak reduction of $t'[x₀:=u']$.
\end{proof}
\begin{lem}\label{nsat2}
  The set $\calW$ is weakly saturated.
\end{lem}
\begin{proof}
  The proof is exactly the same as that of Lemma~\ref{lem:nsat}, except for
  condition (1). In this case, we need to prove that if $H[t[x:=u]] ∈ \calW$
  and $u ∈ \calW$, then $H[(λx.t)\;u] ∈ \calW$. We thus suppose, by
  contradiction, that $H[(λx.t)\;u]$ has an infinite weak reduction. Such
  a reduction must start with $H[(λx.t)\;u] ≻_w^* H'[(λx.t)\;u'] ≻_w^*
  H'[t[x:=u']]$, where $H ≻_w^* H'$ and $u ≻_w^* u'$. It can hence be
  transformed into $H[t[x:=u]] ≻_w^* H'[t[x:=u]]$ and we can use
  Lemma~\ref{nsat2:aux} to obtain an infinite reduction of $H'[t[x:=u]]$
  from the infinite reduction of $H'[t[x:=u']]$. This gives a contradiction
  with $H[t[x:=u]] ∈ \calW$.
\end{proof}

We will now consider the interpretation of terms, types and syntactic
ordinals to handle the fixpoint and the new connectives. However, let
us first give the new domain of interpretation for our types.
\begin{defi}
  The set of every type interpretaton $\formsem$ is now defined as follows.
  $$\formsem = \{Φ⊆\pure \st Φ \text{ weakly saturated},\;\calWZ⊆Φ⊆\calW\}$$
\end{defi}
\begin{defi}
  We modify the definition of the interpretation of terms and formulas given
  in Figure~\ref{interpr} by replacing every occurence of $\calN$ and
  $\calNZS$ with $\calW$ and $\calWZS$ respectively. The new syntactic
  elements are interpreted as follows.

  \centering
  \begin{minipage}{6cm}
    \begin{align*}
      \sem{Yx.t} &= Yx.\sem{t} \\
      \sem{∀α.A} &= \cap_{o \in \ordsem} \sem{A[α:=o]} \\
      \sem{∃α.A} &= \cup_{o \in \ordsem} \sem{A[α:=o]}
    \end{align*}
  \end{minipage}
  \begin{minipage}{6cm}
    \begin{align*}
      \sem{A∧κ}  &=
        \left\{
          \begin{array}{l}
            \!\!\!\sem{A} \text{ if } \sem{κ} ≠ 0 \\
            \!\!\!\calWZS \text{ otherwise}
          \end{array}
        \right. \\
      \sem{A∨κ}  &=
        \left\{
          \begin{array}{l}
            \!\!\!\sem{A} \text{ if } \sem{κ} ≠ 0 \\
            \!\!\!\calW \text{ otherwise}
          \end{array}
        \right.
    \end{align*}
  \end{minipage}
\end{defi}
\begin{thm}\label{th:swnsemantics}
  For every closed parametric term $t \in \parterm$ (resp. ordinal
  $κ ∈ \parord$, resp. type $A ∈ \parform$) we have $\sem{t} ∈ \pure$
  (resp. $\sem{κ} ∈ \ordsem$, resp. $\sem{A} ∈ \formsem$).
\end{thm}
\begin{proof}
  The proof is similar as for Theorem~\ref{th:snsemantics}. The cases for
  the four new type constructors are immediate by induction hypothesis.
\end{proof}

We will now give the adequacy lemma for the new system, which will be similar
to that of Section~\ref{semantics}. For this reason, we will not state all
require lemmas (e.g., the equivalent of Theorem~\ref{lem:substitution}), as
their proof does not change much. We however need a small lemma for handling
the strong normality condition in some of our new typing rules.
\begin{lem}\label{lem:strnorm}
  If $t ∈ Λ$ be a term such that $t↓$ (i.e., $t$ is weakly normal), then
  $\sem{t} ∈ \calW$.
\end{lem}
\begin{proof}
  Immediate by induction, using Theorem~\ref{th:swnsemantics} when
  $t = ε_{x∈A}(t∉B)$.
\end{proof}
\begin{thm}\label{th:swnadequacy}
  Let $γ$ be an ordinal context such that $\sem{τ} > 0$ for all $τ ∈ γ$. 
  \begin{enumerate}
    \item If $\subtype{γ}{t}{A}{B}$ is derivable by a well-founded proof
      and $\sem{t} ∈ \sem{A}$ then $\sem{t} ∈ \sem{B}$.
    \item If $γ \type{t}{A}$ is derivable by a well-founded proof then
      $\sem{t} ∈ \sem{A}$.
  \end{enumerate}
\end{thm}
\begin{proof}
  The proof is similar to that of Theorem~\ref{th:snadequacy}, using
  Theorem~\ref{th:wellfounded}. For the local subtyping rules of
  Figure~\ref{subtypingrules}, the proof remains essentially the same.
  Occurrences of $\calN$ and $\calNZS$ need to be replaced by $\calW$
  and $\calWZS$, and lemmas need to be modified according to the new
  definitions (their proofs are mostly unchanged). Similarly, the
  cases of the ($ε$) and ($×_e$) typing rules are unchanged (up to the
  transmission of the context in the induction hypothesis). Hence, we
  only consider the cases of the remaining typing rules of
  Figure~\ref{typingrules}, and the local subtyping rules of
  Figure~\ref{subtypingrules2}.
  \begin{itemize}
    \item[($→_i$)] We need to show $\sem{λx.t} ∈ \sem{C}$. According to the
      first induction hypothesis, it is enough to show $\sem{λx.t} ∈ \sem{(A
      → B) ∨ γ₀}$. If there is $κ ∈ γ₀$ such that $\sem{κ}=0$ then we have
      $\sem{(A → B) ∨ γ₀} = \calW$ and we can conclude immediately by
      Lemma~\ref{lem:strnorm}. We can thus assume that $\sem{κ} ≠ 0$ for all
      $κ ∈ γ₀$ and that the positivity context of the second induction
      hypothesis is valid to obtain $\sem{t[x:=ε_{x ∈ A}(t ∉ B)]} ∈ \sem{B}$.
      By definition of the choice operator, this means that $\sem{t[x:=u]} ∈
      \sem{B}$ for all $u ∈ \sem{A}$. Hence we can conclude
      $\sem{λx.t} ∈ \sem{A→B} = \sem{(A→B) ∨ γ₀} $ since we know that
      $\sem{A→B}$ is weakly saturated.

    \item[($→_e$)] We need to show $\sem{t\;u} ∈ \sem{B}$. By the first
      induction hypothesis $\sem{t} ∈ \sem{(A → B) ∧ γ₀}$. If $\sem{κ}=0$
      for some $κ ∈ γ₀$, then $\sem{(A → B) ∧ γ₀} = \calWZS$ and thus we
      have $\sem{t} ∈ \calWZS$, which implies $\sem{t\;u} ∈ \calWZS ⊆
      \sem{B}$. Otherwise, we have $\sem{κ} ≠ 0$ for all $κ ∈ γ₀$, and
      thus $\sem{t} ∈ \sem{A → B}$. We can hence use the second induction
      hypothesis to get $\sem{u} ∈ \sem{A}$ and conclude by definition of
      $\sem{A→B}$.

    \item[($\times_i$)] We only need to prove $\sem{\{(l_i : A_i)_{i∈I}\}} ∈
      \sem{\{(l_i:A_i)_{i∈I}\}∨γ₀}$ according to the first induction
      hypothesis. If $\sem{κ} = 0$ for some $κ ∈ γ₀$ and if $t_i ↓$ for all
      $i∈I$, then we can conclude immediately using Lemma~\ref{lem:strnorm}
      as $\sem{\{(l_i:A_i)_{i∈I}\} ∨ γ₀} = \calW$. Otherwise, we can use the
      remaining induction hypotheses to
      get $\sem{t_i} ∈ \sem{A_i}$ for all $i∈I$. From this we obtain
      $\sem{\{(l_i = t_i)_{i∈I}\}} ∈ \sem{\{(l_i : A_i)_{i∈I}\}}$ using
      weak saturation. We can then conclude since
      $\sem{\{(l_i:A_i)_{i∈I}\}∨γ₀} = \sem{\{(l_i : A_i)_{i∈I}\}}$
      by definition.

    \item[($+_i$)] We only need to prove $\sem{C\,t} ∈ \sem{[C \of A]∨γ₀}$
      according to the first induction hypothesis. It $\sem{κ} = 0$ for
      some $κ ∈ γ₀$ and if $t$ is weakly normal, then we can conclude
      immediately using Lemma~\ref{lem:strnorm}. Otherwise, we must
      have $\sem{κ} ≠ 0$ for all $κ ∈ γ₀$. Therefore, we can use the second
      induction hypothesis to get $\sem{t} ∈ \sem{A}$. From this we obtain
      $\sem{C\,t} ∈ \sem{[C \of A]}$ by saturation.

    \item[($+_e$)] We need to show $\sem{\case{t}{(C_i→t_i)_{i∈I}}} ∈
      \sem{B}$. By the first induction hypothesis, we have $\sem{t} ∈
      \sem{[(C_i:A_i)_{i∈I}]∧γ₀}$. If $\sem{κ} = 0$ for some $κ ∈ γ₀$
      then we obtain $\sem{t} ∈ \calWZS$, and thus  $\sem{\case{t}{(C_i
      →t_i)_{i∈I}}} ∈ \calWZS ⊂ \sem{B}$. Otherwise, the result follows
      from the right induction hypotheses and the definition of
      $\sem{[(C_i:A_i)_{i∈I}]}$.

    \item[($Y$)] By definition, we have $(Yx.t) ≻_w t[x:=Yx.t]$. As a
      consequence, the validity of the rule follows from the weak
      saturation condition (5) on $\sem{A}$.

    \item[($∀^o_l$)] If $\sem{t} ∈ \sem{∀α.A}$ then $\sem{t} ∈ \sem{A[α:=
      \sem{κ}]} = \sem{A[α:=κ]}$ by the substitution lemma. Hence, the
      induction hypothesis gives $\sem{t} ∈ \sem{B}$.

    \item[($∀^o_r$)] Let us suppose that $\sem{t} ∈ \sem{A}$ and assume
      $\sem{t} ∉ \sem{∀α.B}$ by contradiction. There must be $o ∈ \ordsem$
      such that $\sem{t} ∉ \sem{B[x:=o]}$. By definition of the choice
      operator, this means means that $\sem{t} ∉ \sem{B[x:=ε_{α<\OMaxi}(
      t ∉ B)]}$. We hence obtain a contradiction with $\sem{t} ∈ \sem{A}$
      using the induction hypothesis.

    \item[($∃^o_r$)] Similar to the ($∀^o_l$) case.

    \item[($∃^o_l$)] Similar to the ($∀^o_r$) case.

    \item[($∧_l$)] We assume that $\sem{t} ∈ \sem{A ∧ κ}$. If $\sem{κ} = 0$
      then $\sem{A ∧ κ} = \calWZS$ and hence $\sem{t} ∈ \sem{B}$. If
      $\sem{κ} ≠ 0$ then $\sem{A ∧ κ} = \sem{A}$ and we can thus conclude
      by induction hypothesis.

    \item[($∧_r$)] Since $κ ∈ γ$ we know that $\sem{κ} ≠ 0$ and thus we
      have $\sem{A ∧ κ} = \sem{A}$. We can thus conclude by induction
      hypothesis.

    \item[($∨_r$)] Similar to the ($∧_l$) case.

    \item[($∨_l$)] Similar to the ($∧_r$) case.
  \end{itemize}
  \vspace{-4.5mm}% HACK FIXME
\end{proof}

\begin{thm}\label{th:finalprops}
  As for the initial system, we get termination (typed terms are normalising
  for every weak reduction strategy), type safety for simple data types and
  consistency.
\end{thm}
\begin{proof}
  The proofs are similar to those of Theorems~\ref{th:strongnorm},
  \ref{th:safety} and \ref{th:consistency} respectively (using the
  results of the current section).
\end{proof}

%%% Local Variables:
%%% ispell-personal-dictionary: "~/.hunspell-en"
%%% ispell-local-dictionary: "british"
%%% End:

\section{Terminating examples}\label{applications}
We will now consider several examples of functions that are typable in our
system, and accepted by our implementation. We will start with examples on
lists, as the usual functions on unary natural numbers are not more difficult
to handle than the recursive identity function of Figure~\ref{fig:id}.

% Map, flatten and insertion sort.

\begin{figure}
  \begin{align*}
    \mathrm{F}(A,X) &= [\mathrm{Nil} \st \mathrm{Cons} \of \{\mathrm{hd} : A; \mathrm{tl} : X\}]\\
    \mathbb{L}_{α}(A) &= μ_{α}X.\mathrm{F}(A, X)\\
    \mathbb{L}(A) &= \mathbb{L}_∞(A)\\
    \mathrm{map} &\,: ∀A.∀B.∀α.(A → B) → \mathbb{L}_{α}(A) → \mathbb{L}_{α}(B)\\ &= Ymap.λf\, l.\casearray{l}{\begin{array}{l} [] → [] \\ x::l → f \; x{::}map \; f \; l\end{array}}\\
    \mathrm{map_2} &\,: ∀A.∀B.∀C.∀α.(A → B → C) → \mathbb{L}_{α}(A) → \mathbb{L}_{α}(B) → \mathbb{L}_{α}(C)\\ &= Ymap_2.λf\, l_1\, l_2.\casearray{l_1}{\begin{array}{l} [] → [] \\ x::l_1 → \casearray{l_2}{\begin{array}{l} [] → [] \\ y::l_2 → f \; x \; y{::}map_2 \; f \; l_1 \; l_2\end{array}}\end{array}}\\
    \mathrm{flatten} &\,: ∀A.\mathbb{L}(\mathbb{L}(A)) → \mathbb{L}(A)\\ &= Yflatten.λl_s.\casearray{l_s}{\begin{array}{l} [] → [] \\ l::l_s → \casearray{l}{\begin{array}{l} [] → flatten \; l_s \\ x::l → x{::}flatten \; (l{::}l_s)\end{array}}\end{array}}\\
    \mathrm{insert} &\,: ∀α.∀A.(A → A → \mathbb{B}) → A → \mathbb{L}_{α}(A) → \mathbb{L}_{α+1}(A)\\ &= Yinsert.λf\, a\, l.\casearray{l}{\begin{array}{l} [] → a{::}[] \\ x::l → \case{f \; a \; x}{ T → a{::}l ∣ F → x{::}insert \; f \; a \; l}\end{array}}\\
    \mathrm{sort} &\,: ∀α.∀A.(A → A → \mathbb{B}) → \mathbb{L}_{α}(A) → \mathbb{L}_{α}(A)\\ &= Ysort.λf\, l.\casearray{l}{\begin{array}{l} [] → [] \\ x::l → \mathrm{insert} \; f \; x \; (sort \; f \; l)\end{array}}
  \end{align*}
  \caption{Examples of functions on lists (map, flatten and insertion sort).}
  \label{fig:map}
\end{figure}

The type of lists of \emph{size} $α$ given at the top of Figure~\ref{fig:map}
is straight forward. It allows us to define the traditional $\mathrm{map}$ function,
which is decorated with the information that it preserves size. Note that its
type does not guarantee that the input and output lists have the same
size, but rather that the output list is at most as long as the input list.
More surprisingly, the $\mathrm{map_2}$ function can also be typed with some size
information. However, the type it is given here is not enough as it forbids
using $\mathrm{map_2}$ on input lists of unrelated sizes, while still preserving
size information about the result. A more precise and useful type for
$\mathrm{map_2}$ would require extending our syntactic ordinals with a $\min$
symbol. Indeed, we could then us the type $∀A.∀B.∀C.∀α.∀β, (A→B→C) →
\mathbb{L}_α(A) → \mathbb{L}_β(B) → \mathbb{L}_{\min(α,β)}(C)$.
Nonetheless, it is important to note that the types of $\mathrm{map}$ and $\mathrm{map_2}$
are subtypes of their usual type (with no size information). For example, we
can derive
$$
  ∀A.∀B.∀α.(A → B) → \mathbb{L}_{α}(A) → \mathbb{L}_{α}(B)
  ⊂ ∀A.∀B.(A → B) → \mathbb{L}(A) → \mathbb{L}(B)
$$
in our system. As a consequence, the $\mathrm{map}$ and $\mathrm{map_2}$ functions of
Figure~\ref{fig:map} are suitable for all applications. In particular, we
do not need to provide two different versions (one with size information,
and one without).

\begin{figure}
  \centering
  {\small $$ Yf.λl_s.\casearray{l_s}{\begin{array}{l} [] → [] \\ l::l_s → \casearray{l}{\begin{array}{l} [] → f \; l_s \\ x::l → x{::}Yg.λl_s.\casearray{l_s}{\begin{array}{l} [] → [] \\ l::l_s → \casearray{l}{\begin{array}{l} [] → f \; l_s \\ x::l → x{::}g \; (l{::}l_s)\end{array}}\end{array}} \; (l{::}l_s)\end{array}}\end{array}} $$}
  \begin{dot2tex}[dot,options=-tmath]
  digraph G {
    N0 [ label = "g" ];
    N1 [ label = "f" ];
    N0 -> N1 [label = "\left(\begin{smallmatrix}\infty&0&\infty\cr
0&\infty&\infty\end{smallmatrix}\right)"]
    N0 -> N0 [label = "\left(\begin{smallmatrix}0&\infty&\infty\cr
\infty&0&\infty\cr
\infty&-1&-1\end{smallmatrix}\right)"]
    N1 -> N1 [label = "\left(\begin{smallmatrix}0&\infty\cr
\infty&-1\end{smallmatrix}\right)"]
    N1 -> N0 [label = "\left(\begin{smallmatrix}\infty&-1\cr
0&\infty\cr
-1&\infty\end{smallmatrix}\right)"]
  }
\end{dot2tex}

  \begin{align*}
    &∀α_{0},α_{1} ( ⊢  \mathrm{f} : ∀A.\mathbb{L}_{α_{1}}(\mathbb{L}_{α_{0}}(A)) → \mathbb{L}(A))\\
    &∀β_{0},β_{1},β_{2} (β_{1} ⊢ β_{2}<β_{1}⇒ \mathrm{g} : F(\mathbb{L}_{β_{2}}(A_{0}), \mathbb{L}_{β_{0}}(\mathbb{L}_{β_{1}}(A_{0}))) → \mathbb{L}(A_{0}))
  \end{align*}
  \caption{Unrolling of flatten and the corresponding call graph.}
  \label{fig:flatten}
\end{figure}

We will now consider the $\mathrm{flatten}$ function, which is also given in
Figure~\ref{fig:map}. On this particular example, proving termination
requires unrolling the fixpoint twice. Indeed, if we only unroll it
once then our algorithm infers the general abstract sequent $∀α_{0},α_{1} ( ⊢  \mathrm{f} : ∀A.\mathbb{L}_{α_{1}}(\mathbb{L}_{α_{0}}(A)) → \mathbb{L}(A))$,
which is not sufficient for proving termination. However,
if we unroll the second recursive call twice we obtain two
different induction hypotheses, and the algorithm succeeds in proving
termination. This amounts to typing the program given at the top
of Figure~\ref{fig:flatten} using the abstract sequents given
at its bottom.
We will now give some explanations about the call graph of the function,
and in particular the size change matrices labeling its edges.
\begin{itemize}
  \item[($f$→$f$)] The loop on $f$ corresponds to the first recursive call.
    The $2×2$ matrix is justified because in this call the size of the
    inner list $α₀$ is constant, while $α₁$ decreases.

  \item[($f$→$g$)] The edge from $f$ to $g$ represents the definition of
    $g$ inside $f$, which must be seen as $f$ calling $g$. In this call,
    the first line of the matrix is justified by $β₀ < α₁$ because $β₀$
    is the size of the tail of the outer list. The second line is justified
    because $β₁$, the size of the inner list, is equal to $α₀$. The last
    line is justified because $β₂$, the size of the first element of the
    outer list decreases (it is smaller than $α₀$).

  \item[($g$→$g$)] The loop on $g$ corresponds to the last recursive call,
    where $β₀$ and $β₁$ are constant (which justifies the first two lines).
    The first element of the list is decreasing, so $β₂$ decreases.
    Moreover, as we keep in the general abstract sequent the information
    that $β₂ < β₁$, we also have a $-1$ in the middle of the last line.

  \item[($g$→$f$)] Finally, the edge from $g$ to $f$ corresponds to the
    third recursive call where we have $α₀ = β₁$, $α₁ = β₀$ and $β₂$
    become useless (hence the two $∞$ on the last column).
\end{itemize}
The size change principle yields a positive answer on this
call graph. This means that the typing derivation is well-founded, and
thus correct.

The last example given in Figure~\ref{fig:map} is insertion sort, for
which our implementation is able to derive both termination and size
preservation. The system is also able to derive the termination of quicksort
and merge sort, but in both cases we are unable to obtain size
preservation. However, it might be possible to obtain size preservation
on such a program by first enriching our language of syntactic ordinals
with an addition symbol for example. For instance, this would allow us
to give a precise type to the partition function required for quicksort.

% Append lists.

\begin{figure}
  \centering
  \begin{align*}
    \mathrm{AList}(A) &= μX.[\mathrm{Nil} \st \mathrm{Cons} \of \{\mathrm{hd} : A; \mathrm{tl} : X\} \st \mathrm{App} \of \{\mathrm{left} : X; \mathrm{right} : X\}]\\
    \mathrm{fromList} &\,: ∀X.\mathrm{List}(X) → \mathrm{AList}(X)\\ &= λl.l\\
    \mathrm{toList} &\,: ∀X.\mathrm{AList}(X) → \mathrm{List}(X)\\ &= YtoList.λl.\casearray{l}{\begin{array}{l} [] → [] \\ e::l → e{::}toList \; l \\ App\record{left=l;right=r} → \mathrm{append} \; (toList \; l) \; (toList \; r)\end{array}}
  \end{align*}
  \caption{Append lists as a supertype of lists.}
  \label{fig:applist}
\end{figure}

To illustrate the use of subtyping, a simple example implementing
\emph{append lists} is provided in Figure~\ref{fig:applist}. Roughly,
an append list is formed like a list, but an additional constructor
is provided for concatenation (we thus obtain constant time
concatenation). Thanks to subtyping, a list is an append list, and
thus the conversion function $\mathrm{fromList}$ is just the identity.
A recursive function $\mathrm{toList}$ is however required in the other
direction to effectively concatenate the lists contained in
$\textrm{App}$ nodes.

% Flows / streams.

\begin{figure}
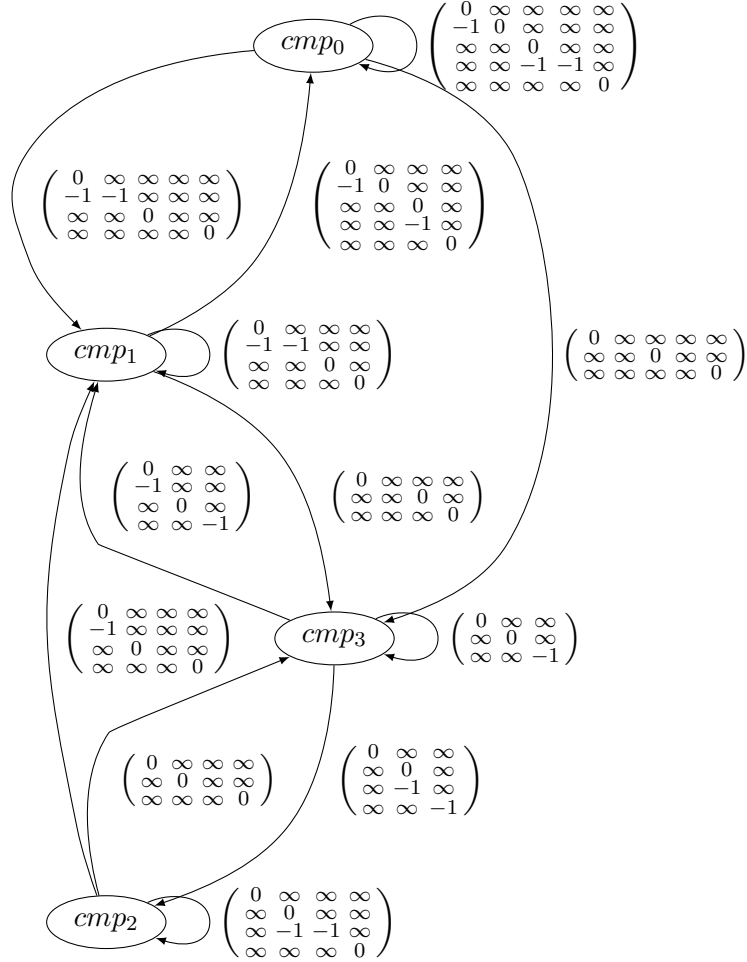

  \centering
  \begin{adjustbox}{max width=0.75\textwidth}
    \begin{dot2tex}[dot,options=-tmath]
  digraph G {
    N0 [ label = "cmp_{0}" ];
    N1 [ label = "cmp_{1}" ];
    N2 [ label = "cmp_{2}" ];
    N3 [ label = "cmp_{3}" ];
    N0 -> N3 [label = "\left(\begin{smallmatrix}0&\infty&\infty&\infty&\infty\cr
\infty&\infty&0&\infty&\infty\cr
\infty&\infty&\infty&\infty&0\end{smallmatrix}\right)"]
    N0 -> N1 [label = "\left(\begin{smallmatrix}0&\infty&\infty&\infty&\infty\cr
-1&-1&\infty&\infty&\infty\cr
\infty&\infty&0&\infty&\infty\cr
\infty&\infty&\infty&\infty&0\end{smallmatrix}\right)"]
    N0 -> N0 [label = "\left(\begin{smallmatrix}0&\infty&\infty&\infty&\infty\cr
-1&0&\infty&\infty&\infty\cr
\infty&\infty&0&\infty&\infty\cr
\infty&\infty&-1&-1&\infty\cr
\infty&\infty&\infty&\infty&0\end{smallmatrix}\right)"]
    N2 -> N3 [label = "\left(\begin{smallmatrix}0&\infty&\infty&\infty\cr
\infty&0&\infty&\infty\cr
\infty&\infty&\infty&0\end{smallmatrix}\right)"]
    N2 -> N1 [label = "\left(\begin{smallmatrix}0&\infty&\infty&\infty\cr
-1&\infty&\infty&\infty\cr
\infty&0&\infty&\infty\cr
\infty&\infty&\infty&0\end{smallmatrix}\right)"]
    N2 -> N2 [label = "\left(\begin{smallmatrix}0&\infty&\infty&\infty\cr
\infty&0&\infty&\infty\cr
\infty&-1&-1&\infty\cr
\infty&\infty&\infty&0\end{smallmatrix}\right)"]
    N1 -> N3 [label = "\left(\begin{smallmatrix}0&\infty&\infty&\infty\cr
\infty&\infty&0&\infty\cr
\infty&\infty&\infty&0\end{smallmatrix}\right)"]
    N1 -> N1 [label = "\left(\begin{smallmatrix}0&\infty&\infty&\infty\cr
-1&-1&\infty&\infty\cr
\infty&\infty&0&\infty\cr
\infty&\infty&\infty&0\end{smallmatrix}\right)"]
    N1 -> N0 [label = "\left(\begin{smallmatrix}0&\infty&\infty&\infty\cr
-1&0&\infty&\infty\cr
\infty&\infty&0&\infty\cr
\infty&\infty&-1&\infty\cr
\infty&\infty&\infty&0\end{smallmatrix}\right)"]
    N3 -> N3 [label = "\left(\begin{smallmatrix}0&\infty&\infty\cr
\infty&0&\infty\cr
\infty&\infty&-1\end{smallmatrix}\right)"]
    N3 -> N1 [label = "\left(\begin{smallmatrix}0&\infty&\infty\cr
-1&\infty&\infty\cr
\infty&0&\infty\cr
\infty&\infty&-1\end{smallmatrix}\right)"]
    N3 -> N2 [label = "\left(\begin{smallmatrix}0&\infty&\infty\cr
\infty&0&\infty\cr
\infty&-1&\infty\cr
\infty&\infty&-1\end{smallmatrix}\right)"]
  }
\end{dot2tex}
  \end{adjustbox}
  \caption{Call-graph of the $\mathrm{cmp}$ function.}
  \label{fig:cmp}
\end{figure}

\begin{figure}
  \begin{align*}
    \mathbb{S}(A) &= νX.(\{\} → A × X)\\
    \mathbb{F}_{α} &= ν_{α}X.μY.(\{\} → [\mathrm{R} \of Y \st \mathrm{K} \of X])\\
    \mathbb{F} &= \mathbb{F}_∞\\
    \mathrm{filter} &\,: ∀A.\mathbb{F} → \mathbb{S}(A) → \mathbb{S}(A)\\ &= Yfilter.λf\, s.(λ(h,t).\casearray{f \; ()}{\begin{array}{l} Rf' → filter \; f' \; t \\ Kf' → λu.(h, filter \; f' \; t)\end{array}}) \; (s \; ())\\
    \mathrm{cmp} &\,: \mathbb{F} → \mathbb{F} → \mathbb{F}\\ &= Ycmp.λf_1\, f_2\, u.\casearray{f_2 \; ()}{\begin{array}{l} Kf_2' → \casearray{f_1 \; ()}{\begin{array}{l} Kf_1' → K (cmp \; f_1' \; f_2') \\ Rf_1' → R (cmp \; f_1' \; f_2')\end{array}} \\ Rf_2' → R (cmp \; f_1 \; f_2')\end{array}}\\
    \mathrm{f2s} &\,: ∀α.\mathbb{F}_{α} → \mathbb{S}_{α}(\mathbb{N})\\ &= Yf2s.λs\, u.\casearray{s \; ()}{\begin{array}{l} Rs → (λ(n,r).(S n, r)) \; (f2s \; s \; ()) \\ Ks → (Z, f2s \; s)\end{array}}\\
    s2f_{aux} &\,: ∀α.\mathbb{F}_{α} → \mathbb{N} → \mathbb{F}_{α+1}\\ &= Ys2f_aux.λs\, n.\casearray{n}{\begin{array}{l} Z → λu.K s \\ Sp → λu.R (s2f_aux \; s \; p)\end{array}}\\
    \mathrm{s2f} &\,: ∀α.\mathbb{S}_{α}(\mathbb{N}) → \mathbb{F}_{α}\\ &= Ys2f.λs\, u.(λ(n,s).s2f_{aux} \; (s2f \; s) \; n \; ()) \; (s \; ())
  \end{align*}
  \caption{Examples with streams and filters on streams.}
  \label{fig:streams}
\end{figure}

To conclude this section, we will now give an example mixing inductive
and coinductive types. We consider the type of streams $\mathbb{S}(A)$
and the type of filter on streams $\mathbb{F}$ defined at the top of
Figure~\ref{fig:streams}. In the type of filters, the variant $R$
indicates that one element of the stream should be \emph{removed}, while
the variant $K$ indicates that one element should be kept. Note that
in the type $\mathbb{F}$, the inner type $μY.(\{\} → [\mathrm{R} \of Y \st \mathrm{K} \of \mathrm{X}])$
imposes that we can only have finitely many $R$ constructors between
$K$ constructors. As a filter must contain infinitely many $K$
constructors, this ensures the productivity of the $\mathrm{filter}$ function,
applying a filter to a stream, and the $\mathrm{cmp}$ function composing two
filters.

As in the example of the $\mathrm{flatten}$ function on lists, both
$\mathrm{filter}$ and $\mathrm{cmp}$ require some unrolling. To avoid this, we may
replace the type $\mathbb{F}$ with $\begin{array}{ll}\mathbb{F}' &= μY.(\{\} → [\mathrm{R} \of Y \st \mathrm{K} \of \mathbb{F}])\end{array}$.
Indeed, although $\mathbb{F} ⊂ \mathbb{F}'$ and $\mathbb{F}' ⊂ \mathbb{F}$ are both
derivable, $\mathbb{F}'$ carries an ordinal representing the initial number
of $R$ constructors in the type. The call-graph for $\mathrm{cmp}$ is given
in Figure~\ref{fig:cmp} and gives an example of a non trivial instance
of the size change principle.

Note also that $\mathbb{F}$ is isomorphic to the type of streams over
natural numbers, and that we can prove the termination of this
isomorphism while keeping size information about the streams. The
isomorphism is given by the $\mathrm{s2f}$ and $\mathrm{f2s}$ functions.

More examples are provided with the implementation of our prototype
\cite{proto}. They contain, for example, the GCD function for binary
natural numbers, and the basic operations for exact real arithmetic
(using the signed digits representation). In particular, all of these
examples are proved terminating by our implementation.

%%% Local Variables:
%%% ispell-personal-dictionary: "~/.hunspell-en"
%%% ispell-local-dictionary: "british"
%%% End:

\section{Type-checking Algorithm}\label{algorithm}
Our system can be implemented by transforming the deduction rule systems
given in this paper into recursive functions. This can be done relatively
easily because the system is mostly syntax-directed. For instance, only
one typing rule applies for each term constructor, and at most two
subtyping rules apply for each pair of type constructors.
It is easy to see that when two subtyping rules may apply (one left rule
and one right rule), then they commute (e.g., quantifier rules). This is
due to the fact that they
do not modify the term carried by the judgment, and that choice operators
are constructed using only the term and the type on the side where it is
applied. This means that the order in which such rules are applied does
not matter.
Moreover, if the rule for implication, product or sum can be applied, then
it is easy to see that no other rule can be applied (except generalisation).

Another important remark about the system is that if we limit the unrolling
depth for fixpoints in typing rules, then the only possible place where an
implementation may loop is in the subtyping function. Indeed, every typing
rule (except fixpoint unrolling) decreases the size of the term, if we
consider choice operators to have size zero (we will come back to this
point when we discuss type errors).

Nonetheless, several subtle details need further discussion. We will here
give some guidelines explaining parts of our implementation. We encourage
the reader to look at the code of our prototype \cite{proto}, which should
be relatively accessible (at least to readers familiar with the
implementation of type systems).
According to the previous remarks, the only implementation freedom is in
the management of the rules introducing unknown types or ordinals (namely
($∀_l$), ($∃_r$), ($∀_l^o$), ($∃_r^o$), ($μ_r$) and ($ν_l$)), in the
management of the ordinal contexts with the $A∧γ$ and $A∨γ$ connective,
and in the construction of circular typing and local subtyping proofs.

\subsection*{Unification variables.}

For handling unknown types and ordinals in subtyping, the natural solution
is to extend their syntax with a set of unification variables. In types,
we will use the letters $U$ and $V$ to denote unification variables, which
correspond to unknown types until their value is inferred. In our prototype
implementation \cite{proto}, unification variables are handled as follows.
\begin{itemize}
  \item If we encounter $\subtype{γ}{t}{U}{U}$ then we use reflexivity.
  \item If we encounter $\subtype{γ}{t}{U}{V}$, then we set $U := V$.
  \item If we encounter $\subtype{γ}{t}{U}{A}$ or $\subtype{γ}{t}{A}{U}$, then
        we decide that $U$ is equal to $A$, provided that it does not occur
        in $A$. Note that it is essential to check occurrence inside choice
        operators for them to be well-defined (i.e., not cyclic). Moreover,
        when $U$ occurs only positively in $A$ we may use $μX.A[U:=X]$ as a
        definition for $U$, thus allowing the system to build new recursive
        types.
\end{itemize}
In fact, this approach is a bit too naive in the case where we have a
projection $t.l_k$ and the type of $t$ is a unification variable. Indeed,
it is usually not sufficient to fix the type of $t$ to be a record type
with only the field $l_k$ (the dual problem arises with variants). To solve
this issue, our unification variables carry a state keeping track of
projected fields (or constructed variants). The state of a unification
variable is initialised or updated when we encounter
$\subtype{γ}{t}{U}{\{ l₁: A₁, \dots, l_n:A_n, \dots\}}$ or
$\subtype{γ}{t}{[ C₁ \of A₁, \dots, C_n \of A_n]}{U}$. This state can
be seen as a subtyping constraint (upper bound for record types,
lower bound for variant types) which is delayed until we have a
subtyping constraint on the other side.

Unification variables are also required for syntactic ordinals to handle
the ($μ_r$), ($ν_l$), ($∀_l^o$) and ($∃_r^o$) rules. In syntactic ordinals,
we will use the letters $O$ and $P$ to denote unifications variables. As
for types, an ordinal unification variable $O$ may carry constraints like
$τ ≤ O < κ$, to delay instantiation until we have a constraint $O ≤ κ$.
Moreover, when we need to prove $\subtype{γ}{t}{A}{μ_O F}$ or
$\subtype{γ}{t}{ν_O F}{B}$ and $O$ is a unification variable, we define
$O$ to be the first ordinal in $γ$ satisfying the constraints on $O$. If
there is none, then we instantiate it with the successor of a unification
variable or with $∞$. We do this because we must fail if there is no
positive solution for $O$. Otherwise, the subtyping procedure would often
loop by building decreasing chains of unification variables.

\subsection*{Circular subtyping proofs.}

The generalisation rule used to build circular proof is the only one that
is not directed by the syntax (or handled by unification variables). As a
consequence, it cannot be implemented directly and requires a special
treatment. In practice, we try to apply the generalisation rule to build
an induction hypothesis each time we encounter a local subtyping judgment
with an inductive or coinductive types on either side. In such an
eventuality, we apply the generalisation rule ($\GP$) by quantifying over
all the ordinals appearing in the types. The produced general abstract
sequent is then looked up in the list of all the encountered induction
hypotheses in an attempt to end the branch of the proof by induction. If
the general abstract sequent has not been encountered before, then it is
registered and the proof proceeds by applying the $\IP{k}$ rule.

Note that when there are no quantifiers, only a finite number of distinct
general abstract sequents can be produced, thus implying the termination
of our algorithm. Indeed, when when proving a subtyping judgement
$\subtype{γ}{t}{A}{B}$, the formulas that appear in the proof can be
uniquely identified by a pointer to a subformula of the original types
$A$ or $B$, and the value of the ordinals. When building a general abstract
sequent, the ordinals are quantified over, and hence the general abstract
sequent only depends on two pointers (for the involved types).
This means that the number of distinct general abstract sequents appearing
in a proof of $\subtype{γ}{t}{A}{B}$ is less than $|A|×|B|$ (where $|C|$
denotes the size of the type $C$). This property is similar to the
finiteness of Kozen's closure for the propositional $μ$-calculus
\cite{kozen}.
When quantification over types is allowed, subtyping may loop by
instantiating unification variables with different types each time a
given quantifier is eliminated. This does not happens very often in
practice.

\subsection*{Circular typing proofs.}

The construction of circular typing proofs follows the same principle as
for circular subtyping proofs. We create a general abstract sequent each
time we encounter a fixpoint $Yx.t$, check whether it was already
encountered before to end the proof, and if not we register the new
hypothesis and continue the proof.
Note however that the generalisation we preform for typing proofs is a
bit more subtle. Indeed, if the type of $Yx.t$ does not contain any
explicit quantifier on ordinals, we generalise infinite ordinals by
decorating negative occurrences of types of the form $μX.A$ (and
positive occurrences of types of the form $νX.A$). For example, this
means that the sequent $\type{Yx.t}{μX A → νY μZ B}$ is generalised to
$∀α ∀β \type{Yx.t}{μ_{α}X A → ν_{β}Y μZ B}$.
However, when the type uses ordinal quantifiers we do not generalise
infinite ordinals and only generalise ordinal variables (as for
subtyping), assuming the given type already carries the proper ordinal
annotation. In other words, if the user has not given explicit size
information in the type of a program, then the first generalisation
will have the effect of eliminating certain occurrences of $∞$,
intuitively replacing them with a smaller, finite ordinal.

\subsection*{Breadth-first search for typing fixpoint.}

As explained in the previous section, unrolling a fixpoint more than
once is often necessary for building typing proofs. When mixed with
unification, this requires a breadth-first proof search strategy. This
means that when typing $Yx.t$, we first finish all the other branches
of the proof, collecting as much as possible information about the
type of $Yx.t$. By doing so, our experimentations have shown that we
have more chances to instantiate unification variable in the expected
way.

To implement the breadth-first strategy we first apply all the typing
rules on the considered term, by delaying all the applications of the
($Y$) rule. In other words, we simply store the typing sequents
corresponding to the ($Y$) rule in a list. We then iterate through all
the stored sequents and first try to apply a possible induction
hypothesis (there are none at the first stage of the search). For all
the remaining sequents we perform a generalisation (as explained above)
and store the general abstract sequent as an
induction hypothesis. Finally, the next stage of breadth-first
search can be launched. It consists in proving all the generalised
sequents by first applying the $\I{k}$ rule on them.

\subsection*{Generalisation and unification variables.}

In practice, the presence of unification variables in general abstract
sequents often leads to failure or non-termination.
Therefore, we instantiate constrained unification variables 
using their own constraints when we generalise a sequent to form a
general abstract sequent. In particular, we fix type unification
variables according to the set of variant constructors or record
fields they carry in their states, and we instantiate ordinal
unification variables with their lower bounds.

Nonetheless, unification variables that are not constrained are still
kept in general abstract sequents. In this case, we need to introduce
second order unification variables that may depend on the value of
generalised ordinals. This is required as otherwise the unification
variables would not be able to use the ordinals that are quantified
over by the generalisation.
For example, if a unification variable $U$ occurs in a sequent
$\type{Yx.t}{μX.A → νY. μZ. B}$, then we introduce a new second
order unification variable $V$ with two ordinal parameters.
The general abstract sequent is then
$∀α ∀β \type{Yt}{(μ_{α}X.A→ν_{β}Y.μZ.B)[U:=V(α,β)]}$, and $U$ is
instantiated with $V(∞,∞)$.
Second order unification variables are dealt with in a very simple way,
using projection when possible and imitation (i.e. constant value) when
projection is not possible. For example, if we need to solve a constraint
$γ ⊢ V(τ,κ) ≤ τ$ then we will only try to set $V$ to the first projection
and hence $V(τ,κ) = τ$.

\subsection*{Dealing with type errors}

In our implementation, there are two different kinds of type errors:
\emph{clashes} which immediately stop the proof search, and loops that
can be interrupted by the user. As only subtyping may loop, we can
display the last encountered typing judgment in both cases, as well as
the subtyping instance that failed to be proved. We can thus obtain a
message like ``$t$ has type $A$ and is used with type $B$''.

For readability, it is important to note that it is never required to
display choice operators in full. Indeed, we can limit ourselves to the
name of the variable they bind, and the position of the variable it was
substituted to in the source code. Note however that the error messages
of the current prototype are not optimal. They have been optimised for
the debugging of the prototype itself rather than for debugging programs
written using the prototype. We believe that we could improve error
messages for it to be as easy (or as difficult) to debug type errors with
our algorithm than with mainstream ML implementations. However, proving
termination requires an extra effort for advanced examples.

%%% Local Variables:
%%% ispell-personal-dictionary: "~/.hunspell-en"
%%% ispell-local-dictionary: "british"
%%% End:

\section{Type annotations and dot notation.}

\begin{figure}
  \centering
  \begin{align*}
    \mathrm{C}(O,M) &= \{\mathrm{dom} : M → O; \mathrm{cod} : M → O; \mathrm{cmp} : M → M → M\} \\
    \mathrm{Cat} &= ∃O.∃M.\mathrm{C}(O, M) \\
    \mathrm{dual} &: \mathrm{Cat} → \mathrm{Cat}\\
      &= λc.\left\{\setlength{\arraycolsep}{0.2em}\begin{array}{ll}\mathrm{dom} : c.M → c.O &= c.cod;\\
\mathrm{cod} : c.M → c.O &= c.dom;\\
\mathrm{cmp} : c.M → c.M → c.M &= λx\, y.c.cmp \; y \; x\end{array}\right\}\\
    \mathrm{dual2} &: \mathrm{Cat} → \mathrm{Cat}\\
      &= λc.\begin{array}[t]{l}\LET O,M \ST c:\mathrm{C}(O, M) \IN\\ \left\{\setlength{\arraycolsep}{0.2em}\begin{array}{ll}\mathrm{dom} : M → O &= c.cod;\\
\mathrm{cod} : M → O &= c.dom;\\
\mathrm{cmp} : M → M → M &= λx\, y.c.cmp \; y \; x\end{array}\right\}\end{array}
  \end{align*}
  \caption{Example involving dot projection (dual category).}
  \label{fig:dualcat}
\end{figure}

Using the guidelines provided in the previous section, it is possible to
build a satisfactory implementation. However, since the system is likely
to be undecidable, we need to provide a way of annotating complex programs.

As we are considering a Curry style language, type annotations are not
completely natural. Simple type coercions like $t : A$ can be added to
the system without difficulty using the following rule.
\begin{prooftree}
\AxiomC{$\type{t}{A}$}
\AxiomC{$\subtype{}{t}{A}{B}$}
\BinaryInfC{$\type{t : A}{B}$}
\end{prooftree}
However, such type annotations are often required to reference bound
type variables, and a type abstraction constructor $ΛX.t$ is only
natural in Church style calculi. A simple idea to solve the annotation problem
in Curry style is to write annotations like the following.
$$\LET \vec{X} \ST x : A(\vec{X}) \IN t$$
They allow the user to name a type (most of the time a choice operator)
by pattern matching the type of the bound variable $x$. During type
checking, $x$ is replaced by a choice operator which carries its type
$T$. It is thus possible to pattern match $T$ against $A(\vec{X})$ to
obtain the value of the variables of $\vec{X}$ (this is relatively simple
to implement). For example, a fully annotated identity function can be
written as follows.
$$λx. \LET X \ST x:X \IN x:X.$$
%Although this syntax may seem strange at first, it is extremely convenient
%in practice. Note that we also allow the user to name ordinals and to give
%the type of the whole term instead of the type of a variable. For example,
%we can write the following.
%$$\LET \vec{α},\vec{X} \ST \_ : A(\vec{α},\vec{X}) \IN t$$

Moreover, this kind of annotations may be used to define dot notation on
existential types. It may be used to replace the usual dot notation for
abstract types. Indeed, if a $λ$-variable $x$ has type
$∃X ∃Y A(X,Y)$ then we can access $X$ and $Y$ using the following.
$$\LET X,Y \ST x : A(X,Y) \IN t$$
As we use local subtyping when matching type, the implementation can
easily search $X₀$ and $Y₀$ such that $γ ⊢ t ∈ A(X₀,Y₀) ⊂ ∃X ∃Y A(X,Y)$.
This will leads to $X₀ = ε_X t : ∃Y A(X,Y)$ and $Y₀ = ε_Y t : A(X₀,Y)$.
Yet, this notation style is too heavy and in this particular case, we
prefer writing $x.X$ and $x.Y$, which rely on the name of bound variables
to build the same witnesses as above from the type of $x$, or more
precisely from the type of the term witness that will be substituted to
$x$. It is important to remark that the implementation never needs to
rename a bound variable because we substitute closed terms, types or
ordinals to variables and renaming is never necessary in this case.
As an example, we can define a type for categories using two abstract
types $O$ and $M$ for objects and morphisms. We can then use
both ways to annotate the definition of a function ``$\mathrm{dual}$''
computing the opposite of a category (see Figure~\ref{fig:dualcat}).

Note that the syntactic sugar defined here for dot notation is limited
as it only applies to variables. A more general dot notation such as
$(f\;t).X$ would be more difficult to obtain (in particular in presence
of effects), because it denotes a type that may contain a computation.
Nonetheless, it is always possible to name $f\;t$ using a let-binding.

%%% Local Variables:
%%% ispell-personal-dictionary: "~/.hunspell-en"
%%% ispell-local-dictionary: "british"
%%% End:

\section{Perspectives and Future Work}
Our experiments show that our framework based on system F, subtyping,
circular proofs and choice operators is practical and can be implemented
easily. However, a lot of work remains to explore combinations of our system
with several common programming features and to transform it into a
real programming language.

\subsection*{Higher-order types.}
In our system, only types and ordinals can be quantified over.  We had
to introduce second order unification variables and the implementation
might be more natural with higher-order types.  The main difficulty
for extending our system to higher-order is purely practical. The
handling of unification variables needs to be generalised into a form
of higher-order pattern matching. However, our system allows us to
avoid computing the variance of higher-order expressions (which is not
completely trivial), thanks to the absence of syntactic covariance
condition on our inductive and coinductive types.

\subsection*{Dependent types and proofs of programs.}
One of our motivations for this work is the integration of subtyping
to the realisability models defined in a previous work by Rodolphe
Lepigre~\cite{lepigre}. To achieve this goal, the system needs to be
extended with a first-order layer having terms as individuals. Two new
type constructors $t ∈ A$ (singleton types) and
$A \restriction t ≡ u$ (meaning $A$ when $t$ and $u$ are
observationally equal and $∀X.X$ otherwise) are then required to
encode dependent products and program specifications. These two
ingredients would be a first step toward program proving in our
system.

\subsection*{Extensible sums and products.}
The proposed system is relatively expressive, however it lacks
flexibility for records and pattern-matching. A form
of inheritance allowing extensible records and sums is desirable.
Moreover, features like record opening are required to recover the
full power of ML modules and functors. We also expect that such a feature
will allow for a better type inference, and thus simplify the
development of complex programs.

\subsection*{Completeness without quantifiers.}

Our algorithm seems terminating for the fragment without $∀$ and $∃$
quantifiers. We are actually able to prove its completeness if we also
remove the function type, but a few problems remain when dealing with
arrow types, mainly the mere sense of completeness. Various
possibilities exist, for instance depending if we want to have
$⊢ A ⊂ ([] → B)$ for any types $A$ and $B$.

\subsection*{A larger complete Subsystem.}

If we succeed in proving the completeness of the fragment of the system
without quantifiers, the next step would be to see if we can gain
completeness with some restriction on quantification (like ML style
polymorphism).
More generally, the cases leading to non-termination of subtyping
should be better understood to avoid it as much as possible and
try to produce better error messages when the system is interrupted.

%%% Local Variables:
%%% ispell-personal-dictionary: "~/.hunspell-en"
%%% ispell-local-dictionary: "british"
%%% End:

\bibliographystyle{plain}
\bibliography{biblio.bib}

%%% Local Variables:
%%% ispell-personal-dictionary: "~/.hunspell-en"
%%% ispell-local-dictionary: "british"
%%% End:

\end{document}